\documentclass[twocolumn]{autart}    

\usepackage{graphicx}          
\usepackage[T1]{fontenc}
\usepackage{arydshln}
\usepackage{tikz}
\usepackage[english]{babel}
\usepackage{amsmath,amssymb,amsfonts}
\usepackage{algorithm}
\usepackage[algo2e,lined,norelsize]{algorithm2e}
\usepackage{dsfont}
\usepackage[dvipsnames]{xcolor}

\newtheorem{theorem}{Theorem}[section]

\newtheorem{proposition}{Proposition}[section]
\newtheorem{remark}{Remark}[section]

\begin{document}
	
	\begin{frontmatter}
		
		\title{\vspace{-8mm}Network-Realised Model Predictive Control\\ Part II: Distributed Constraint Management\\{\color{white}(Andrei's Version)}\vspace{-12mm}} 
		
		
		\author[UPB]{Andrei Speril\u a}\ead{andrei.sperila@upb.ro},    
		\author[L2S]{Alessio Iovine}\ead{alessio.iovine@centralesupelec.fr},               
		\author[L2S]{Sorin Olaru}\ead{sorin.olaru@centralesupelec.fr},  
		\author[L2S]{Patrick Panciatici}\ead{patrick.panciatici@ieee.org}  
		
		\address[UPB]{Department of Automatic Control and Systems Engineering, ``Politehnica'' University of Bucharest, Bucharest, Romania\vspace{-1mm}}
		\address[L2S]{Laboratoire des Signaux et Syst\`emes, CentraleSup\'elec, Paris-Saclay University, Gif-sur-Yvette, France\vspace{-9mm}}  

		\begin{keyword}                           
			Distributed control; convex optimisation; model predictive control; networked systems; scalable implementations.   \vspace{-2mm}               
		\end{keyword}                             

		\begin{abstract}                          
			A two-layer control architecture is proposed, which promotes scalable implementations for model predictive controllers. The top layer acts as both a reference governor for the bottom layer and as a feedback controller for the regulated network. By employing set-based methods, global theoretical guarantees are obtained by enforcing local constraints upon the network's variables and upon those of the first layer's implementation. The proposed technique offers recursive feasibility guarantees as one of its central features, and the expressions of the resulting predictive strategies bear a striking resemblance to classical formulations from model predictive control literature, allowing for flexible and easily customisable implementations.\vspace{-5mm}
		\end{abstract}
		
	\end{frontmatter}
	
	{\color{black}
		
		\section{Introduction}\label{sec:intro}\vspace{-3mm}
		
		\subsection{Setting and Context}\vspace{-3mm}
		
		Model Predictive Control (MPC, see \cite{RMD} and also \cite{BBM}) is one of the most popular control-based frameworks in both the industrial setting and in system-theoretical literature. One of the key features that has generated its current popularity is its affinity for enforcing set-based constraints in the centralised setting \cite{STMC}. On the other hand, as technical applications become increasingly interconnected and widely distributed, the synthesis of distributed control laws has emerged as a focus of contemporary research in the control community.\vspace{-3mm}
		
		Among the myriad approaches developed for the MPC paradigm (see \cite{DMPCB} for a comprehensive survey on the topic), some of the most promising solutions are those inspired by robust MPC (see \cite{RMPC} for a popular centralised solution), among which we highlight the techniques proposed in \cite{R8A,R8B,R8C,R8D} for the distributed setting. A key aspect of this particular framework is given by the use of an ancillary control law, based upon static state feedback, which operates in tandem with its MPC policy.\vspace{-3mm}
		
		Distinct from the latter approach, the procedure formalised in \cite{DMPC1,DMPC2} leverages the powerful design framework from \cite{SLS} to address the intricacies of the distributed setting. One such intricacy, which arises chiefly in formation control, is the problem of attenuating the propagation of disturbance through said formation (see \cite{SS_rev}). This problem has received a great deal of attention in MPC-oriented literature, with multiple solutions having been proposed in both the distributed setting \cite{plat_MPC_1,plat_MPC_2,plat_MPC_3,plat_MPC_4}, and also in the centralised one \cite{plat_MPC_5}.\vspace{-3mm}
		
		\subsection{Motivation}\vspace{-4mm}
		
		Despite the valuable contributions made in the previously cited works, certain limitations inherent in these approaches serve as the chief motivation for the developments presented in this paper. More precisely, the control laws proposed in \cite{R8A,R8B,R8D} notably require a set of \emph{static and stabilising} state-feedback strategies, which are constrained to having block-diagonal sparsity structures. Furthermore, the procedure described in \cite{R8C} revolves around intricate set-theoretical notions, such as \emph{joint set invariance}, while implicitly assuming that the controlled network is bipartite and that the state matrix of its (linear) dynamics satisfies a restrictive spectral condition. Finally, the design framework presented in \cite{DMPC1,DMPC2} tackles both dynamical disturbance decoupling and distributed constraint satisfaction jointly, \emph{in a single design phase}, while the formation control-oriented approaches in \cite{plat_MPC_1,plat_MPC_2,plat_MPC_3,plat_MPC_4,plat_MPC_5} require \emph{restrictive or computationally demanding} procedures to reach their goals.\vspace{-4mm}
		
		\subsection{Scope of Work}\vspace{-4mm}
		
		In contrast to the techniques from \cite{R8A,R8B,R8C,R8D}, the MPC-based control laws discussed in this paper do not focus on network regulation in the presence of constraints, but rather on \emph{closed-loop supervision} of both the plant and its control laws, which based are upon the Network Realisation Function (NRF, see the companion paper \cite{part1}, along with \cite{NRF}) representation. In our framework, disturbance rejection and reference tracking are handled predominantly by the distributed (and dynamical) controllers discussed in \cite{part1}. By exploiting the closed-loop properties ensured by these NRF-based subcontrollers \cite{plutonizare,NRF,aug_sparse}, the MPC-inspired procedures described in the sequel are more akin to distributed versions of the so-called reference governors (see \cite{refgov} for a broad discussion of these policies). Indeed, a careful inspection of the survey presented in \cite{arch_rev} reveals that our framework does not neatly fit into any of the architectures addressed therein.\vspace{-4mm}
		
		\subsection{Contribution}\vspace{-3mm}
		
		From a system-theoretical perspective, the chief contributions of the work presented in this paper can be listed via the following three features:\vspace{-3mm}
		\begin{enumerate}
			\item[a)] ensuring command constraint satisfaction, in an MPC-based framework, for a plant in closed-loop configuration with a \emph{dynamical controller};\vspace{0.5mm}
			
			\item[b)] coordinating the control authority of our MPC-based policies with the NRF-based controllers from \cite{NRF,aug_sparse}, while splitting the MPC's actuation efforts between the latter and the controlled network;\vspace{0.5mm}
			
			\item[c)] enabling constraint-compliant and NRF-based control, in which the MPC-based subcontrollers are designed to act only as a safeguard mechanism against constraint violation.
		\end{enumerate}\vspace{-3mm}
		A further point to emphasise is the fact that all of the above features are achieved in a fully \emph{distributed setting}. Moreover, we point out that the mechanism by which a dynamical controller's output is reliably constrained (for command limitation) has not, to the best of our knowledge, been adequately addressed in system-theoretical literature \emph{even in the centralised setting}.
		
		\vspace{-3mm}
		
		\subsection{Paper Structure}\vspace{-3mm}
		
		The rest of the paper is structured as follows. Section~\ref{sec:prelims} contains a set of preliminary notions, while Section~\ref{sec:arch} discusses our two-layer architecture and its overarching objectives, focusing on the MPC layer. Section~\ref{sec:transform} presents the means of obtaining tractable prediction models for an area-partitioned control scheme, and Section~\ref{sec:MPC_des} handles the technicalities of constraint management in the distributed setting. Section~\ref{sec:MPC_imp} holds the main theoretical results and implementation procedures, while Section~\ref{sec:examp} showcases a numerical example based upon the vehicle platooning application discussed in \cite{plutonizare} and Section~\ref{sec:outro} offers a set of concluding remarks. In addition to this, we include a pair of appendices: one focusing on the key concepts of NRF-based control laws (Appendix~\ref{app:NRF}), as discussed in our companion paper \cite{part1}, and one which holds the proofs of this paper's main results (Appendix~\ref{app:proofs}).\vspace{-3mm}
		
		We now move on to introducing the various preliminary notions which are necessary to state our main results.\vspace{-3mm}
		
	}

	\section{Preliminaries}\vspace{-3mm}
	\label{sec:prelims}
	
	\subsection{Notation and Definitions}\vspace{-3mm}
	\label{subsec:not}
	
	Let $\mathbb{N}$, $\mathbb{R}$ and $\mathbb{C}$ denote, respectively, the set of natural, real and complex numbers. Let $\mathcal{M}^{p\times m}$ be the set of all $p\times m$ matrices whose entries are scalar elements that are part of a set denoted $\mathcal{M}$. Similarly, $\mathcal{M}^{p}$ denotes the set of vectors with dimension $p$ and entries in $\mathcal{M}$. For any matrix $M$, $\mathrm{row}_i(M)$ is its $i^\text{th}$ row, $\mathrm{col}_j(M)$ stands for its $j^\text{th}$ column and $\mathrm{elm}_{ij}(M):=\mathrm{row}_i(\mathrm{col}_j(M))$. Moreover, the \emph{transpose} of an arbitrary matrix $M$ is denoted $M^\top$. We denote by $\mathcal{R}(z)$ the set of real-rational functions of indeterminate $z\in\mathbb{C}$. A matrix with entries in $\mathcal{R}(z)$ is called a Transfer Function Matrix (TFM), and all such TFMs will be denoted using boldface letters.\vspace{-3mm}
	
	The vector $e_i$ stands for the $i^\text{th}$ column belonging to the identity matrix of compatible dimensions. For any matrix $M\in\mathbb{R}^{p\times m}$ and any set $\mathcal{X}\subseteq\mathbb{R}^m$, we define $M\mathcal{X}:=\{Mx\ \vert\ x\in\mathcal{X}\}$ and $(-\mathcal{X}):=\{-x\ \vert\ x\in\mathcal{X}\}$. In addition to this, for any two sets $\mathcal{X}_1\subseteq\mathbb{R}^m$ and $\mathcal{X}_2\subseteq\mathbb{R}^m$, we will henceforth denote their Minkowski sum as follows $\mathcal{X}_1\oplus\mathcal{X}_2:=\{x_1+x_2\ \vert\  x_1\in\mathcal{X}_1,\ x_2\in\mathcal{X}_2\}$, while the operation $\mathcal{X}_1\ominus\mathcal{X}_2:=\{y\in\mathbb{R}^m\ \vert\ \{y\}\oplus\mathcal{X}_2\subseteq\mathcal{X}_1\}$ will represent the Pontryagin difference between the two sets.\vspace{-3mm}
	
	\subsection{System Representations}
	\label{subsec:sys_nots}\vspace{-3mm}
	
	The class of systems considered in this paper are represented in the time domain by the set of equations\vspace{-2mm}
	\begin{subequations}
		\begin{align}
			x[k+1] =&\ Ax[k]+B_u\, u[k]+B_d\,d[k],\label{eq:ss_a}\\
			y[k]=&\ Cx[k]+D_u\, u[k]+D_d\,d[k]\label{eq:ss_b},
		\end{align}
	\end{subequations}\phantom{ }\vspace{-8mm}
	
	for any \textcolor{black}{$k\in\mathbb{Z}$ with $k\geq k_0\in\mathbb{Z}$}, where $k_0$ represents the initial time instant. For the representations of type \eqref{eq:ss_a}-\eqref{eq:ss_b}, which are referred to as \emph{state-space realisations}, $u$ is the realisation's \emph{controlled input vector}, $d$ its \emph{disturbance vector}, $y$ its \emph{output vector} and $x$ its \emph{state vector}.\vspace{-3mm}
	
	{\color{black}
		The type of network that our procedure aims to control is described by a more particular class of realisations than \eqref{eq:ss_a}-\eqref{eq:ss_b}. These systems also satisfy the fact that\vspace{-3mm}
		\begin{equation}\label{eq:ss_c}\tag{1c}
			C=I,\ D_u = O,\ D_d = O,\vspace{-3mm}
		\end{equation}
		implying that the states are available for measurement.}\vspace{-3mm} 
	
	{\color{black}
		\begin{remark}\label{rem:dist_pres}
			The disturbance vector $d[k]$ along with the noise affecting the state measurements \eqref{eq:ss_b}-\eqref{eq:ss_c} are key components of our framework. One of the main goals of the distributed MPC-based procedure proposed in this paper is to ensure constraint satisfaction for $x[k]$ and $u[k]$ in the presence of the aforementioned exogenous signals (along with additional ones, introduced in the sequel).\vspace{-3mm}
		\end{remark}
	}
	
	For systems of type \eqref{eq:ss_a}-\eqref{eq:ss_c}, we denote by $x_{c}\in\mathbb{R}^{n_x}$ the initial condition of the state vector and we also consider $A\in\mathbb{R}^{n_x\times n_x}$, $B_u\in\mathbb{R}^{n_x\times n_u}$, $B_d\in\mathbb{R}^{n_x\times n_d}$, $C\in\mathbb{R}^{n_y\times n_x}$, while, more generically, we consider  $D_u\in\mathbb{R}^{n_y\times n_u}$ and $D_d\in\mathbb{R}^{n_y\times n_d}$. The positive integer $n_x$ will be referred to as the \emph{order} of the realisation, and the link between a system described by \eqref{eq:ss_a}-\eqref{eq:ss_b} and its TFM is given by\vspace{-3mm}
	\begin{multline}\label{eq:TFM_def}
		\mathbf{G}(z)=\left[\scriptsize\begin{array}{c|cc}
			A-z I_{n_x} & B_u & B_d \\\hline C & D_u & D_d
		\end{array}\right]:=\\:={\scriptsize\begin{array}{l}
				C(z I_{n_x}-A)^{-1}
		\end{array}}\hspace{-1mm}\scriptsize\begin{bmatrix}
			B_u & B_d
		\end{bmatrix}+\scriptsize\begin{bmatrix}
			D_u & D_d
		\end{bmatrix}.
	\end{multline}\phantom{ }\vspace{-6mm}
	
	Finally, given any $\mathbf{H}\in\mathcal{R}(z)^{n_y\times n_u}$ which can be described by \eqref{eq:ss_a}-\eqref{eq:ss_b} and \eqref{eq:TFM_def}, the shorthand notation $y[k]=\mathbf{H}(z)\star u[k]$ represents the time-response of $\mathbf{H}(z)$ to an input signal vector $u[k]$, which can be computed as\vspace{-3mm}
	\begin{equation}\label{eq:forced_resp}
		\scriptsize\begin{array}{l}
			y[k]=\mathbf{H}(z)\star u[k] = D_Hu[k] + \sum_{\textcolor{black}{i\geq1}} C_H^{}A_H^{i-1}B_H^{}u[k-i],
		\end{array}\hspace{-1mm}\vspace{-3mm}\normalsize
	\end{equation}
	for any realisation $(A_H,B_H,C_H,D_H)$ of $\mathbf{H}(z)$. \textcolor{black}{Additionally, in accordance with the dynamics from \eqref{eq:ss_a}-\eqref{eq:ss_b}, we point out that all of the signal vectors considered in this paper are taken to be equal to zero for all $k<k_0$.}\vspace{-3mm}

	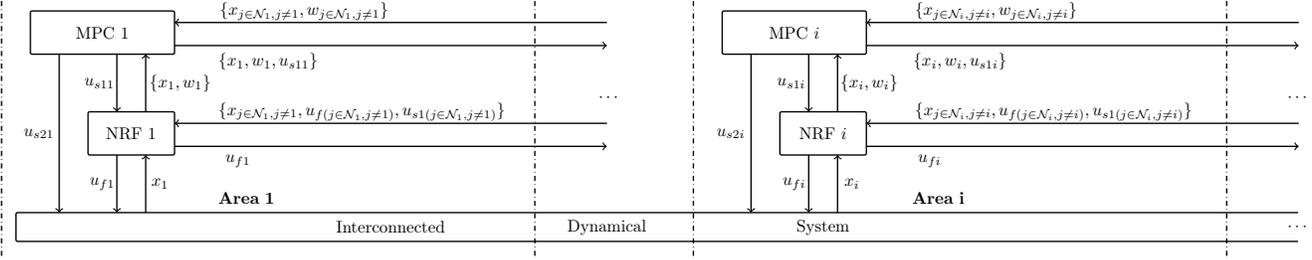
\begin{figure*}[t]
		\centering
		\resizebox{.975\textwidth}{!}{
			\begin{tikzpicture}[scale=0.6]
				\draw [thick,rounded corners=1]  (-15,-0.5) rectangle +(45,1);
				\draw [thick]  (-2,0)   node {Interconnected};
				\draw [thick]  (5.5,0)   node {Dynamical};
				\draw [thick]  (13,0)   node {System};
				
				\draw [thick,rounded corners=1]  (-1.5-11,2.5) rectangle +(3,1.5);
				\draw [thick]  (0-11,3.25)   node {NRF $1$};
				\node at (4-11,1) {\textbf{Area} $\mathbf 1$};
				\draw [thick,rounded corners=1]  (-3.5-11,6) rectangle +(5,1.5);
				\draw [thick]  (-1-11,6.75)   node {MPC $1$};
				\draw[->,thick] (0.5-11,0.5)--(0.5-11,2.5);
				\node at (1-11,1.5) {$x_1$};
				\draw[->,thick] (-0.5-11,2.5)--(-0.5-11,0.5);
				\node at (-1-11,1.5) {$u_{f1}$};
				\draw[->,thick] (0.5-11,4)--(0.5-11,6);
				\node at (1.7-11,5) {$\{x_1,w_1\}$};
				\draw[->,thick] (-0.5-11,6)--(-0.5-11,4);
				\node at (-1.1-11,5) {$u_{s11}$};
				\draw[->,thick] (-2.5-11,6)--(-2.5-11,0.5);
				\node at (-3.2-11,3.25) {$u_{s21}$};
				\draw[->,thick] (1.5-11,2.8)--(16.5-11,2.8);
				\node at (3.7-11,2.3) {$u_{f1}$};
				\draw[->,thick] (16.5-11,3.6)--(1.5-11,3.6);
				\node at (8-11,4) {$\{x_{j\in\mathcal{N}_1,j\neq 1},u_{f(j\in\mathcal{N}_1,j\neq 1)},u_{s1(j\in\mathcal{N}_1,j\neq 1)}\}$};
				\draw[->,thick] (1.5-11,6.3)--(16.5-11,6.3);
				\node at (4.75-11,5.75) {$\{x_1,w_1,u_{s11}\}$};
				\draw[->,thick] (16.5-11,7.1)--(1.5-11,7.1);
				\node at (6-11,7.5) {$\{x_{j\in\mathcal{N}_1,j\neq 1},w_{j\in\mathcal{N}_1,j\neq 1}\}$};
				\draw[dash dot,thick] (-4.5-11,-1)--(-4.5-11,8);
				\draw[dash dot,thick] (14-11,-1)--(14-11,8);
				
				\node at (5.6,4.5) {$\cdots$};
				
				\draw [thick,rounded corners=1]  (-1.5+13,2.5) rectangle +(3,1.5);
				\draw [thick]  (0+13,3.25)   node {NRF $i$};
				\node at (4+13,1) {\textbf{Area} $\mathbf i$};
				\draw [thick,rounded corners=1]  (-3.5+13,6) rectangle +(5,1.5);
				\draw [thick]  (-1+13,6.75)   node {MPC $i$};
				\draw[->,thick] (0.5+13,0.5)--(0.5+13,2.5);
				\node at (1+13,1.5) {$x_i$};
				\draw[->,thick] (-0.5+13,2.5)--(-0.5+13,0.5);
				\node at (-1+13,1.5) {$u_{fi}$};
				\draw[->,thick] (0.5+13,4)--(0.5+13,6);
				\node at (1.6+13,5) {$\{x_i,w_i\}$};
				\draw[->,thick] (-0.5+13,6)--(-0.5+13,4);
				\node at (-1.1+13,5) {$u_{s1i}$};
				\draw[->,thick] (-2.5+13,6)--(-2.5+13,0.5);
				\node at (-3.2+13,3.25) {$u_{s2i}$};
				\draw[->,thick] (1.5+13,2.8)--(16.5+13,2.8);
				\node at (3.7+13,2.3) {$u_{fi}$};
				\draw[->,thick] (16.5+13,3.6)--(1.5+13,3.6);
				\node at (8+13,4) {$\{x_{j\in\mathcal{N}_i,j\neq i},u_{f(j\in\mathcal{N}_i,j\neq i)},u_{s1(j\in\mathcal{N}_i,j\neq i)}\}$};
				\draw[->,thick] (1.5+13,6.3)--(16.5+13,6.3);
				\node at (4.75+13,5.75) {$\{x_i,w_i,u_{s1i}\}$};
				\draw[->,thick] (16.5+13,7.1)--(1.5+13,7.1);
				\node at (6+13,7.5) {$\{x_{j\in\mathcal{N}_i,j\neq i},w_{j\in\mathcal{N}_i,j\neq i}\}$};
				\draw[dash dot,thick] (-4.5+13,-1)--(-4.5+13,8);
				\draw[dash dot,thick] (14+13,-1)--(14+13,8);
				
				\node at (29.5,4.5) {$\cdots$};
				
				\draw [thick,color=white,fill=white]  (29.5,-0.6) rectangle +(1,1.2);
				
				\node at (29.5,0) {$\cdots$};
		\end{tikzpicture}}\vspace{-2mm}
		\caption{High-level implementation scheme depicting the proposed distributed control strategy}\vspace{-1mm}
		\label{fig:scheme}
		\hrulefill\vspace{-1mm}
	\end{figure*}

	\subsection{Distributed Networks}
	\label{subsec:distrib_part}\vspace{-3mm}
	
	Although we consider that the vectors $x$ and $u$ are completely accessible, in terms of measurement and actuation, respectively, we do not assume that manipulating \emph{all} of these variables is possible from any one location in our physical network. The latter fact naturally separates the network into $N\in\mathbb{N}$ distinct areas, with $N>1$.\vspace{-3mm}
	
	To each of the network's $N$ areas, we now assign two sets of indices, which denote the entries of $x$ and $u$ that are (uniquely) accessible for measurement or actuation in that area. Without loss of generality, we assume that these indices are assigned to the sets in ascending order, since the rows and columns of the matrices expressed in \eqref{eq:ss_a}-\eqref{eq:ss_c} may be permuted at will. Therefore, each area will be represented by the pair\vspace{-2mm}
	\begin{equation}\label{eq:trip}
		\mathcal{A}_i:=(\mathcal{A}_{xi},\mathcal{A}_{ui}),\,\forall\,i\in\{1:N\},\vspace{-3mm}
	\end{equation}
	which collects the two aforementioned index sets.\vspace{-3mm}
	
	To streamline the definition of the sets given in \eqref{eq:trip}, we will provide generic expressions via the placeholder subscript $\bullet$, which stands for one of the subscripts $\{x,u\}$. Let each set contain a number of $n_{\bullet i}\in\mathbb{N}$ indices, where $n_{\bullet i}>0,\,\forall\,i\in\{1:N\}$, such that we define\vspace{-2mm}
	\begin{equation}\label{eq:net_part}
		\hspace{-1mm}\left\{\begin{aligned}
			\mathcal{A}_{\bullet i}:=\{\alpha_{\bullet i}+j\ \vert\ j\in1:n_{\bullet i}\},\,\forall\,i\in\{1:N\},\\
			\alpha_{\bullet 1}:=0,\, \alpha_{\bullet \ell}:=\alpha_{\bullet (\ell-1)}+n_{\bullet (\ell-1)},\,\forall\,\ell\in\{2:N\}.
		\end{aligned}\right.\vspace{-2mm}
	\end{equation}
	To each partition\footnote{As an example, a network with 12 states and 7 inputs could be partitioned as in \eqref{eq:trip}-\eqref{eq:net_part} via the following collection of sets $\scriptsize\begin{array}{l}
			\{(\{1:2\},\{1\}),(\{3:5\},\{2:3\}),(\{6:11\},\{4:5\}),(\{12\},\{6:7\})\}
		\end{array}$} in \eqref{eq:net_part}, we attach the selection matrix\vspace{-3mm}
	\begin{equation}\label{eq:sel_mat}
		S_{\bullet i}:=\left[\scriptsize\begin{array}{ccc}
			e_{\alpha_{\bullet i}+1}&\dots&e_{\alpha_{\bullet i}+n_{\bullet i}}
		\end{array}\right]\in\mathbb{R}^{n_{\bullet}\times n_{\bullet i}},\,\forall\,i\in\{1:N\},\vspace{1mm}
	\end{equation}
	for which $x_i[k]:=S^\top_{{xi}}x[k]$, $x_{c{i}}:=S^\top_{{xi}}x_c$, $u_i[k]:=S^\top_{{ui}}u[k]$.
	\vspace{-7mm}
	
	\section{The Control Architecture}\label{sec:arch}\vspace{-3mm}
	
	\subsection{Global Perspective and Functioning}\label{subsec:global_arch}\vspace{-3mm}
	
	The aim of this manuscript and of its companion paper \cite{part1} is to formalise the design framework of the two-layer control architecture for the interconnected dynamical system depicted in Figure~\ref{fig:scheme}. By identifying the subset of areas $\mathcal{N}_i\subseteq\{1:N\}$ which can transmit information to the network's $i^\text{th}$ area (and which trivially satisfy $i\in\mathcal{N}_i$), for all $i\in\{1:N\}$, the communication graph is clearly designated and the fundamental operating principle of our control scheme can be expressed as follows:\vspace{-3mm}
	\begin{enumerate}
		\item[a)] the state-space-implemented and NRF-based subcontrollers receive network state feedback and the command signals of other first-layer subcontrollers from their areas' neighbourhoods, using these values to compute the signals denoted $u_{fi}$ in Figure~\ref{fig:scheme};\smallskip
		
		\item[b)] the MPC subcontrollers also receive network state information along with the state variables of the NRF subcontroller implementations, denoted $w_i$ in Figure~\ref{fig:scheme}, from their areas' neighbours to compute:\smallskip
		
		\begin{enumerate}
			\item[b1)] the command signals denoted by $u_{s1i}$ in Figure~\ref{fig:scheme}, which \emph{are broadcast} and combined additively with the state vectors $x_i$, before the later are fed to the NRF subcontrollers;\smallskip
			
			\item[b2)] the command signals denoted by $u_{s2i}$ in Figure~\ref{fig:scheme}, which \emph{remain local} and are combined additively with the NRF outputs $u_{fi}$, to obtain the control signals being sent to the actuators.\vspace{-3mm}
		\end{enumerate}
	\end{enumerate}
	
	\begin{figure*}
		{\color{black}
			\begin{equation}\label{eq:area_prob}\tag{8}
				\begin{aligned}
					&\min\limits_{u_{s1i}[k],\dots,u_{s1i}[k+T_i-1],u_{s2i}[k],\dots,u_{s2i}[k+T_i+\overline T_i-1]}\textstyle\sum_{t=1}^{T_i+\overline T_i}g_{t}(\xi_i[k+t],u_{s1i}[k+t-1],u_{s2i}[k+t-1]),\\
					&\text{subject to}\left\{
					\begin{aligned}
						&\xi_i[k+t] = A_{si} \xi_i[k+t-1] + B_{s1i} u_{s1i}[k+t-1] + B_{s2i} u_{s2i}[k+t-1],\,\forall\,t\in\{1:T_i+\overline T_i\},\\
						&\xi_i[k+t]\in\Xi_{it},\,\forall\,t\in\{1:T_i\},\ \xi_i[k]=O,\\
						&u_{s1i}[k+t-1]\in{\mathcal{U}}_{s1i},\, u_{s2i}[k+t-1]\in{\mathcal{U}}_{s2i},\,\forall\,t\in\{1:T_i\}.
					\end{aligned}\right.
				\end{aligned}
			\end{equation}
		}\hrulefill\vspace{-3mm}
	\end{figure*}
	
	{\color{black}
		\begin{remark}\label{rem:split_obj}
			In terms of control architecture design, it should be clear that both the first and the second layers employ feedback and feedforward components. However, their goals and their communication-based mechanisms are completely different. Where the first layer focuses on dynamical decoupling and disturbance rejection, the second layer concerns itself with constraint satisfaction and feasibility, in the sense of robust control invariance.\vspace{-3mm}
	\end{remark}}
	
	{\color{black}
		As previously mentioned in Remark~\ref{rem:dist_pres}, disturbance plays a key role in our chosen framework. Thus, we recall the disturbance vector $d[k]$ from \eqref{eq:ss_a}-\eqref{eq:ss_c} and we denote the measurement noise mentioned in said remark as $\zeta[k]\in\mathbb{R}^{n_x}$. Moreover, we also introduce $\beta_f[k]\in\mathbb{R}^{n_u}$, for the first layer's communication disturbance, along with $\beta_{s1}[k]\in\mathbb{R}^{n_x}$ and $\beta_{s2}[k]\in\mathbb{R}^{n_u}$, the communication disturbance associated with the second layer. The latter three signal vectors act as sources of additive disturbance on the command signals depicted in Figure~\ref{fig:scheme}, namely $u_{f}[k]:=\tiny\begin{bmatrix}
			u_{f1}^\top[k] & \dots & u_{fN}^\top[k]
		\end{bmatrix}^\top$, $u_{s1}[k]:=\tiny\begin{bmatrix}
			u_{s11}^\top[k] & \dots & u_{s1N}^\top[k]
		\end{bmatrix}^\top$ and $u_{s2}[k]:=\tiny\begin{bmatrix}
			u_{s21}^\top[k] & \dots & u_{s2N}^\top[k]
		\end{bmatrix}^\top$, respectively. All of the exogenous signals mentioned in this paragraph may be combined into the global disturbance vector defined as\vspace{-3mm}
		\begin{equation}\label{eq:MPC_exo}
			d_s[k]:=\begin{bmatrix}
				(\zeta[k]+\beta_{s1}[k])^\top & \beta_{s2}^\top[k] & \beta_f^\top[k] & d^\top[k]
			\end{bmatrix}^\top,\vspace{-3mm}
		\end{equation}
		with respect to which we now formally define the objectives of our MPC layer.\vspace{-3mm}
		
	}
	
	{\color{black}
		
		\subsection{Objectives of the MPC Layer}\label{subsec:MPC_obj}\vspace{-3mm}
		
		The second layer of the control architecture described in Section~\ref{subsec:global_arch} must be formed by $N$ separate MPC-based subcontrollers, which accomplish the following goals:\vspace{-3mm}
		\begin{enumerate}
			\item[I)] Given two sets $\mathcal{X}\subseteq\mathbb{R}^{n_x}$ and $\mathcal{U}\subseteq\mathbb{R}^{n_u}$, the second layer's global control signals $u_{s1}[k]$ and $u_{s2}[k]$ \emph{robustly satisfy} the constraints $x[k]\in\mathcal{X}$ and $u[k]\in\mathcal{U}$ in the presence of $d_s[k]$ from \eqref{eq:MPC_exo};\smallskip
			
			\item[II)] The area-based components of $u_{s1}[k]$ and $u_{s2}[k]$ are computed using  \emph{receding horizon-based optimisation} (see, for example, Chapter 1 in \cite{RMD}), with each area admitting tractable formulations of type \eqref{eq:area_prob}, located at the top of this page, where:\smallskip
			\begin{enumerate}
				\item[a)] $g_t(\cdot)$ is the stage cost at time instant $k+t$,\vspace{0.5mm}
				\item[b)] $\xi_i$ denotes the local state vector of the $i^\text{th}$ area's closed-loop system, to be addressed (along with its particular initialisation) in the sequel,\vspace{0.5mm}
				\item[c)] the scalar values $T_i\in\mathbb{N}\setminus\{0\}$ and $\overline T_i\in\mathbb{N}$ are, respectively, the (set-)constrained and the unconstrained prediction horizons of each area,\vspace{0.5mm}
				\item[d)] the sets denoted ${\mathcal{U}}_{s1i}$ along with ${\mathcal{U}}_{s2i}$ constrain the local second-layer commands and $\Xi_{it}$ is the set linking the local constraints with the global ones, to be constructed in the sequel;\vspace{0.5mm}
			\end{enumerate}
			
			\item[III)] For all $i\in\{1:N\}$, the $i^\text{th}$ area's MPC-based subcontroller from \eqref{eq:area_prob} computes $u_{s1i}[k]$ and $u_{s2i}[k]$ using \emph{only information} available in $\mathcal{N}_i$;\vspace{0.5mm}
			
			\item[IV)] Ensuring the feasibility of each optimisation problem from \eqref{eq:area_prob} at some $k=k_0$ ensures all of them \emph{remain feasible} for all $k>k_0$.\vspace{-3mm}
		\end{enumerate}
		
		{\color{black}
			\begin{remark}
				As previously noted in Remark~\ref{rem:split_obj}, the goals stated above indicate that our second layer is solely responsible for enforcing constraints in a distributed manner. Thus, all objectives related to reference tracking or robust stabilisation are to be handled by the NRF-based first layer, so as to take full advantage of the latter's remarkable closed-loop guarantees (see, for example, the results in \cite{plutonizare,NRF,aug_sparse}). In our framework, the MPC layer primarily serves as a safeguard for the NRF-based control laws, further highlighting the specific nature of our approach. As shown via the numerical example in the sequel, we design our second layer with the aim of minimising its impact on the first-layer-based closed-loop system, generating command signals of non-negligible amplitude only when constraint violation becomes imminent.\vspace{-3mm}
			\end{remark}
		}
		
		To achieve all of the above-stated objectives, however, tractable and area-based models will have to be derived for the first-layer-based closed-loop system.\vspace{-3mm}

		\section{Obtaining an MPC-Oriented Formulation}\vspace{-3mm}
		\label{sec:transform}
		
		This section is dedicated to discussing the derivation and the manipulation of the models which form the equality constraints of the problems stated in \eqref{eq:area_prob}. Readers familiar with the \emph{robust MPC framework} (see Chapter 3 in \cite{RMD} for a comprehensive treatment of the topic) will undoubtedly recognise many of the formulations and conventions adopted in the sequel. Nevertheless, the use of \emph{dynamical} feedback rather than classical static gains in the so-called \emph{tube-based MPC} framework (see, for example, Section~3.5 in \cite{RMD}) makes our problem distinct from other approaches in system-theoretical literature. Therefore, given this distinct element of novelty, we proceed to lay out our expressions in an explicit manner.\vspace{-3mm}
		
	}
	
	\subsection{Interfacing with the First Layer}\label{subsec:interf}\vspace{-3mm}
	
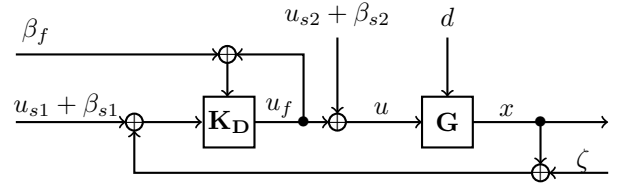
\begin{figure}[t]
		\centering
		\begin{tikzpicture}[scale=0.225]
			\draw [thick] [->] (-3,0) -- (3.5,0);
			\draw [thick]  (4,0) circle(0.5);
			\draw [thick]  (4,0) node {$+$};
			\draw [thick] (3,1)   node {\bf{ }} (0,1) node {$u_{s1}+\beta_{s1}$};
			\draw [thick] [->] (4.5,0) -- (8,0);
			\draw [thick, rounded corners=1]  (8,-1.5) rectangle +(3,3);
			\draw [thick] (9.5,0)   node {{$\,\mathbf{K}_{\mathbf{D}}$}} ;
			\draw [thick] [->] (11.1,0) -- (15.5,0);
			\draw [thick]  (16,0) circle(0.5cm);
			\draw [thick]  (16,0) node {$+$};
			\draw [thick] (12.6,0.8)   node {$u_f$} ;
			\draw [thick] [->] (14,0) -- (14,4)--(10,4);
			\draw [thick] (18.6,0.8)   node {$u$} ;
			\draw [thick] (14,0)   node {$\bullet$} ;
			\draw [thick]  (14.8,0.7)   node {\bf{ }};
			\draw [thick] [->] (16,5) -- (16,0.5);
			\draw [thick]  (-1.8,4)  node[anchor=south] {$\beta_f$};
			\draw [thick]  (9.5,4) circle(0.5);
			\draw [thick]  (9.5,4) node {$+$};
			\draw [thick] [->] (9.5,3.5) -- (9.5,1.5);
			\draw [thick] [->] (-3,4) -- (9,4);
			\draw [thick]  (16,5.1)  node[anchor=south] {$u_{s2}+\beta_{s2}$};
			\draw [thick] [->] (16.5,0) -- (21,0);
			\draw [thick, rounded corners=1]  (21,-1.5) rectangle +(3,3) ;
			\draw [thick]  (22.5,0)   node {{${\bf G}$}} ;
			\draw [thick]  (26,0.7)   node {$x$};
			\draw [thick] [->] (24,0) -- (32,0);
			\draw [thick] (28,0)   node {$\bullet$} ;
			\draw [thick] [->] (28,0) -- (28,-2.5);
			\draw [thick] [->] (27.5,-3) -- (4,-3)-- (4, -0.5);
			\draw [thick]  (28,-3) circle(0.5);
			\draw [thick]  (28,-3) node {$+$};
			\draw [thick] [->] (32,-3) -- (28.5,-3);
			\draw [thick]  (30.5,-2.25)  node {$\zeta$};
			\draw [thick] [->] (22.5,5) -- (22.5,1.5);
			\draw [thick]  (22.5,5.1)  node[anchor=south] {$d$};
		\end{tikzpicture}\vspace{-2mm}
		\caption{Feedback loop of a network's model ${\bf G}(z)$ with the NRF-based distributed implementation $\mathbf{K}_{\bf \mathbf{D}}(z)$}\vspace{-1mm}
		\label{fig:NRF_implem}
	\end{figure}

	We begin by briefly summarising several key concepts from our companion paper \cite{part1}. Our architecture's first layer is implemented as depicted in Figure~\ref{fig:NRF_implem}, located on the next page, in which the distributed control laws are represented by the following TFM:
	\vspace{-3mm}
	\begin{equation}\label{eq:Kd_def}\tag{9}
		\mathbf{K}_{\mathbf{D}}(z):=\begin{bmatrix}
			\mathbf{K}_{\mathbf{D}1}^\top(z) & \dots & \mathbf{K}_{\mathbf{D}N}^\top(z)
		\end{bmatrix}^\top,\vspace{-3mm}
	\end{equation}
	where each $\mathbf{K}_{\mathbf{D}i}(z)=S_{ui}^\top\mathbf{K}_{\mathbf{D}}(z)$ represents the TFM of the $i^\text{th}$ area's local subcontroller.\vspace{-3mm}
	
	\begin{remark}\label{rem:com}
		We assume that the communication infrastructure of our control architecture is equipped with error-detection/correction mechanisms. Thus, all communication errors affecting transmitted information are modelled as additive disturbance signals which originate at the source of transmission, not at the receiving ends. {\color{black}Consequently, the resulting signal diagram of the first layer's closed-loop system will be precisely the one shown in Fig.~\ref{fig:NRF_implem}, even when all $\mathbf{K}_{\mathbf{D}i}(z)$ are implemented separately, and the closed-loop mappings will have the beneficial properties discussed in \cite{NRF,part1,aug_sparse}.}\vspace{-3mm}
	\end{remark}
	
	\begin{figure*}
		\begin{equation}\label{eq:area_NRF}\tag{10}
			\mathbf{K}_{\mathbf{D}i}(z)=\left[\scriptsize\begin{array}{c|c}
				A_{wi}-z I_{n_{wi}} & B_{wi} \\ \hline C_{wi} & D_{wi}
			\end{array}\right]:= \left[\tiny\begin{array}{ccc|c}
				A_{r(\alpha_{ui}+1)}-z I_{n_{r(\alpha_{ui}+1)}} & & & B_{r(\alpha_{ui}+1)} \\
				& \ddots & & \vdots \\
				& & A_{r(\alpha_{ui}+n_{ui})}-z I_{n_{r(\alpha_{ui}+n_{ui})}} & B_{r(\alpha_{ui}+n_{ui})} \\ \hline 
				C_{r(\alpha_{ui}+1)} & & & O\\
				& \ddots & & \vdots\\
				& & C_{r(\alpha_{ui}+n_{ui})} & O\\
			\end{array}\right].\vspace{2mm}
		\end{equation}
		\hrulefill\vspace{-1mm}
	\end{figure*}
	
	Crucially, these subcontrollers are implemented via the state-space realisations given in \eqref{eq:area_NRF}, located at the top of the next page, whose component matrices (see Section~4.1 in \cite{part1} for more details) are given, $\forall\,\ell\in\mathcal{A}_{ui}$, by 
	\vspace{-4mm}\stepcounter{equation}\stepcounter{equation}\stepcounter{equation}
	\begin{equation}\label{eq:mat_coef}
		\left\{\begin{array}{ll}
			\widetilde{A}_{r\ell}:=\tiny\begin{bmatrix}
				-a_{1\ell} &\dots& -a_{(n_{r\ell}-1)\ell}
			\end{bmatrix}^\top,& A_{r\ell}:=\ \tiny\begin{bmatrix}
				\widetilde{A}_{r\ell} & I_{n_{r\ell}-1}\\
				-a_{n_{r\ell}\ell} & O
			\end{bmatrix},\\
			B_{r\ell}:=\tiny\begin{bmatrix}
				K_{1\ell}^\top &\dots& K_{n_{r\ell}\ell}^\top
			\end{bmatrix}^\top,& C_{r\ell}:=\tiny\ \ \begin{bmatrix}
				1 & 0 & \dots & 0
			\end{bmatrix}.
		\end{array}\right.
	\end{equation}
	Henceforth, we define the orders $n_{wi}:=\textstyle\sum_{\ell=1+\alpha_{ui}}^{\,n_{ui}+\alpha_{ui}} n_{r\ell}$ and $n_w:=\textstyle\sum_{i=1}^N n_{wi}$. We also denote the state vectors of the implementations from \eqref{eq:area_NRF} as $w_i[k]$ and their initial conditions as $w_{ci}\in\mathbb{R}^{n_{wi}}$. Moreover, we concatenate the latter into $w_c:=\tiny\begin{bmatrix}
		w_{c1}^\top & \dots & w_{cN}^\top
	\end{bmatrix}^\top\in\mathbb{R}^{n_w}$, which are the initial conditions of the first layer's \emph{global} implementation.\vspace{-3mm}
	
	One of the main features of this control scheme is the fact that it can \emph{independently} compute the control signals of the first layer as follows:\vspace{-4mm}
	\begin{equation}\label{eq:uf_implem}
		u_{fi}[k]:=S_{{ui}}^\top u_f[k] = \mathbf{K}_{\mathbf{D}i}(z)\star\tiny\begin{bmatrix}
			u_f[k]+\beta_f[k] \\ x[k]+\zeta[k]+u_{s1}[k]+\beta_{s1}[k]
		\end{bmatrix},\normalsize\vspace{1mm}
	\end{equation}
	for all $i\in\{1:N\}$, while employing only information from each area's neighbourhood. This reduces to the fact that the implementations from \eqref{eq:area_NRF}-\eqref{eq:mat_coef} must satisfy\vspace{-3mm}
	\begin{equation}\label{eq:dist_implem}
		B_{r\ell}\,\mathrm{diag}(S_{uj},S_{xj})=O,\,\forall\,\ell\in\mathcal{A}_{ui},\,j\in\{1:N\}\setminus\,\mathcal{N}_i.\vspace{-3mm}
	\end{equation}
	\begin{remark}
		Although we consider the first layer's control laws to be obtained through an NRF-based framework, as in our companion paper \cite{part1}, note that any set of distributed subcontrollers implemented akin to \eqref{eq:Kd_def}-\eqref{eq:dist_implem} may be employed for the procedures described in the sequel.\vspace{-3mm}
	\end{remark}

	By using the signal vector introduced in \eqref{eq:MPC_exo}, it is possible to express the first layer's closed-loop dynamics as\vspace{-3mm}
	\begin{multline}\label{eq:cl_dyn}
		\scriptsize\begin{bmatrix}
			x[k]  \\  u_f[k]
		\end{bmatrix}=\left(\mathbf{F}_{\mathbf{Q}}(z)\scriptsize\begin{bmatrix}
			I_{(n_x+n_u)} \\ O
		\end{bmatrix}\right)\star\scriptsize\begin{bmatrix}
			u_{s1}[k] \\ u_{s2}[k]
		\end{bmatrix}+\\+\mathbf{F}_{\mathbf{Q}}(z)\star d_s[k]+\mathcal{I}_{\mathbf{Q}}[k]
		{\scriptsize\begin{bmatrix}
				x_c \\  w_c
		\end{bmatrix}}
		\,,
	\end{multline}
	\phantom{}
	
	\vspace{-8mm}
	for all $k\geq k_0$, where $\mathbf{F}_{\mathbf{Q}}(z)$ is a strictly proper TFM, \emph{i.e.}, $\lim_{|z|\rightarrow\infty}\mathbf{F}_{\mathbf{Q}}(z)=O$. Notably, $\mathbf{F}_{\mathbf{Q}}(z)$ models the closed-loop's response to exogenous inputs, while $\mathcal{I}_{\mathbf{Q}}[k]$ is a time-varying matrix which governs the closed-loop system's evolution with respect to its initial conditions. \vspace{-3mm}
	{\color{black}
		\begin{remark}\label{rem:dist_clarif}
			While the signal vectors $d[k]$ and $\zeta[k]$ in \eqref{eq:MPC_exo} and (implicitly) in \eqref{eq:cl_dyn} possess classical interpretations in system-theoretical literature, the three vectors denoted generically as $\beta_{\bullet}[k]$ in \eqref{eq:MPC_exo} are caused primarily by encoding errors in the employed communication infrastructure. Due to finite-precision representation, the absolute values of the signals in $\beta_{\bullet}[k]$ may be considered to be upper bounded and, unless stated otherwise, the samples making up $\beta_{\bullet}[k]$ are implicitly assumed to be uniformly distributed within their respective bounds. In addition to this, we point out that boundedness (not necessarily in terms of amplitude) also extends to both $d[k]$ and $\zeta[k]$, for most practical applications. Indeed, due to the stabilising properties of NRF-based control, the effect of all these signals upon the closed-loop system can be made bounded in some sense, by an appropriate choice of systemic norm optimisation (see Chapter 4 of \cite{zhou}, and Section~5.2 of \cite{part1}). Moreover, the results from \cite{PUB} can be neatly incorporated into our considered approach if one wishes to delve further into the stochastic setting.\vspace{-2mm}
	\end{remark}}
	The main result of our companion paper \cite{part1} shows that $\mathbf{F}_{\mathbf{Q}}(z)$ and $\mathcal{I}_{\mathbf{Q}}[k]$ are easily tunable via a freely chosen $\mathbf{Q}$-parameter (hence, the employed notation), thus enabling the \emph{complete design} of the first layer's closed-loop response. The full details regarding the synthesis and tuning of these closed-loop maps are given in our companion paper \cite{part1}, yet, for the benefit of the reader, we have included a summary of the key properties ensured by NRF-based control in Appendix~\ref{app:NRF}.\vspace{-2mm}

	\subsection{Nominal Area-based Prediction Models}\vspace{-2mm}
	
	To formulate tractable prediction models for our architecture's second layer, we need to account for the state variables of the implementations given in \eqref{eq:area_NRF}. Therefore, aside from the $x$ and $u$ subscripts appearing in \eqref{eq:net_part}-\eqref{eq:sel_mat}, we introduce the additional subscript $w$ to refer to the state vector of each area's NRF subcontroller, and we extend the index collection defined initially in \eqref{eq:trip} to\vspace{-2mm}
	\begin{equation*}
		\mathcal{A}_i:=(\mathcal{A}_{xi},\mathcal{A}_{ui},\mathcal{A}_{wi}),\,\forall\,i\in\{1:N\}.\vspace{-2mm}
	\end{equation*}
	We begin our reformulation of the first layer's closed-loop dynamics by pointing out the fact that, since the TFM $\mathbf{F}_{\mathbf{Q}}(z)$ from \eqref{eq:cl_dyn} is strictly proper, it is always possible to obtain state-space realisations of type\vspace{-1mm}
	\begin{equation}\label{eq:disc_real}
		\left[\tiny\begin{array}{cc}
			S^\top_{{xi}} & O \\\hdashline
			O & S^\top_{{ui}}
		\end{array}\right]\mathbf{F}_{\mathbf{Q}}(z)\left[\tiny\begin{array}{c:c}
			S_{{xi}} & O\\O & S_{{ui}}\\ O & O
		\end{array}\right]=\left[\tiny\begin{array}{c|c:c}
			A_{si}-zI_{n_{si}} & B_{s1i} & B_{s2i}\\\hline C_{xi} & O & O\\\hdashline C_{ui} & O & O
		\end{array}\right],\vspace{1mm}
	\end{equation}
	for all $i\in\{1:N\}$. By defining now $Z_i:=\mathrm{diag}(S_{xi},S_{ui})$ along with ${Z}_{ci}:=\mathrm{diag}(S_{xi},S_{wi})$, we employ the realisations from \eqref{eq:disc_real} to describe the MPC-oriented area model of the $i^\text{th}$ area, which can be expressed as\vspace{-2mm}
	\begin{subequations}
		\begin{align}\label{eq:disc_cl_dyn_a}
			\xi_i[k+1] =A_{si} \xi_i[k] + &\ B_{s1i} u_{s1i}[k]+B_{s2i} u_{s2i}[k],\\
			\label{eq:disc_cl_dyn_b}
			x_i[k] =C_{xi} \xi_i[k] + &\ (\psi_{x i}[k] + \theta_{x i}[k] + \delta_{x i}[k]),\\
			u_{fi}[k] =C_{ui} \xi_i[k] + &\ (\psi_{ui}[k] + \theta_{ui}[k] + \delta_{ui}[k]),\label{eq:disc_cl_dyn_c}\\
			\label{eq:disc_cl_dyn_d}
			&\hspace{-28mm}{\color{black}k\in\mathbb{Z},\ k\geq k_0\in\mathbb{Z},}\ \xi_i[k_0]=O,\,\forall\,i\in\{1:N\},
		\end{align}
	\end{subequations}\newpage
	where we have that:\vspace{-2mm}
	\begin{enumerate}
		\item[a)] $\xi_i[k]$ is the state vector of the realisation obtained in \eqref{eq:disc_real} and which \emph{always} has zero initial conditions at initial time $k_0$ (recall the identity from \eqref{eq:forced_resp});\smallskip
		
		\item[b)] the contribution of exogenous disturbance to the area dynamics is given by \vspace{-3mm}
	\end{enumerate}
	\begin{equation}\label{eq:disc_cl_dyn_1}
		\left\{
		\begin{aligned}
			\psi_i[k]:=&\ Z_i^\top\mathbf{F}_{\mathbf{Q}}(z)\star d_s[k],\\
			\psi_{x i}[k]:=&\ \tiny\begin{bmatrix}
				I_{n_{xi}} & O
			\end{bmatrix}\psi_i[k],\ 
			\psi_{ui}[k]:=\ \tiny\begin{bmatrix}
				O & I_{n_{ui}}
			\end{bmatrix}\psi_i[k],
		\end{aligned}
		\right.\vspace{-3mm}\hspace{-1mm}
	\end{equation}
	\begin{enumerate}
		\item[c)] the contribution of the \emph{neighbouring} areas' initial conditions to the area dynamics is given by\vspace{-3mm}
	\end{enumerate}
	\begin{equation}\label{eq:disc_cl_dyn_2}
		\left\{\begin{aligned}
			&\theta_i[k]:=\textstyle\sum_{j\in\mathcal{N}_i}Z_i^\top \mathcal{I}_{\mathbf{Q}}[k]Z_{cj}\tiny\begin{bmatrix}
				x_{cj}\\w_{cj}
			\end{bmatrix},\normalsize\\
			&\theta_{x i}[k]:=\tiny\begin{bmatrix}
				I_{n_{xi}} &O
			\end{bmatrix}\theta_i[k],\ \theta_{ui}[k]:=\tiny\begin{bmatrix}
				O & I_{n_{ui}}
			\end{bmatrix}\theta_i[k],
		\end{aligned}
		\right.\vspace{-3mm}
	\end{equation}
	\begin{enumerate}
		\item[d)] the \emph{residual} contribution of initial conditions from outside of the $i^\text{th}$ area's neighbourhood and of cross-coupling with other areas' MPC subcontrollers is\vspace{-3mm}
	\end{enumerate}\begin{equation}\label{eq:disc_cl_dyn_3}
		\hspace{-2mm}\left\{\begin{aligned}
			&\delta_i[k]:=\hspace{-2mm}\sum_{j\in\{1:N\}\setminus\{i\}}\hspace{-2mm}Z_i^\top\mathbf{F}_{\mathbf{Q}}(z)\tiny\begin{bmatrix}
				Z_j\\O
			\end{bmatrix}\star\begin{bmatrix}
				u_{s1j}[k] \\ u_{s2j}[k]
			\end{bmatrix}+\\&\qquad\qquad\quad\,+\sum_{j\in\{1:N\}\setminus\,\mathcal{N}_i}Z_i^\top \mathcal{I}_{\mathbf{Q}}[k]Z_{cj}\tiny\begin{bmatrix}
				x_{cj}\\w_{cj}
			\end{bmatrix},\\
			&\delta_{x i}[k]:=\tiny\begin{bmatrix}
				I_{n_{xi}} & O
			\end{bmatrix}\delta_i[k],\ \delta_{ui}[k]:=\tiny\begin{bmatrix}
				O & I_{n_{ui}}
			\end{bmatrix}\delta_i[k],
		\end{aligned}\right.\vspace{-3mm}
		\normalsize
	\end{equation}
	
	The signals from \eqref{eq:disc_cl_dyn_1}-\eqref{eq:disc_cl_dyn_3} separate naturally into two classes, in the form of information that is:\vspace{-3mm}
	\begin{enumerate}
		\item[i)]  \emph{available} to a second-layer subcontroller, represented by $\theta_{xi}[k]$ and $\theta_{ui}[k]$;\smallskip
		
		\item[ii)] \emph{unavailable} to a second-layer subcontroller, represented by $(\psi_{xi}[k]+\delta_{xi}[k])$ and $(\psi_{ui}[k]+\delta_{ui}[k])$.\vspace{-3mm}
	\end{enumerate}
	
	\begin{remark}\label{rem:decup}
		The quantities described in \eqref{eq:disc_cl_dyn_3} are called \emph{residual} because it is the primary goal of the first layer to attenuate the effect that these signals have on the area dynamics from \eqref{eq:disc_cl_dyn_a}-\eqref{eq:disc_cl_dyn_d}. Since these terms, along with those in \eqref{eq:disc_cl_dyn_1}, represent the contribution of variables which are not available (in terms of measurement or prediction) to the $i^\text{th}$ area, the only sensible recourse is to minimise their influence (see Section~5 in \cite{part1}).  \vspace{-3mm}
	\end{remark}
	
	Thus, each subcontroller must employ the information that is encoded in the signals from \eqref{eq:disc_cl_dyn_2} to compute a pair of commands $u_{s1i}[k]$ and $u_{s2i}[k]$ which robustly account for the effects of the signals from \eqref{eq:disc_cl_dyn_1} and \eqref{eq:disc_cl_dyn_3}. Yet, before translating this objective into a set-theoretical formulation, we investigate the effect of \emph{unreliable measurements} for $x_c$ and $w_c$ in the distributed setting.\vspace{-3mm}
	
	\subsection{Accounting for Inaccurate Initial Conditions}\label{subsec:ci_dist}\vspace{-3mm}
	
	Given some pre-specified set $\mathcal{V}\subseteq\mathbb{R}^{n_x+n_w}$, we consider the additively disturbed initial conditions defined as\vspace{-3mm}
	\begin{equation}\label{eq:noise_init}
		\tiny\begin{bmatrix}
			\widetilde{x}_c \\ \widetilde{w}_{c}
		\end{bmatrix}:=\tiny\begin{bmatrix}
			{x}_c+\nu_x \\{w}_{c}+\nu_w
		\end{bmatrix},\quad 	\tiny\begin{bmatrix}
			\widetilde{x}_{c{i}} \\ \widetilde{w}_{ci}
		\end{bmatrix}:=\tiny\begin{bmatrix}
			{x}_{ci}+S^\top_{{xi}}\nu_x \\ {w}_{ci}+S^\top_{{wi}}\nu_w
		\end{bmatrix},\normalsize\vspace{-3mm}
	\end{equation}
	for any index $i\in\{1:N\}$, along with any $\tiny\begin{bmatrix}
		\nu_x^\top \ \,\nu_w^\top
	\end{bmatrix}^\top\in\mathcal{V}$.\vspace{-3mm}
	\begin{remark}\label{rem:ci_dist_cause}
		In light of the arguments that were made in Remark~\ref{rem:com}, notice that the type of disturbance discussed in this section can always be modelled as in \eqref{eq:noise_init}. Practical causes for these unwanted terms include noisy measurement of the network's state variables and faulty reading of the NRF-based controller's state vector.\vspace{-3mm}
	\end{remark}
	
	The main consequence of taking into account these terms is that the prediction models from \eqref{eq:disc_cl_dyn_b}-\eqref{eq:disc_cl_dyn_c} become\vspace{-3mm}
	\begin{subequations}
		\begin{align}
			\label{eq:noise_pert_1}
			\widetilde x_i[k] =&\ C_{xi} \xi_i[k] + (\psi_{x i}[k] + \widetilde\theta_{x i}[k] + \widetilde\delta_{x i}[k]),\\
			\widetilde u_{fi}[k] =&\ C_{ui} \xi_i[k] + (\psi_{u i}[k] + \widetilde\theta_{u i}[k] + \widetilde\delta_{u i}[k]),\label{eq:noise_pert_2}
		\end{align}
	\end{subequations}
	\phantom{}
	
	\vspace{-10mm}
	when considering the noise-affected initial conditions introduced in \eqref{eq:noise_init} instead of the real ones, where\vspace{-3mm}
	\begin{equation}\label{eq:disc_cl_dyn_2p}
		\hspace{-2mm}\left\{\begin{aligned}
			\widetilde \theta_i[k]:=&\ \theta_i[k] + \sum_{j\in\mathcal{N}_i}Z_i^\top \mathcal{I}_{\mathbf{Q}}[k]Z_{cj}\tiny\begin{bmatrix}
				S^\top_{{xj}}\nu_x \\ S^\top_{{wj}}\nu_w
			\end{bmatrix},\\
			\widetilde \theta_{x i}[k]:=&\ \tiny\begin{bmatrix}
				I_{n_{xi}} & O
			\end{bmatrix}\widetilde \theta_i[k],\ 
			\widetilde \theta_{ui}[k]:=\ \tiny\begin{bmatrix}
				O & I_{n_{ui}}
			\end{bmatrix}\widetilde \theta_i[k],
		\end{aligned}
		\right.\hspace{-3mm}\vspace{-3mm}
	\end{equation}
	along with\vspace{-3mm}
	\begin{equation}\label{eq:disc_cl_dyn_3p}
		\hspace{-2mm}\left\{\begin{aligned}
			\widetilde \delta_i[k]:=&\ \delta_i[k]+\sum_{j\in\{1:N\}\setminus\,\mathcal{N}_i}Z_i^\top \mathcal{I}_{\mathbf{Q}}[k]Z_{cj}\tiny\begin{bmatrix}
				S^\top_{{xj}}\nu_x \\ S^\top_{{wj}}\nu_w
			\end{bmatrix},\\
			\widetilde \delta_{x i}[k]:=&\ \tiny\begin{bmatrix}
				I_{n_{xi}} & O
			\end{bmatrix}\widetilde \delta_i[k],\ 
			\widetilde \delta_{ui}[k]:=\ \tiny\begin{bmatrix}
				O & I_{n_{ui}}
			\end{bmatrix}\widetilde \delta_i[k],
		\end{aligned}\right.\hspace{-2mm}\vspace{-3mm}
		\normalsize
	\end{equation}
	represent the noise-affected versions of the signals from \eqref{eq:disc_cl_dyn_2} and \eqref{eq:disc_cl_dyn_3}, respectively, for all $i\in\{1:N\}$.\vspace{-3mm}
	
	\subsection{Connecting the Augmented Prediction Models to the Nominal Ones}\label{subsec:aug_pred}\vspace{-3mm}
	
	Since the vector $\tiny\begin{bmatrix}
		\nu_x^\top & \nu_w^\top
	\end{bmatrix}^\top$ cannot be determined \emph{a priori}, our MPC-based subcontrollers will have to use the initial conditions in \eqref{eq:noise_init} instead of the actual ones. Consequently, our control policies will employ the prediction models affected by additive disturbance \eqref{eq:noise_pert_1}-\eqref{eq:noise_pert_2}, instead of the ones given in \eqref{eq:disc_cl_dyn_b}-\eqref{eq:disc_cl_dyn_c}. However, a straightforward computation yields the fact that the original area dynamics, given by the identities obtained in \eqref{eq:disc_cl_dyn_b}-\eqref{eq:disc_cl_dyn_c}, can be retrieved from \eqref{eq:noise_pert_1}-\eqref{eq:noise_pert_2} via\vspace{-2mm}
	\begin{equation}\label{eq:noise_pert_3}
		x_i[k]=\widetilde x_i[k]-\eta_{x i}[k],\quad u_{fi}[k]=\widetilde{u}_{fi}[k]-\eta_{ui}[k],\vspace{-2mm}
	\end{equation}
	for all $i\in\{1:N\}$, where $\eta_{x i}[k]$ and $\eta_{ui}[k]$ are defined as\vspace{-2mm}
	\begin{equation}\label{eq:disc_cl_dyn_4}
		\left\{\begin{aligned}
			\eta_i[k]:=&\ \sum_{j=1}^{N}Z_i^\top \mathcal{I}_{\mathbf{Q}}[k]Z_{cj}\tiny\begin{bmatrix}
				S^\top_{{xj}}\nu_x \\ S^\top_{{wj}}\nu_w
			\end{bmatrix},\\
			\eta_{x i}[k]:=&\ \tiny\begin{bmatrix}
				I_{n_{xi}} & O
			\end{bmatrix}\eta_i[k],\ 
			\eta_{ui}[k]:=\ \tiny\begin{bmatrix}
				O & I_{n_{ui}}
			\end{bmatrix}\eta_i[k].
		\end{aligned}\right.
		\normalsize\vspace{-2mm}
	\end{equation}
	Having obtained the area-based prediction models in \eqref{eq:disc_cl_dyn_a}-\eqref{eq:disc_cl_dyn_d} and having augmented them as shown in \eqref{eq:noise_pert_1}-\eqref{eq:noise_pert_2}, we now proceed to discuss constraint formulation for the optimisation problems described in \eqref{eq:area_prob}.
	\vspace{-6mm}

	\section{Constraints in the Distributed Setting}\vspace{-3mm}
	\label{sec:MPC_des}
	
	The most important step in breaking down the second layer's distributed synthesis problem into locally solvable subproblems is to assign independent control objectives to each of the network's $N$ areas. Recalling the sets $\mathcal{X}$ and $\mathcal{U}$, introduced in Section~\ref{subsec:MPC_obj}, we begin by defining the pairs of sets $(\mathcal{X}_i,\mathcal{U}_i)$, with $i\in\{1:N\}$, which satisfy\vspace{-2mm}
	\begin{equation}\label{eq:set_equiv}
		\left\{
		\begin{aligned}
			x[k]\in\mathcal{X}&\Longleftrightarrow x_i[k]\in\mathcal{X}_i,\,\forall\,i\in\{1:N\},\\
			u[k]\in\mathcal{U}&\Longleftrightarrow u_i[k]\in\mathcal{U}_i,\,\forall\,i\in\{1:N\}.
		\end{aligned}
		\right.\vspace{-2mm}
	\end{equation}
	
	\begin{remark}\label{rem:set_decomp}
		The assumption that the pair $(\mathcal{X},\mathcal{U})$ can be decomposed as in \eqref{eq:set_equiv} is not only required, when attempting to break down the global problem into any configuration of local subproblems, but it is also a mild one (see its equivalent counterpart in \cite{DMPC1,DMPC2}). Indeed, the conditions located on the left-hand side of \eqref{eq:set_equiv} are often represented (in practice) by box constraints, for which said decomposition may always be performed.\vspace{-3mm}
	\end{remark}
	
	The chief aim of the subcontrollers which form the second layer (recall \eqref{eq:area_prob} and Figure~\ref{fig:scheme}) is to ensure the \emph{global}, left-hand side inclusions from \eqref{eq:set_equiv} by independently satisfying the \emph{local} ones, on the right-hand side of \eqref{eq:set_equiv}. With respect to the closed-loop dynamics from \eqref{eq:cl_dyn}, this problem can be broken down into five components:\vspace{-3mm}
	\begin{enumerate}
		\item[i)] accounting (in a set-theoretical framework) for the effect of the exogenous disturbance from \eqref{eq:MPC_exo};\smallskip
		
		\item[ii)] translating the command-related objectives from \eqref{eq:set_equiv} into first-layer command constraints;\smallskip
		
		\item[iii)] linking command constraints to the state variables of the implementations from \eqref{eq:area_NRF};\smallskip
		
		\item[iv)] accounting for the inaccurate measurements modelled in \eqref{eq:noise_init};\smallskip
		
		\item[v)] accounting for area cross-coupling.\vspace{-3mm}
	\end{enumerate}
	We now proceed to tackle each of these subproblems, in the order in which they are indicated above.\vspace{-3mm}
	
	\subsection{Placing Constraints on Exogenous Disturbance}\vspace{-3mm}
	
	We begin by addressing the components of the signal vector from \eqref{eq:MPC_exo}. Recalling the measurement noise set from \eqref{eq:noise_init} and the arguments made in Remark~\ref{rem:ci_dist_cause}, we begin by considering that $\zeta[k]\in\tiny\begin{bmatrix}
		I_n & O
	\end{bmatrix}\mathcal{V}$. We contain the rest of the exogenous signals from \eqref{eq:MPC_exo} by introducing\vspace{-3mm}
	\begin{equation}\label{eq:dist_sets}
		\left\{\hspace{-1mm}
		\scriptsize\begin{array}{ll}
			\mathcal D_{d}\subseteq\mathbb{R}^{n_d}\,\text{s.t. }d[k]\in \mathcal D_{d},
			&\mathcal D_{\beta_f}\subseteq\mathbb{R}^{n_u}\,\text{s.t. }\beta_{f}[k]\in \mathcal D_{\beta_f},\\
			\mathcal D_{\beta_{s1}}\subseteq\mathbb{R}^{n_x}\,\text{s.t. }\beta_{s1}[k]\in \mathcal D_{\beta_{s1}},
			&\mathcal D_{\beta_{s2}}\subseteq\mathbb{R}^{n_x}\,\text{s.t. }\beta_{s2}[k]\in \mathcal D_{\beta_{s2}}.
		\end{array}
		\right.
	\end{equation}
	By doing so, it becomes possible to construct the set $
	\mathcal{D}_s\subseteq\mathbb{R}^{2n_u+n_x+n_d}$ which satisfies $d_s[k]\in\mathcal{D}_s$ and which, in tandem with the expressions from \eqref{eq:disc_cl_dyn_1}, enables us to compute the sets containing the effects of $d_s[k]$, with $k\geq k_0$ and $t\in\mathbb{N}$, on each area's dynamics, denoted as\vspace{-3mm}
	\begin{equation}\label{eq:pert_sets}
		\hspace{-2mm}\left\{
		\begin{aligned}
			\Psi_{x i t}\subseteq\mathbb{R}^{n_{xi}}\text{ s.t. }\psi_{x i}[k_0+t]\in\Psi_{x i t},\,\forall\,d_s[k]\in\mathcal{D}_s,\\
			\Psi_{u i t}\subseteq\mathbb{R}^{n_{ui}}\text{ s.t. }\psi_{ui}[k_0+t]\in\Psi_{u i t},\,\forall\,d_s[k]\in\mathcal{D}_s.
		\end{aligned}
		\right.\hspace{-1mm}\vspace{-8mm}
	\end{equation}
	{\color{black}
		\begin{remark}\label{rem:NRF_benef}
			One of the key benefits of our two-layer approach is the fact that, from the point of view of the NRF-based layer, the signals $u_{s1}[k]$ and $u_{s2}[k]$ are perceived in the same manner as the all the other components of the vector from \eqref{eq:MPC_exo}, namely as additive and exogenous disturbance. Consequently, the \emph{special disturbance attenuation properties} (such as string stability, see \cite{plutonizare} and \cite{SS_rev}, along with the references therein) of our control system are preserved when adding the second layer on top of the first. Note that such special properties are particularly challenging to ensure using exclusively MPC-based techniques, a topic which has generated a great deal of research in system-theoretical literature \cite{plat_MPC_1,plat_MPC_2,plat_MPC_3,plat_MPC_4,plat_MPC_5}. One of the main contributions of the present work is that it combines the native benefits of two separate techniques, disturbance attenuation for NRF-based control and constraint management for MPC, into a single framework.\vspace{-3mm}
		\end{remark}
	}
	Moving onward, we address the variables associated with the first-layer subcontrollers and their implementations.\vspace{-3mm}
	
	\begin{figure*}
		\begin{multline}\label{eq:feedback_set}\tag{33}
			{\mathcal{D}}_{wi}\subseteq\mathbb{R}^{n_{ui}+n_{xi}}\text{ s.t. }
			\tiny\begin{bmatrix}
				u_{fi}[k]+S^\top_{{ui}}\beta_f[k]\\  x_i[k]+S^\top_{{xi}}\zeta[k]+u_{s1i}[k]+S^\top_{{xi}}\beta_{s1}[k]
			\end{bmatrix}\in{\mathcal{D}}_{wi},\\
			\forall\ 
			u_{fi}[k]\in{\mathcal{U}}_{fi},\ 
			(S^\top_{{ui}}\beta_f[k])\in S^\top_{{ui}}\mathcal{D}_{\beta_f},\ 
			x_{i}[k]\in{\mathcal{X}}_{i},
			(S^\top_{{xi}}\zeta[k])\in\tiny\begin{bmatrix}
				S^\top_{{xi}}\ O
			\end{bmatrix}\mathcal{V},\ u_{s1i}[k]\in\mathcal{U}_{s1i},\  
			(S^\top_{{xi}}\beta_{s1}[k])\in S^\top_{{ui}}\mathcal{D}_{\beta_{s1}}.
		\end{multline}\vspace{-3mm}
		\phantom{ }
		
		\vspace{-6mm}
		\hrulefill\vspace{-3mm}
		\begin{subequations}
			\begin{align}\label{eq:NRF_state_sets_a}\tag{34a}
				\mathcal{W}_{\ell n_{r\ell}}:=&\ (-a_{n_{r\ell}\ell})\mathcal{W}_{\ell 1} \oplus\left(\textstyle\bigoplus_{p\in\mathcal{N}_i}\left(K_{n_{r\ell}\ell}^{}Z_{rp}^{}Z_{rp}^\top\right){\mathcal{D}}_{wp}\right),\,\forall\ \ell\in\mathcal{A}_{ui},\\
				\mathcal{W}_{\ell j}:=&\ (-a_{j\ell})\mathcal{W}_{\ell 1} \oplus\mathcal{W}_{\ell (j+1)}\oplus\left(\textstyle\bigoplus_{p\in\mathcal{N}_i}\left(K_{j\ell}^{}Z_{rp}^{}Z_{rp}^\top\right){\mathcal{D}}_{wp}\right),\,\forall\,\ell\in\mathcal{A}_{ui},\,j\in\{2:n_{r\ell}-1\}.\label{eq:NRF_state_sets_b}\tag{34b}
			\end{align}
		\end{subequations}\vspace{-3mm}
		\phantom{ }
		
		\vspace{-6mm}
		\hrulefill\vspace{-3mm}
	\end{figure*}
	
	\subsection{Placing Constraints on First Layer Commands}\vspace{-4mm}
	
	A cursory inspection of Figure~\ref{fig:NRF_implem} reveals the identity\vspace{-3mm}
	\begin{equation}\label{eq:app_cmd}\tag{29}
		u_i[k]=u_{fi}[k]+u_{s2i}[k]+S^\top_{{ui}}\beta_{s2}[k],\,\forall\,i\in\{1:N\}.\vspace{-3mm}
	\end{equation}
	For the state dynamics obtained in Section~\ref{sec:transform}, the problem of constraining the command signals sent to the network reduces to judiciously constraining the control effort of both the first and the second layer. Thus, for each area, we choose the pair of sets introduced in \eqref{eq:area_prob}, namely \vspace{-3mm}
	\begin{subequations}
		\begin{align}\label{eq:cmd_sets_1}
			&{\mathcal{U}}_{s1i}\subseteq\mathbb{R}^{n_{xi}}\text{ s.t. }u_{s1i}[k]\in{\mathcal{U}}_{s1i},\,\forall\,i\in\{1:N\},\\
			&{\mathcal{U}}_{s2i}\subseteq\mathbb{R}^{n_{ui}}\text{ s.t. }u_{s2i}[k]\in{\mathcal{U}}_{s2i},\,\forall\,i\in\{1:N\}.\label{eq:cmd_sets_2}
		\end{align}
	\end{subequations}
	\vspace{-12mm}
	
	\noindent
	and we employ the ones from \eqref{eq:cmd_sets_2} in order to compute\vspace{-3mm}
	\begin{equation}\label{eq:NRF_cmd_sets}
		{\mathcal{U}}_{fi}:=\mathcal{U}_i\ominus(\,\mathcal{U}_{s2i}\oplus(S^\top_{{ui}}\mathcal{D}_{\beta_{s2}})),\,\forall\,i\in\{1:N\}.\vspace{-3mm}
	\end{equation}
	Notice that we have $u_{fi}[k]\in{\mathcal{U}}_{fi}\Longrightarrow u_{i}[k]\in\mathcal{U}_{i}$, which allows us to translate the objectives given in \eqref{eq:set_equiv} to the output vectors of our area dynamics from \eqref{eq:disc_cl_dyn_a}-\eqref{eq:disc_cl_dyn_d}. \vspace{-3mm}
	
	\begin{remark}\label{rem:cmd_sets}
		\hspace{-1mm}The \emph{freely chosen sets} given in \eqref{eq:cmd_sets_1}-\eqref{eq:cmd_sets_2} are the \emph{main design parameters} of the second layer. They can be assimilated with the extent to which the second layer is permitted to interfere in the dynamics of the first layer's closed-loop system, so as to ensure that \eqref{eq:set_equiv} is satisfied. The manner in which these sets may be chosen, and the resulting trade-off, will be addressed in the sequel.\vspace{-4mm}
	\end{remark}

	\subsection{Placing Constraints on First-Layer States}\vspace{-4mm}
	
	Since the initial conditions of the first layer's state-space implementation explicitly impact the signals located on the left-hand side of \eqref{eq:cl_dyn}, it follows that the state vectors of the NRF-based subcontrollers from \eqref{eq:area_NRF} should also possess a set-based description, which will be employed to ensure the satisfaction of the inclusions from \eqref{eq:set_equiv}.\vspace{-3mm}
	
	{\color{black}
		\begin{remark}
			The results presented in this section represent one of the key theoretical novelties introduced in this paper. In the classical tube MPC setting (see, for example, \cite{R8A,R8B,R8D}), the command constraints on the signals analogous to $u_{fi}[k]$ are straightforward to compute by means of constant matrices. In our framework, however, it becomes necessary to pass through the dynamics in \eqref{eq:area_NRF}, which significantly complicates the resulting expressions. Given the delicate nature of this undertaking, said expressions are presented, within the current section, in as explicit a manner as possible, regardless of the apparent similarity with standard tube MPC (recall the first paragraph of Section~\ref{sec:transform}). Nevertheless, the benefits of NRF-based control (as stated in Remark~\ref{rem:NRF_benef}) more than make up for the increase in complexity when formulating the synthesis procedure (and not, we point out, when implementing the control laws). Indeed, a further advantage of our approach is the fact that, instead of limiting ourselves to block-diagonal and constant stabilising feedback (see Assumption~1 in \cite{R8A,R8B,R8D}), we leverage a greater degree of freedom in considering dynamical stabilising feedback \eqref{eq:area_NRF} endowed with generic sparsity patterns \eqref{eq:mat_coef}.
			\vspace{-3mm}
		\end{remark}
	}
	
	We begin by inspecting the distinctive structure of the matrices which make up the realisations from \eqref{eq:area_NRF}. The latter enables us to naturally partition each of the state vectors belonging to the first-layer subcontrollers into $w_i[k]=\tiny\begin{bmatrix}
		w_{r(\alpha_{ui}+1)}^\top[k] & \dots & w_{r(\alpha_{ui}+n_{ui})}^\top[k]
	\end{bmatrix}$, in which the vectors $w_{r\ell}[k]$ have $n_{r\ell}$ rows for all $\ell\in\{1:n_u\}$, \emph{i.e.}, the same as the matrices $A_{r\ell}$ and $B_{r\ell}$ from \eqref{eq:mat_coef}. Recalling \eqref{eq:mat_coef}, the key observation which underpins the entire mathematical mechanism presented in this section is the fact that
	$\mathrm{row}_\ell(u_f[k])=C_{r\ell}w_{r\ell}[k]=\mathrm{row}_1(w_{r\ell}[k])$, for all $\ell\in\{1:n_u\}$. Thus, our current objective becomes apparent: we aim to employ the bounds placed upon the partitions of $u_f[k]$ to satisfy further set inclusions with respect to $w_{r\ell}[k]$, in a distributed manner.\vspace{-3mm}

	To state the main result of this section, we introduce the sets ${\mathcal{W}}_{\ell1}\subseteq\mathbb{R}$, with $\ell\in\{1:n_u\}$, which satisfy\vspace{-3mm}
	\begin{equation}\label{eq:Wi_sets}\tag{32}
		\mathrm{row}_\ell({u}_{f}[k])\in \mathcal{W}_{\ell1},\,\forall\,\ell\in\mathcal{A}_{ui} \Longrightarrow  u_{fi}[k]\in{\mathcal{U}}_{fi},\vspace{-3mm}
	\end{equation}
	for all $i\in\{1:N\}$. The following result employs the newly introduced sets to constrain the state variables of the first-layer subcontrollers.\vspace{-3mm}
	
	\begin{proposition}\label{prop:NRF_state_manip}
		Let the first-layer subcontrollers be implemented as in \eqref{eq:Kd_def}-\eqref{eq:dist_implem}. For each $i\in\{1:N\}$, define ${Z}_{ri}:=\mathrm{diag}(S_{ui},S_{xi})$ along with the sets from \eqref{eq:feedback_set} and \eqref{eq:NRF_state_sets_a}-\eqref{eq:NRF_state_sets_b}, which are located at the top of this page. Assume, moreover, that:\vspace{-3mm}
		\begin{enumerate}
			\item[A1)] the inclusions given in \eqref{eq:dist_sets}, \eqref{eq:cmd_sets_1}-\eqref{eq:cmd_sets_2} and \eqref{eq:feedback_set} hold for $k=k_0$,\smallskip
			
			\item[A2)] $\mathrm{row}_j(w_{r\ell}[k_0])\in\mathcal{W}_{\ell j},\,\forall\,\ell\in\mathcal{A}_{ui},\,j\in\{1:n_{r\ell}\}$,\smallskip
			
			\item[A3)] $\mathrm{row}_\ell({u}_{f}[k_0+1])\in \mathcal{W}_{\ell1},\,\forall\,\ell\in\mathcal{A}_{ui},\,i\in\{1:N\}$,\vspace{-3mm}
		\end{enumerate}
		given some $k_0\in\mathbb{Z}$. Then, $\forall\,i\in\{1:N\}$, we have that:\vspace{-3mm}
		\begin{enumerate}
			\item[i)] $\mathrm{row}_j(w_{r\ell}[k_0+1])\in\mathcal{W}_{\ell 1},\,\forall\,\ell\in\mathcal{A}_{ui},\,j\in\{1:n_{r\ell}\}$;\smallskip
			
			\item[ii)] $u_{i}[k_0]\in{\mathcal{U}}_{i}$.\vspace{-6mm}
		\end{enumerate}
	\end{proposition}
	\begin{pf}
		See Appendix~\ref{app:proofs}.\vspace{-3mm}
	\end{pf}

	\begin{remark}\label{rem:comp_NRF_sets}
		Note that the $i^\text{th}$ area's sets given in \eqref{eq:NRF_state_sets_a}-\eqref{eq:NRF_state_sets_b} are computed efficiently, using only for those indices belonging to $\mathcal{N}_i$. This efficient computation is owed to the sparsity structure displayed in \eqref{eq:dist_implem}, which incentivises the synthesis of sparse control laws for the first layer (see Appendix~\ref{app:NRF} and \cite{part1}). {\color{black}Note that the bounds associated with these sets could also be tightened by adapting the ideas presented in \cite{R8B}, through the sequential reduction of those $\mathcal{X}_j$, $\mathcal{U}_{j}$, $\mathcal{U}_{s1j}$ and $\mathcal{U}_{s2j}$ for which $j\in\mathcal{N}_i$.}\vspace{-3mm}
	\end{remark}
	
	The result formalised via Proposition~\ref{prop:NRF_state_manip} will prove instrumental in guaranteeing the recursive feasibility of the second layer and, in retrospect, the choice to impose constraints upon $u_{fi}[k]$, instead of the $u_i[k]$ vectors in \eqref{eq:set_equiv}, becomes justified. Not only do the \emph{specialised implementations} presented in \eqref{eq:area_NRF} allow us to construct set-based descriptions \eqref{eq:NRF_state_sets_a}-\eqref{eq:NRF_state_sets_b} for the states belonging to the first-layer implementations, but they also enable us to \emph{explicitly link} the sets mentioned above to the second layer's design objectives. By eliminating \emph{the need to individually select} each of these sets (which is an intricate and highly empirical problem, in the distributed setting), the design procedure becomes \emph{greatly simplified}.\vspace{-3mm}
	
	{\color{black}
		\begin{remark}\label{rem:tradeoff}
			By inspecting Figure~\ref{fig:scheme} along with the various sets defined in this section, the intended use of $u_{s1i}[k]$ and $u_{s2i}[k]$ becomes clear. Indeed, just as $u_{s2i}[k]$ is primarily meant to {ensure that $x_i[k]\in\mathcal{X}_i$}, so too is $u_{s1i}[k]$ chiefly responsible for {ensuring that $\mathrm{row}_j(w_{r\ell}[k])\in\mathcal{W}_{\ell j}$ and, implicitly, that $u_i[k]\in\mathcal{U}_i$ (recall Proposition~\ref{prop:NRF_state_manip})}.\vspace{-3mm}
		\end{remark}
	}

	\subsection{Addressing Uncertainty in the Initial Conditions}\vspace{-3mm}
	
	Having obtained the means by which we will satisfy the inclusions given in \eqref{eq:set_equiv}, we now turn our attention to the disturbance discussed in Sections~\ref{subsec:ci_dist} and~\ref{subsec:aug_pred}. In order to account for the inaccuracies modelled in \eqref{eq:noise_init}, we begin by exploiting the expressions from \eqref{eq:disc_cl_dyn_4} to compute \vspace{-3mm}\stepcounter{equation}\stepcounter{equation}\stepcounter{equation}
	\begin{equation}\label{eq:noise_sets}
		\left\{
		\begin{aligned}
			\mathcal{H}_{x i t}\subseteq\mathbb{R}^{n_{xi}}\text{ s.t. }\eta_{x i}[k_0+t]\in\mathcal{H}_{x i t},\,\forall\,\tiny\begin{bmatrix}
				\nu_x^\top \  \nu_w^\top
			\end{bmatrix}^\top\in\mathcal{V},\\
			\mathcal{H}_{u i t}\subseteq\mathbb{R}^{n_{ui}}\text{ s.t. }\eta_{ui}[k_0+t]\in\mathcal{H}_{u i t},\,\forall\,\tiny\begin{bmatrix}
				\nu_x^\top \  \nu_w^\top
			\end{bmatrix}^\top\in\mathcal{V},
		\end{aligned}
		\right.
	\end{equation}
	which allows us to define, for all $i\in\{1:N\}$, the sets\vspace{-3mm}
	\begin{equation}\label{eq:tilde_sets}
		\widetilde{\mathcal{X}}_{it}:={\mathcal{X}}_i\ominus(-\mathcal{H}_{x i t}),\vspace{-5mm}
	\end{equation}
	and\vspace{-2mm}
	\begin{equation}\label{eq:Wi_tilde_sets}
		\widetilde{\mathcal{W}}_{\ell1t}:=\mathcal{W}_{\ell1}\ominus\big(-e^\top_{(\ell-\alpha_{ui})}\mathcal{H}_{u i t}\big),\,\forall\,\ell\in\mathcal{A}_{ui}.\vspace{-3mm}
	\end{equation}
	Thus, our problem reduces to employing the dynamics from \eqref{eq:noise_pert_1}-\eqref{eq:noise_pert_2} with the aim of constraining $\widetilde x_i[k]$ and $\widetilde{u}_{fi}[k]$ to the sets obtained in \eqref{eq:tilde_sets} and \eqref{eq:Wi_tilde_sets}, respectively. Recalling the identities from \eqref{eq:noise_pert_3} directly yields the desired inclusions stated in \eqref{eq:set_equiv}. All that remains is to compute containing sets for the signals from \eqref{eq:disc_cl_dyn_2p}-\eqref{eq:disc_cl_dyn_3p}. \vspace{-3mm}
	
	\begin{figure*}
		\begin{equation}\label{eq:state_sets}\tag{44}
			\Xi_{it}:=\left\{\phi\in\mathbb{R}^{n_{si}}\ \left|{\ \scriptsize\begin{bmatrix}
					H_{1it}C_{xi}\\H_{2\left(\alpha_{ui}+1\right)t}\mathrm{row}_1(C_{ui})\\\vdots\\H_{2\left(\alpha_{ui}+n_{ui}\right)t}\mathrm{row}_{n_{ui}}(C_{ui})
			\end{bmatrix}}\right.\phi\leq\scriptsize\begin{bmatrix}
				h_{1it}-H_{1it}\widetilde{\theta}_{x i}[k_0+t]\\h_{2\left(\alpha_{ui}+1\right)t}-H_{2\left(\alpha_{ui}+1\right)t}\mathrm{row}_1\left(\widetilde{\theta}_{ui}[k_0+t]\right)\\
				\vdots\\h_{2\left(\alpha_{ui}+n_{ui}\right)t}-H_{2\left(\alpha_{ui}+n_{ui}\right)t}\mathrm{row}_{n_{ui}}\left(\widetilde{\theta}_{ui}[k_0+t]\right)
			\end{bmatrix}\right\},\forall\,t\in\{1:T_i\}.\vspace{-1mm}
		\end{equation}\vspace{-2mm}
		\hrulefill\vspace{-1mm}
	\end{figure*}
	
	\subsection{Placing Constraints on Area Cross-coupling}\vspace{-3mm}
	
	To provide compact formulations in the sequel, we first construct a collection of sets $\mathcal{W}_i\subseteq\mathbb{R}^{n_{wi}}$ which satisfy\vspace{-3mm}
	\begin{equation}\label{eq:Wi_suf_sets}
		\hspace{-2mm}\scriptsize\begin{array}{l}
			\mathrm{row}_j(w_{r\ell}[k])\in\mathcal{W}_{\ell j},\forall j\in\{1:n_{r\ell}\},\ell\in\mathcal{A}_{ui}\Rightarrow w_i[k]\in\mathcal{W}_i,
		\end{array}\normalsize\hspace{-1mm}\vspace{-3mm}
	\end{equation}
	for all $i\in\{1:N\}$, by simply taking the Cartesian product of the $\mathcal{W}_{\ell j}$ sets appearing in \eqref{eq:Wi_suf_sets}. Since the chief goal of the second layer is to constrain the state variables of the network and those of the implementations given in \eqref{eq:area_NRF} to the sets $\mathcal{X}_i$ and $\mathcal{W}_i$, respectively, we consider that the initial conditions of the first layer's closed-loop system satisfy $x_{ci}\in\mathcal{X}_i$ along with $w_{ci}\in\mathcal{W}_i$, for all $i\in\{1:N\}$. In this case, the definitions from \eqref{eq:noise_init} yield\vspace{-3mm}
	\begin{subequations}
		\begin{align}\label{eq:pert_ci_sets_a}
			&\hspace{-1mm}\widetilde{x}_{ci}={x}_{ci}+S^\top_{{xi}}\nu_x\in\widetilde{\mathcal{X}}_{ci}:=\mathcal{X}_i\oplus\tiny\begin{bmatrix}
				S^\top_{{xi}} & O
			\end{bmatrix}\mathcal{V},\hspace{-1mm}\\
			&\hspace{-1mm}\widetilde{w}_{ci}={w}_{ci}+S^\top_{{wi}}\nu_w\in\widetilde{\mathcal{W}}_{ci}:=\mathcal{W}_i\oplus\tiny\begin{bmatrix}
				O & S^\top_{{wi}}
			\end{bmatrix}\mathcal{V},\hspace{-1mm}\label{eq:pert_ci_sets_b}
		\end{align}
	\end{subequations}
	\phantom{ }
	
	\vspace{-9mm}
	for all $i\in\{1:N\}$, which allows us to construct\vspace{-3mm}
	\begin{equation}\label{eq:ci_sets}
		\left\{
		\begin{aligned}
			&\widetilde\Theta_{x i t}\subseteq\mathbb{R}^{n_{xi}}\text{ s.t. }\widetilde\theta_{x i}[k_0+t]\in\widetilde\Theta_{x i t},\\
			&\qquad\qquad\qquad\qquad\quad\forall\,\widetilde{x}_{cj}\in\widetilde{\mathcal{X}}_{cj},\ \widetilde{w}_{cj}\in\widetilde{\mathcal{W}}_{cj},\ j\in\mathcal{N}_i,\\
			&\widetilde\Theta_{u i t}\subseteq\mathbb{R}^{n_{ui}}\text{ s.t. }\widetilde\theta_{ui}[k_0+t]\in\widetilde\Theta_{u i t},\\
			&\qquad\qquad\qquad\qquad\quad\forall\,\widetilde{x}_{cj}\in\widetilde{\mathcal{X}}_{cj},\ \widetilde{w}_{cj}\in\widetilde{\mathcal{W}}_{cj},\ j\in\mathcal{N}_i,
		\end{aligned}
		\right.\normalsize\vspace{-2mm}
	\end{equation}
	along with\vspace{-3mm}
	\begin{equation}\label{eq:cross_sets}
		\left\{
		\begin{aligned}
			&\widetilde\Delta_{x i t}\subseteq\mathbb{R}^{n_{xi}}\text{ s.t. }\widetilde\delta_{x i}[k_0+t]\in\widetilde\Delta_{x i t},\\
			&\qquad \forall\,u_{s1j}[k]\in{\mathcal{U}}_{s1j},\,u_{s2j}[k]\in{\mathcal{U}}_{s2j},\,j\in\{1:N\}\setminus\{i\},\\
			&\qquad \forall\,\widetilde{x}_{c\ell}\in\widetilde{\mathcal{X}}_{c\ell},\ \widetilde{w}_{c\ell}\in\widetilde{\mathcal{W}}_{c\ell},\ j\in\{1:N\}\setminus\,\mathcal{N}_i,\\
			&\widetilde\Delta_{u i t}\subseteq\mathbb{R}^{n_{ui}}\text{ s.t. }\widetilde\delta_{ui}[k_0+t]\in\widetilde\Delta_{u i t},\\
			&\qquad \forall\,u_{s1j}[k]\in{\mathcal{U}}_{s1j},\,u_{s2j}[k]\in{\mathcal{U}}_{s2j},\,j\in\{1:N\}\setminus\{i\},\\
			&\qquad \forall\,\widetilde{x}_{c\ell}\in\widetilde{\mathcal{X}}_{c\ell},\ \widetilde{w}_{c\ell}\in\widetilde{\mathcal{W}}_{c\ell},\ \ell\in\{1:N\}\setminus\,\mathcal{N}_i.
		\end{aligned}
		\right.\normalsize\vspace{-3mm}
	\end{equation}

	{\color{black}
		\begin{remark}
			In choosing the sets from \eqref{eq:cmd_sets_2}, one negotiates a command-budgeting balance (as limited by $\mathcal{U}_i$) between the first and second layers. This choice induces a trade-off associated with selecting the sets from both \eqref{eq:cmd_sets_1} and\eqref{eq:cmd_sets_2}, since the more freedom to act $u_{s1i}[k]$ is given, via the set from \eqref{eq:cmd_sets_1}, its effect will expand throughout the sets from \eqref{eq:feedback_set} and \eqref{eq:NRF_state_sets_a}-\eqref{eq:NRF_state_sets_b}. Similarly, if the network's areas are not all \emph{perfectly decoupled} (recall Remark~\ref{rem:decup}), then too much leeway given to $u_{s1i}[k]$ and to $u_{s2i}[k]$ may result in significant area cross-coupling. In this case, one may derive an extended version of the technique from \cite{R8C} (which demonstrates its effectiveness for only two collaborating areas), so as to mitigate these particular aspects. Section~4 in \cite{R8C} also provides valuable insight into negotiating this particular type of trade-off.\vspace{-3mm}
		\end{remark}
	}

	With all of these sets now defined according to their corresponding areas, we now move on to the following section, in which we present guarantees for the (recursive) satisfaction of the second layer's constraints.\vspace{-3mm}
	
	\section{\hspace{-2mm}Distributed Implementation of the MPC Layer}
	\label{sec:MPC_imp}\vspace{-3mm}
	
	In this section, we proceed to show how to ensure \emph{global} theoretical guarantees for the first layer's closed-loop system, by \emph{independently} implementing an MPC controller for each area's dynamics described in \eqref{eq:disc_cl_dyn_a}-\eqref{eq:disc_cl_dyn_d}. The results presented here are inspired by the core idea investigated in \cite{AI}, and our proposed technique represents the MPC-oriented analogue of the indicated work.
	\vspace{-3mm}
	
	{\color{black}
		\begin{remark}\label{rem:SLS}
			As discussed in this manuscript's companion paper \cite{part1}, our approach exhibits several notable similarities with the SLS-based procedure from \cite{DMPC1,DMPC2}. Indeed, in both instances, the focus rests upon designing the network's closed-loop response in accordance with a number of structural restrictions. Additionally, the \emph{d-locality constraints} from \cite{DMPC1,DMPC2} closely match our own area-based sparsity patterns discussed in Remark~\ref{rem:decup}, while the feasibilities of these two approaches are also closely intertwined, since both the SLS \cite{SLS} and the Youla Parametrisation (upon which the first layer is based, see \cite{NRF}) yield the set of all achievable and stable closed-loop maps (see also \cite{Luca3}). On the other hand, our procedure leverages the fact that the NRF-based layer can \emph{always be designed and implemented offline}, thus relegating all online computation to the significantly less resource-demanding MPC layer, as shown in the sequel. This feature stands in stark contrast to the framework presented in \cite{DMPC1,DMPC2}, in which sparsity constraints must be factored into the online computation, alongside all of the MPC-like constraints.\vspace{-3mm}
		\end{remark}
	}
	
	{
		\color{black}
		\subsection{Distributed Enforcement of Constraint Satisfaction}\vspace{-3mm}
		
		We begin by addressing the final and most important component of the formulations introduced in \eqref{eq:area_prob}. Recall that, in stating the objectives of our control architecture's second layer, the optimisation problems from \eqref{eq:area_prob} make use of the sets $\Xi_{it}$ to ensure the satisfaction of the constraints related to $x[k]$ and $u[k]$. The aforementioned sets, whose precise expressions have been deferred up to this point, may now finally be stated in a formal manner. To do so, we first compute the following sets\vspace{-3mm}
		\begin{equation}\label{eq:inner_appr_1}
			\left\{\begin{aligned}
				&\mathcal{P}_{1it}:=\{v\in\mathbb{R}^{n_{xi}}\ \vert\ H_{1it}v\leq h_{1it}\},\\
				&\mathcal{P}_{1it}\subseteq\widetilde{\mathcal{X}}_{it}\ominus(\Psi_{x i t}\oplus\widetilde{\Delta}_{x i t}),
			\end{aligned}
			\right.\vspace{-4mm}
		\end{equation}
		and\vspace{-2mm}
		\begin{equation}\label{eq:inner_appr_2}
			\hspace{-2mm}\left\{\begin{aligned}
				&\mathcal{P}_{2\ell t}:=\{v\in\mathbb{R}\ \vert\ H_{2\ell t}v\leq h_{2\ell t}\},\\
				&\mathcal{P}_{2\ell t}\subseteq\widetilde{W}_{\ell  1t}\ominus\big(\big(e^\top_{(\ell-\alpha_{ui})}\Psi_{u i t}\big)\oplus\big(e^\top_{(\ell-\alpha_{ui})}\widetilde{\Delta}_{u i t}\big)\big),
			\end{aligned}
			\right.\hspace{-2mm}\vspace{-3mm}
		\end{equation}
		for every index $\ell\in\mathcal{A}_{ui}$ and any set of real-valued tuples $\tiny\begin{array}{c}
			(H_{1it},h_{1it})\in\mathbb{R}^{q_{1it}\times n_{xi}}\times\mathbb{R}^{q_{1it}}
		\end{array}$ and $\tiny\begin{array}{c}
			(H_{2\ell t},h_{2\ell t})\in\mathbb{R}^{q_{2\ell t}}\times\mathbb{R}^{q_{2\ell t}}
		\end{array}$ which satisfy \eqref{eq:inner_appr_1}-\eqref{eq:inner_appr_2}. These tuples allow us to define $\Xi_{it}$ as in \eqref{eq:state_sets}, located at the top of this page.\vspace{-3mm}
		
		{\color{black}
			\begin{remark}\label{rem:NRF_feas}
				We implicitly assume that the first-layer closed-loop system has been designed so as to ensure that the sets from \eqref{eq:inner_appr_1}-\eqref{eq:state_sets} are \emph{non-empty}. If this is not the case, the reason is that the first layer has not been appropriately tuned to address cross-coupling and exogenous disturbance. Provided that the sets $\mathcal{X}$ and $\mathcal{U}$ (along with their area-based decompositions) are available during the first layer's design phase, a judicious choice of systemic norm optimisation in Section~5.2 of \cite{part1} (tied to the nature of these sets) can be used to ensure non-emptiness. Indeed, given that our first layer is specifically designed to improve upon the open-loop response characteristics of the network described by \eqref{eq:ss_a}-\eqref{eq:ss_c}, our assumption is by no means restrictive. This fact also highlights one of the key benefits of our approach: constraint feasibility is certified after first-layer synthesis, with any potential infeasibility being detected early on in the design procedure.
				\vspace{-7mm}
			\end{remark}
		}
		
		With all of their components clarified, and assuming that every problem described in \eqref{eq:area_prob} is feasible for each time instant $k\geq k_0$ (a key property that will be rigorously addressed in the sequel), each MPC-based subcontroller in our second layer can be implemented as per Algorithm~\ref{alg:MPC_implem}.\vspace{-3mm}
	}

	\begin{algorithm}[ht]
		
		\KwData{The $i^\text{th}$ network's area model and the design parameters of the optimisation problem from \eqref{eq:area_prob}}\smallskip
		
		\KwResult{The values of $u_{si1}[k]$ and $u_{si2}[k]$}\smallskip
		
		\textbf{Step 1a:} Measure the noise-affected initial condition vector $\begin{bmatrix}
			\widetilde{x}_{ci}^\top & \widetilde{w}_{ci}^\top
		\end{bmatrix}^\top$ and transmit it to the subcontrollers of all network areas of index $j\neq i$, for which $i\in\mathcal{N}_j$;\smallskip
		
		\textbf{Step 1b:} Receive the noise-affected initial condition vectors $\begin{bmatrix}
			\widetilde{x}_{cj}^\top & \widetilde{w}_{cj}^\top
		\end{bmatrix}^\top$ from the subcontrollers of all network areas of index $j\in\mathcal{N}_i\setminus\{i\}$;\smallskip
		
		\textbf{Step 2:} Form the vectors $\widetilde{\theta}_{x i}[k+t]$ and $\widetilde{\theta}_{u i}[k+t]$ for all $t\in\{1:T_i\}$ and use them to construct the sets in \eqref{eq:state_sets};\smallskip
		
		\textbf{Step 3:} Solve \eqref{eq:area_prob} for two sequences of optimal command signal vectors $u^\star_{s1i}[k+t-1]$ and $u^\star_{s2i}[k+t-1]$, in which $t\in\{1:T_i+\overline T_i\}$;\smallskip
		
		\textbf{Step 4a:} Set $u_{s1i}[k]\leftarrow u^\star_{s1i}[k]$ and $u_{s2i}[k]\leftarrow u^\star_{s2i}[k]$, then transmit $u_{s1i}[k]$ to the subcontrollers of all network areas of index $j\neq i$, for which $i\in\mathcal{N}_j$;\smallskip
		
		\textbf{Step 4b:} Receive $u_{s1j}[k]$ from the subcontrollers of all network areas of index $j\in\mathcal{N}_i\setminus\{i\}$, then pass them along to the local first-layer subcontroller;\smallskip
		
		\textbf{Step 5:} Wait for the start of the following sampling period and return to \textbf{Step 1};
		
		\caption{Implementation of the $i^\text{th}$ subcontroller}
		\label{alg:MPC_implem}
	\end{algorithm}\vspace{-1mm}
	
	\begin{remark}\label{rem:benef}
		It is straightforward to notice that the tuples formed by $(H_{1it},h_{1it})$ and $(H_{2\ell t},h_{2\ell t})$ characterise polyhedral inner-approximations of the sets on the right-hand sides of \eqref{eq:inner_appr_1}-\eqref{eq:inner_appr_2}. Crucially, these pairs can always be determined \emph{offline}, which greatly simplifies the online construction of the sets from \eqref{eq:state_sets} in \textbf{\emph{Step 3}} of Algorithm~\ref{alg:MPC_implem}. {\color{black} In addition to this computational benefit, note that no explicit construction of any invariant sets occurs in our setting. For especially large or highly interconnected networks, the type of invariant sets discussed in \cite{R8A} or \cite{R8C} may be highly complex, or even numerically intractable for practical implementations. Our framework allows us to bypass explicit invariance-related concerns, and it allows us to control the complexity of the sets in \eqref{eq:state_sets}, by means of the inner approximations in \eqref{eq:inner_appr_1}-\eqref{eq:inner_appr_2}.}\vspace{-3mm}
	\end{remark}
	
	Although the implementations proposed in Algorithm~\ref{alg:MPC_implem} show that the MPC-based subcontrollers interact with their counterparts from neighbouring areas, with the aim of exchanging local state measurements, we highlight the fact that all of the \emph{individual} command signals are computed \emph{separately}, in a distributed manner.\vspace{-3mm}
	
	{\color{black}\begin{remark}\label{rem:unc_hor}
			Regarding the computation of said command signals, the MPC formulations provided in \eqref{eq:area_prob} are particularly flexible, in the context of (distributed) constraint satisfaction. The two prediction horizons, introduced in point c) of Section~\ref{subsec:MPC_obj} and associated with these problems, permit the careful management of the computational burdens which come with solving \eqref{eq:area_prob}. Since we are chiefly interested in satisfying the latter's constraints only for the sake of recursive feasibility (see the numerical example presented in the sequel), it is possible to improve the closed-loop performance of the MPC-based layer by increasing the unconstrained prediction horizon without significantly impacting the overall computational cost.\vspace{-3mm}
	\end{remark}}
	
	The following result shows that the \emph{local} efforts of these subcontrollers ensure \emph{global} theoretical guarantees.\vspace{-3mm}
	
	\begin{proposition}\label{prop:one_step_feas}
		For each of the area dynamics described in \eqref{eq:disc_cl_dyn_a}-\eqref{eq:disc_cl_dyn_d}, let an MPC subcontroller be implemented as in Algorithm~\ref{alg:MPC_implem} and let the communicated information between these subcontrollers be affected as in \eqref{eq:noise_init}. Assume that:\vspace{-3mm}
		\begin{enumerate}
			\item[A1)] The first layer's initial condition at time $k=k_0$ satisfies $x_c \in\mathcal{X}$ along with $\mathrm{row}_j(w_{cr\ell})\in\mathcal{W}_{\ell j}$, for all $\ell\in\{1:n_u\}$ and all $j\in\{1:n_{r\ell}\}$;\smallskip
			
			\item[A2)] All of the optimisation problems described in \eqref{eq:area_prob} are feasible at time $k=k_0$.\vspace{-3mm}
		\end{enumerate}
		Then, by computing all partitions of $u_{s1}[k_0]$ and $u_{s2}[k_0]$ as in Algorithm~\ref{alg:MPC_implem} and by applying these two vectors as shown in Figure~\ref{fig:scheme}, it follows that $x[k_0+1]\in\mathcal{X}$, $u[k_0]\in\mathcal{U}$ and $\mathrm{row}_j(w_{r\ell}[k_0+1])\in\mathcal{W}_{\ell j},\,\forall\,\ell\in\{1:n_u\},\,j\in\{1:n_{r\ell}\}$.\vspace{-6mm}
	\end{proposition}
	\begin{pf}
		See Appendix~\ref{app:proofs}.\vspace{-6mm}
	\end{pf}
	
	{\color{black}\begin{remark}
			Note that Assumption~A1) from Proposition~\ref{prop:one_step_feas} hinges upon the initial condition of the controlled network and upon the sets $\mathcal{X}_i$ in \eqref{eq:set_equiv}, since the initial conditions of the implementations in \eqref{eq:area_NRF} can be imposed at will. Should the system from \eqref{eq:ss_a}-\eqref{eq:ss_c} be initialised in some $x_{nf}\not\in\mathcal{X}$, the procedure presented in this paper and in \cite{part1} can also be used as a form of recovery mechanism. The key is to relax the state constraints given by $\mathcal{X}$ into some feasible and connected set $\mathcal{X}_{r}\supseteq(\mathcal{X}\cup\{x_{nf}\})$, while keeping the same $\mathcal{U}$ and resynthesising the second layer in accordance with the relaxed constraints. Since, in our framework, the first layer is responsible for disturbance rejection and reference tracking, the NRF-based control laws can then be used to drive the network's state within $\mathcal{X}$, while the second layer handles command limitation. The first layer's subcontrollers can then be reinitialised so as to satisfy Assumption~A1), and nominal operation may commence using the original second layer.\vspace{-3mm}
	\end{remark}}
	
	When starting from a point inside the state-constraining set $\mathcal{X}$, Proposition~\ref{prop:one_step_feas} shows that our control scheme is able to maintain the first layer's state variables inside $\mathcal{X}$ while also satisfying the command constraints encoded by $\mathcal{U}$, provided that all of the MPC-based problems showcased in \eqref{eq:area_prob} are feasible. Thus, given the critical importance of satisfying Assumption~A2) from Proposition~\ref{prop:one_step_feas}, we now proceed to investigate a set of conditions which offer strong theoretical guarantees regarding the feasibility of the optimisation problems from \eqref{eq:area_prob}.\vspace{-4mm}

	\subsection{Guaranteeing Recursive Feasibility}\vspace{-4mm}
	
	In order to extend the theoretical guarantees obtained in Proposition~\ref{prop:one_step_feas} beyond the initial time instant, we must ensure that each of the MPC-based control laws from \eqref{eq:area_prob} is feasible for every time instant ${\color{black}k\in\mathbb{Z}}$ with $k\geq k_0\in\mathbb{Z}$. This property, known as \emph{recursive feasibility} (see Section~12.3 in \cite{BBM}), plays a key role in our chosen approach and, consequently, the following theorem (which directly builds upon the results obtained in Proposition~\ref{prop:one_step_feas}) represents the paper's main theoretical result.\vspace{-3mm}
	
	{\color{black}
		
		\begin{theorem}\label{thm:ongoing_feas}
			For each of the area dynamics described in \eqref{eq:disc_cl_dyn_a}-\eqref{eq:disc_cl_dyn_d}, let an MPC subcontroller be implemented as in Algorithm~\ref{alg:MPC_implem} and let the communicated information between these subcontrollers be affected as in \eqref{eq:noise_init}. For all $i\in\{1:N\}$, define also the following sets:\stepcounter{equation}\vspace{-3mm}
			\begin{subequations}
				\begin{align}\nonumber
					\widetilde{\Theta}_{it}:=&\left\{\begin{bmatrix}
						\theta_x^\top & \theta_{u1}^\top & \dots & \theta_{un_{ui}}^\top
					\end{bmatrix}^\top \Big\vert\ \theta_x \in \widetilde{\Theta}_{xit},\right.\\
					&\qquad\quad\left.\theta_{uj}\in \left(e^\top_{j}\widetilde{\Theta}_{uit}\right),\, j\in\{1:n_{ui}\}\right\},\label{eq:feas_sets_def_1}\\\nonumber
					\widetilde{\mathcal{P}}_{it}:=&\left\{\begin{bmatrix}
						p_x^\top & p_{u1}^\top & \dots & p_{un_{ui}}^\top
					\end{bmatrix}^\top \Big\vert\ p_x \in {\mathcal{P}}_{1it},\right.\\
					&\qquad\quad\left.p_{uj}\in {\mathcal{P}}_{2(j+\alpha_{ui})t},\, j\in\{1:n_{ui}\}\right\},\label{eq:feas_sets_def_2}
				\end{align}
			\end{subequations}
			and assume that:\vspace{-3mm}
			\begin{enumerate}
				\item[A1)] The first layer's initial condition at time $k=k_0$ satisfies $x_c \in\mathcal{X}$ along with $\mathrm{row}_j(w_{cr\ell})\in\mathcal{W}_{\ell j}$, for all $\ell\in\{1:n_u\}$ and all $j\in\{1:n_{r\ell}\}$;\smallskip
				
				\item[A2)] For all $i\in\{1:N\}$, there exists $\rho_i\in\mathbb{N}$ with $\rho_i\geq 1$, such that the following inclusions hold:\vspace{-3mm}
				\begin{equation}\label{eq:feas_sets_include}
					\hspace{-10mm}\scriptsize\begin{array}{l}
						\widetilde{\Theta}_{it}\subseteq\hspace{-0.5mm}\left(\hspace{-0.5mm}\begin{bmatrix}
							-C_{xi}\\-C_{ui}
						\end{bmatrix}\displaystyle\bigoplus_{j=1}^{t}A_{si}^{j-1}\Big(\hspace{-0.5mm}(B_{s1i}\hspace{0.5mm}\mathcal{U}_{s1i})\hspace{-0.5mm}\oplus\hspace{-0.5mm} (B_{s2i}\hspace{0.5mm}\mathcal{U}_{s2i})\hspace{-0.5mm}\Big)\hspace{-0.5mm}\right)\hspace{-0.5mm}\oplus\widetilde{\mathcal{P}}_{it},
					\end{array}\normalsize\hspace{-2mm}\vspace{-2mm}
				\end{equation}
				for all $t\in\{1:\rho_i\}$.\vspace{-3mm}
			\end{enumerate}
			Then, by setting $T_i=\rho_i,\,\forall\,i\in\{1:N\}$, we have that:\vspace{-3mm}
			\begin{enumerate}
				\item[i)] Each of the optimisation problems described in \eqref{eq:area_prob} becomes recursively feasible;\smallskip
				
				\item[ii)] By successively computing all partitions of $u_{s1}[k]$ and $u_{s2}[k]$ as in Algorithm~\ref{alg:MPC_implem} and by applying these two vectors as shown in Figure~\ref{fig:scheme}, it follows that $x[k+1]\in\mathcal{X}$, $u[k]\in\mathcal{U}$ and $\mathrm{row}_j(w_{r\ell}[k+1])\in\mathcal{W}_{\ell j}$, for all $\ell\in\{1:n_u\}$, $j\in\{1:n_{r\ell}\}$ and $k\geq k_0$.\vspace{-6mm}
			\end{enumerate}
		\end{theorem}
		\begin{pf}
			See Appendix~\ref{app:proofs}.\vspace{-6mm}
		\end{pf}
		
	}
	
	{\color{black}\begin{remark}\label{rem:horizon}
			Notice the fact that it is possible to retrieve the above-mentioned guarantees by taking $\rho_i=1$, for every $i\in\{1:N\}$, resulting in exceptionally efficient implementations (see the numerical example from the sequel) from a computational standpoint. Indeed, as the values of $\rho_i$ increase, not only does \eqref{eq:feas_sets_include} become more challenging to satisfy, but the additional constraints associated with taking $T_i=\rho_i$ may dramatically increase the cost of solving \eqref{eq:area_prob}, depending on the complexity of the sets in \eqref{eq:state_sets}. As discussed in Remark~\ref{rem:unc_hor}, taking larger values for $\overline{T}_i$ may yield similar degrees of performance as increasing $T_i$, without significantly impacting computational costs.
			\vspace{-2mm}
	\end{remark}}
	
	To summarise the design workflow of our architecture's second layer, we provide a high-level algorithm (located at the top of the next page) that condenses the myriad topics discussed in this paper into a sequence of steps.\vspace{-2mm}
	
	{\color{black}\begin{remark}\label{rem:relax_codep}
			Performing \textbf{\emph{Steps 3a-3d}} in Algorithm~\ref{alg:MPC_des} makes the paramount importance of our architecture's first layer readily apparent. It is precisely due to the proper tuning of the NRF-based control laws that area cross-coupling may be diminished (and potentially rendered negligible, for all practical purposes) to such a degree that the MPC-based subcontrollers may be tuned individually via \eqref{eq:cmd_sets_1}-\eqref{eq:cmd_sets_2}. While this fact effectively bypasses the need to carefully construct the sets from \eqref{eq:cmd_sets_1}-\eqref{eq:cmd_sets_2} and it crucially prevents the design procedure of the MPC-based subcontrollers from being co-dependent in any significant way, we point out that said procedure does indeed require information related to the first-layer response of other areas. However, this information becomes fully available once the NRF-based control laws have been obtained, and it is precisely the role of these distributed controllers to ensure the feasibility (recall Remark~\ref{rem:NRF_feas}) 
			and relaxed co-dependency of the second layer's design.\vspace{-3mm}
		\end{remark}
	}

	\begin{algorithm}[!ht]
		
		\KwData{The first layer's closed-loop dynamics from \eqref{eq:cl_dyn}, the objective sets $\mathcal{X}$ and $\mathcal{U}$, along with the exogenous signal sets $\mathcal{V}$, $\mathcal{D}_s$ and the ones from \eqref{eq:dist_sets}}\smallskip
		
		\KwResult{MPC policies obtained by solving \eqref{eq:area_prob}}\smallskip
		
		\textbf{Initialisation:} The state-space dynamics \eqref{eq:disc_cl_dyn_a}-\eqref{eq:disc_cl_dyn_d};\smallskip
		
		\textbf{Step 1:} Split the sets $\mathcal{X}$ and $\mathcal{U}$ into their area-wise components, as shown in \eqref{eq:set_equiv};\smallskip
		
		\textbf{Step 2a:} Form the noise propagation sets given in \eqref{eq:noise_sets} and compute the sets from \eqref{eq:tilde_sets} and \eqref{eq:pert_ci_sets_a};\smallskip
		
		\textbf{Step 2b:} Form the disturbance propagation sets in \eqref{eq:pert_sets};\smallskip
		
		\textbf{Step 3a:} Choose the sets from \eqref{eq:cmd_sets_1}-\eqref{eq:cmd_sets_2};\smallskip
		
		\textbf{Step 3b:} Form the sets in \eqref{eq:NRF_cmd_sets}-\eqref{eq:feedback_set}, \eqref{eq:NRF_state_sets_a}-\eqref{eq:NRF_state_sets_b}, those from \eqref{eq:Wi_tilde_sets}-\eqref{eq:Wi_suf_sets} and the ones from \eqref{eq:pert_ci_sets_b}-\eqref{eq:cross_sets};\smallskip
		
		\textbf{Step 3c:} Compute the polyhedral inner-approximations given in \eqref{eq:inner_appr_1}-\eqref{eq:inner_appr_2} and the sets from \eqref{eq:feas_sets_def_1}-\eqref{eq:feas_sets_def_2};\smallskip
		
		\textbf{Step 3d:} Set $\rho_i=1$ and check \eqref{eq:feas_sets_include} for all $i\in\{1:N\}$;\smallskip
		
		\eIf{\eqref{eq:feas_sets_include} \emph{holds for all} $i\in\{1:N\}$}{
			go to \textbf{Step 4a};
		}{
			go to \textbf{Step 3a} and retune the sets from \eqref{eq:cmd_sets_1}-\eqref{eq:cmd_sets_2} in accordance with Remark~\ref{rem:tradeoff};
		}\smallskip
		
		\textbf{Step 4a:}  For all $i\in\{1:N\}$, try increasingly larger values for $\rho_i\geq 2$ and then check \eqref{eq:feas_sets_include};\smallskip
		
		\textbf{Step 4b:} For all $i\in\{1:N\}$, set $T_i$ as the largest found value of $\rho_i$ which satisfies \eqref{eq:feas_sets_include};\smallskip
		
		\textbf{Step 4c:} For all $i\in\{1:N\}$, select an appropriate $\overline T_i$ and the weighting matrices of the cost function in \eqref{eq:area_prob};
		
		\caption{MPC layer design procedure}
		\label{alg:MPC_des}
	\end{algorithm}
	
	{\color{black}\subsection{Computational Considerations}\label{subsec:compute}\vspace{-2mm}
		
		The following discussion revolves around the three core concepts that underpin the computational burden of our procedure: synthesis costs, design scalability, and implementation complexity.\vspace{-3mm}
		
		\subsubsection{Synthesis Costs}\vspace{-3mm}
		
		One of the main features associated with Algorithm~\ref{alg:MPC_des} is the fact that all of its constituent steps (along with the design phase of the first layer, see our companion paper \cite{part1}) can be performed \emph{offline}, and that any reconfiguration of the second layer's constraints or cost functions \emph{does not impact} the first layer's control laws (in contrast to the technique from \cite{DMPC1,DMPC2}, recalling Remark~\ref{rem:SLS}
		). The chief computational burden associated with our solution is the cost of computing the sets in \eqref{eq:NRF_state_sets_a}-\eqref{eq:NRF_state_sets_b}, along with those in \eqref{eq:inner_appr_1}-\eqref{eq:inner_appr_2}, for large-scale applications. Although the measures described in Remarks~\ref{rem:comp_NRF_sets} and~\ref{rem:benef} can alleviate some of the complexity associated with these operations, the latter remain heavily dependent on the shape and the structure of the involved sets, making it challenging to come up with a one-size-fits-all computational cost. In spite of this fact, switching to a homothetic \cite{homot} or even zonotope-based \cite{zonot} framework (while employing various approximations, either in terms of the operation or of the operands themselves) is guaranteed to have a significant, positive impact on the complexity of the synthesis procedure. For a practical and specific example of the costs associated with the design of the second layer, see the numerical example presented in the sequel, and the source files from its GitHub repository.\vspace{-3mm}
		
		\subsubsection{Design Scalability}\label{subsubsec:scal}\vspace{-3mm} While all of the steps which form Algorithm~\ref{alg:MPC_des} can, indeed, be iterated in a fully offline setting, yet another appealing feature is the fact that once the second layer's command budgeting has been negotiated via the selection of the sets from \eqref{eq:cmd_sets_1}-\eqref{eq:cmd_sets_2}, the subsequent set-based computations from Algorithm~\ref{alg:MPC_des} can be \emph{performed in parallel}. Thus, the design procedure of each area can proceed independently of the others, once the aforementioned choice has been made. This fact is a direct consequence of not employing any notions of joint set invariance (see, for example, \cite{R8C}) and of relying, instead, upon the decoupling-focused control laws discussed in \cite{part1}.\vspace{-3mm}
		
		\subsubsection{Implementation Complexity}\label{subsubsec:implem}\vspace{-3mm}
		Due to the separate implementations described in \eqref{eq:area_NRF} and in Algorithm~\ref{alg:MPC_implem}, a key feature of our proposed solution is the fact that the runtime cost of our architecture is \emph{divided} among the network's areas. Moreover, said computational burden is appealingly modest, since it reduces to several inexpensive operations, namely: four matrix-vector multiplications and two vector-vector additions for an area's first-layer subcontroller \eqref{eq:area_NRF}, along with the solving of the optimisation problem from \eqref{eq:area_prob} for each area's second-layer subcontroller. In addition to this, we point out that each of the latter problems boils down to a \emph{standard Quadratic Programming} (QP) optimisation problem, whenever the cost in \eqref{eq:area_prob} has a quadratic (and symmetrical) expression and the sets from \eqref{eq:cmd_sets_1}-\eqref{eq:cmd_sets_2} are chosen to be polyhedral in nature (which is always possible, as per Remark~\ref{rem:cmd_sets}). Although the overall complexity of this problem is highly dependent upon the number of constraints it includes (recall Remark~\ref{rem:benef}), note that numerous solvers and optimisation-based toolboxes (such as \cite{mosek} and \cite{YALMIP}, respectively) are widely available for the efficient solving of such standard problems.\vspace{-3mm}
	}
	
	\section{Numerical Example}
	\label{sec:examp}\vspace{-3mm}
	
	\subsection{Network Model and Control Objectives}
	\label{subsec:app}\vspace{-3mm}
	
	\subsubsection{The Dynamics of the Application}\vspace{-3mm}
	
	To showcase the effectiveness of the proposed design framework, we now focus on a numerical example centred on the vehicle platoon application discussed in \cite{plutonizare}. We consider a formation of $N=10$ identical vehicles, in which the continuous-time models from (22) in \cite{plutonizare} are configured with the following parameters: $m_i = 1$, $\tau_i = 0.1$ and $\sigma_i = 1,\,\forall\,i\in\{1:N\}$. The resulting zero-order-hold-discretised dynamics of each car, for a sampling period of $T_s = 0.1$ \textcolor{black}{and $\forall\,k\geq k_0=0$}, are given by\vspace{-3mm}
	\begin{equation*}
		\scriptsize\begin{bmatrix}
			p_i[k+1] \\ v_i[k+1] \\ \mu_i[k+1]
		\end{bmatrix}= \underbrace{\left[\scriptsize\begin{array}{rrr}
				1    &   0.1  & -0.0331\\
				0    &     1  & -0.5689\\
				0    &     0  &  0.3679
			\end{array}\right]}_{A_{car}}\scriptsize\begin{bmatrix}
			p_i[k] \\ v_i[k] \\ \mu_i[k]
		\end{bmatrix}+\underbrace{\scriptsize\begin{bmatrix}
				0.0381\\
				0.6689\\
				0.6321\\
		\end{bmatrix}}_{B_{car}}u_i[k],\normalsize\vspace{-3mm}
	\end{equation*}
	where $u_i[k]$ is the control signal of the $i^\text{th}$ car, $p_i[k]$ denotes its (absolute) position, $v_i[k]$ is its speed and $\mu_i[k]$ is the state of the $i^\text{th}$ car's actuator (the $\Phi$ terms in \cite{plutonizare}).\vspace{-3mm}
	
	In addition to these variables, we also consider the "position" of the $0^\text{th}$ car (which acts as a reference signal to the lead car in the platoon) denoted $p_0[k]$, the latter's increment in one time instant, denoted $\Delta p_0[k]$, and the length of the vehicle located in front of the platoon's $i^\text{th}$ car, denoted $\ell_i[k]$. Forming $\ell_v^\top[k]:=\begin{bmatrix}
		\ell_1[k]&\dots&\ell_N[k]
	\end{bmatrix}$, the dynamics of these variables are captured by\vspace{-3mm}
	\begin{equation*}
		\scriptsize\begin{bmatrix}
			p_0[k+1] \\ \ell_v[k+1]
		\end{bmatrix} = I_{N+1} \scriptsize\begin{bmatrix}
			p_0[k] \\ \ell_v[k]
		\end{bmatrix} + e_1 \Delta p_0[k].\normalsize\vspace{-3mm}
	\end{equation*}
	For the homogeneous platoon considered in this application, we consider $\ell_i[k] = 5$ for all $i\in\{2:N\}$ along with $\ell_1[k] = 0$, since the $0^\text{th}$ (reference) car is a virtual one.\vspace{-3mm}
	
	\subsubsection{First-Layer Objectives and Platoon Model}\vspace{-3mm}
	
	In keeping with the control laws proposed in \cite{plutonizare}, we wish to designate each car as an individual area and to obtain a first layer in which the control signals are computed (in the absence of exogenous or second-layer signals) as\vspace{-3mm}
	\begin{equation}\label{eq:NRF_plat}
		\mathrm{row}_i(u_{f}[k]) = \left\{\tiny\begin{array}{l}
			\mathbf{\Gamma}_1(z)\star (y_1[k]+hv_1[k]),\ i=1,\\
			\mathbf{\Gamma}_i(z)\star (y_i[k]\,+\,hv_i[k]) + \mathbf{\Phi}_i(z)\star \mathrm{row}_{i-1}(u_{f}[k]),
		\end{array}\right.\normalsize
	\end{equation}
	for all $i\in\{2:N\}$, where $y_i[k]:=p_i[k]+\ell_i[k]-p_{i-1}[k]$ stands for the $i^\text{th}$ car's length-based interdistance and $h>0$ is the platoon's time headway (see \cite{plutonizare} and the references therein for its theoretical justification), which has been chosen as $h=5$ for the considered application.\vspace{-3mm}
	
	Given the control laws of type \eqref{eq:NRF_plat} we wish to obtain, and the fact that we will want to impose constraints on the $y_i$ signals (to prevent traffic collisions), we will model our platoon in such a way as to have the headway-based interdistances as state variables. Thus, we form\vspace{-3mm}
	\begin{equation*}
		T=\scriptsize\begin{bmatrix}
			I_{N+1}    & O          & O          & O      & \dots  & O          & O \\
			T_{\ell1}  & I_3        & O          & O      & \dots  & O          & O \\
			T_{\ell2}  & T_{\ell d} & I_3        & O      & \dots  & O          & O \\
			T_{\ell3}  & O          & T_{\ell d} & I_3    & \dots  & O          & O \\
			\vdots     & \vdots     &            & \ddots & \ddots & \vdots     & \vdots \\
			T_{\ell N} & O          & O          & O      & \dots  & T_{\ell d} & I_3
		\end{bmatrix}\in\mathbb{R}^{(4N+1)\times(4N+1)},\vspace{-3mm}
	\end{equation*}
	where $T_{\ell d} = \scriptsize\begin{bmatrix}-e_1 & O & O\end{bmatrix}\in\mathbb{R}^{3\times 3}$ and \vspace{-3mm}
	\begin{equation*}
		T_{\ell i} = \left\{\scriptsize\begin{array}{l}
			\scriptsize\begin{bmatrix}
				e_2 - e_1 & O & O
			\end{bmatrix}^\top\in\mathbb{R}^{3\times(N+1)},\ i=1,\\ \vspace{-4mm} \\
			\scriptsize\begin{bmatrix}
				e_{i+1} & O & O
			\end{bmatrix}^\top\in\mathbb{R}^{3\times(N+1)},\,\forall\,i\in\{2:N\}.
		\end{array}\right.\vspace{-3mm}
	\end{equation*}
	to obtain the platoon's model of type \eqref{eq:ss_a}-\eqref{eq:ss_c}, given by\vspace{-3mm}
	\begin{equation}\label{eq:plat_ss}
		x[k+1] = A x[k] + B u[k] + e_1 \Delta p_0[k],\vspace{-3mm}
	\end{equation}
	where we have that:\vspace{-3mm}
	\begin{enumerate}
		\item[a)] $A:=T\cdot\mathrm{diag}(I_{N+1}\ ,\ \underbrace{A_{car}\ ,\ \dots\ ,\ A_{car}}_{N\text{ times}})\cdot T^{-1}$;
		
		\item[b)]$B:=T\cdot\begin{bmatrix}
			O & I_{3N}
		\end{bmatrix}^\top\cdot\mathrm{diag}(\underbrace{B_{car}\ ,\ \dots\ ,\ B_{car}}_{N\text{ times}})$;
		
		\item[c)] the platoon's state vector is expressed as follows $x[k] := \begin{bmatrix}
			p_0[k] & \ell_v^\top[k] & x_1^\top[k] & \dots & x_N^\top[k] 
		\end{bmatrix}^\top$, where $x_i[k]:=\,\,\begin{bmatrix}
			y_i[k] & v_i[k] & \mu_i[k]
		\end{bmatrix}^\top$ denotes the vector containing the platoon states of the $i^\text{th}$ area;
		
		\item[d)] the platoon's input vector is given explicitly by $u[k] := \begin{bmatrix}
			u_1^\top[k] & \dots & u_N^\top[k] 
		\end{bmatrix}^\top$;
		
		\item[e)] $B_d := e_1\in\mathbb{R}^{4N+1}$ and $d[k]:=\Delta p_0[k]$, by direct correspondence with \eqref{eq:ss_a}-\eqref{eq:ss_c}.\vspace{-3mm}
	\end{enumerate}

	\begin{figure*}
		\begin{subequations}
			\begin{align}\label{eq:ex_cl_lead}\tag{53a}
				\scriptsize\begin{bmatrix}
					x_1[k+1]\\w_1[k+1]
				\end{bmatrix}=&\scriptsize\begin{bmatrix}
					A_{car} & B_{car} \\ B_{\mathbf{\Gamma}1} & a_1
				\end{bmatrix}\scriptsize\begin{bmatrix}
					x_1[k]\\w_1[k]
				\end{bmatrix}+\scriptsize\begin{bmatrix}
					B_{car} & O\\O & B_{\mathbf{\Gamma}1}
				\end{bmatrix}\scriptsize\begin{bmatrix}
					u_{s21}[k] \\ u_{s11}[k]
				\end{bmatrix}+\scriptsize\begin{bmatrix}
					-e_1 & B_{car} & O\\0 & O & B_{\mathbf{\Gamma}1} 
				\end{bmatrix}\tiny\begin{bmatrix}
					\Delta p_0[k] \\ \beta_{s21}[k] \\ \zeta_1[k]+\beta_{s11}[k]
				\end{bmatrix},\\
				\scriptsize\begin{bmatrix}
					x_i[k+1]\\w_i[k+1]
				\end{bmatrix}=&\scriptsize\begin{bmatrix}
					A_{car} & B_{car} \\ B_{\mathbf{\Gamma}i} & a_i
				\end{bmatrix}\scriptsize\begin{bmatrix}
					x_i[k]\\w_i[k]
				\end{bmatrix}+\scriptsize\begin{bmatrix}
					B_{car} & O\\O & B_{\mathbf{\Gamma}i}
				\end{bmatrix}\scriptsize\begin{bmatrix}
					u_{s2i}[k] \\ u_{s1i}[k]
				\end{bmatrix}+\scriptsize\begin{bmatrix}
					A_w & B_w\\O & b_{\mathbf{\Phi}i}
				\end{bmatrix}\scriptsize\begin{bmatrix}
					x_{i-1}[k]\\w_{i-1}[k]
				\end{bmatrix}+\nonumber\\
				&\hspace{50mm}+\scriptsize\begin{bmatrix}
					B_{w} & B_{car} & O & O \\ O & O & b_{\mathbf{\Phi}i} & B_{\mathbf{\Gamma}i}
				\end{bmatrix}\tiny\begin{bmatrix}
					u_{s2(i-1)}[k] + \beta_{s2(i-1)}[k] \\ \beta_{s2i}[k] \\ \beta_{f(i-1)}[k] \\ \zeta_i[k]+\beta_{s1i}[k]
				\end{bmatrix},\,\forall\,i\in\{2:N\}.\label{eq:ex_cl_fol}\tag{53b}
			\end{align}
		\end{subequations}
		\phantom{ }
		
		\vspace{-6mm}
		\hrulefill\vspace{-2mm}
		\begin{equation}\label{eq:ex_X_hat}\tag{54}
			\scriptsize\begin{array}{l}
				\widehat{\mathcal{X}}:=\left\{\widehat{x}\in\mathbb{R}^4\,\big\vert\,-360\leq e_1^\top\widehat{x}\leq0,\ 0\leq e_2^\top\widehat{x}\leq36,\ \left|e_3^\top\widehat{x}\right|\leq10,\ \left|e_4^\top\widehat{x}\right|\leq 4.99,\ 
				5.01\left|e_1^\top B_{car}\right| \leq W_x\widehat{x} \leq 36T_s-5.01\left|e_1^\top B_{car}\right|\right\}.
			\end{array}\normalsize\vspace{-4mm}
		\end{equation}\hrulefill\vspace{-2mm}
	\end{figure*}
	
	By choosing $\mathbb{C}_g = \{z\in\mathbb{C}\ \vert\ |z|<1\}$, the first layer will aim to stabilise platoon's closed-loop dynamics, in the sense that, by implementing the control laws of type \eqref{eq:NRF_plat} as detailed in our companion paper \cite{part1}, the closed-loop\newpage\noindent dynamics will be described by\vspace{-3mm}
	\begin{equation}\label{eq:first_layer_cl}\tag{49}
		\scriptsize\begin{bmatrix}
			p_0[k+1] \\ \ell_v[k+1] \\ x_{f}[k+1]
		\end{bmatrix} = \scriptsize\begin{bmatrix}
			1 & O & O\\ O & I_N & O \\ O & O & A_{f}
		\end{bmatrix}\scriptsize\begin{bmatrix}
			p_0[k] \\ \ell_v[k] \\ x_{f}[k]
		\end{bmatrix} + \scriptsize\begin{bmatrix}
			B_0 \\ B_\ell \\ B_{f}
		\end{bmatrix} d_s[k],\normalsize\vspace{-3mm}
	\end{equation}
	in which:\vspace{-3mm}
	\begin{enumerate}
		\item[a)] the eigenvalues of $A_{f}$ are all in $\mathbb{C}_g$;
		
		\item[b)] $x_{f}[k] = \begin{bmatrix}
			x_1^\top[k] & w_1^\top[k] & \dots & x_N^\top[k] & w_N^\top[k]
		\end{bmatrix}^\top$, with $w_i[k]$ being the state vectors belonging to the NRF implementations of type \eqref{eq:area_NRF}, for all $i\in\{1:N\}$;\smallskip
		
		\item[c)] $d_s[k]$ is the signal vector from \eqref{eq:MPC_exo}. \vspace{-3mm}
	\end{enumerate}
	
	\begin{remark}
		Since $p_0[k]$ and $\ell_i[k]$, for any $i\in\{1:N\}$, cannot be influenced by the $u_i[k]$ signals, the first layer will be unable to stabilise their dynamics. However, given the nature of our application and the physical interpretation of the variables, the aforementioned stabilisation is of no practical interest. The second layer will focus only on the identity $x_f[k+1]=A_fx_f[k]+B_fd_s[k]$, which will represent our first layer's closed-loop dynamics.\vspace{-3mm}
	\end{remark}
	
	\subsubsection{Second-Layer Objectives}\vspace{-3mm}
	
	The main goal of the second layer is to ensure constraint satisfaction, with respect to the variables that make up $x_i[k]$ and $u_i[k]$, in the presence of recursive feasibility. Thus, for all $i\in\{1:N\}$ and all $k\geq k_0$, we wish to have:\vspace{-3mm}
	\begin{equation}\label{eq:ex_con}
		y_i[k]\in[-360,0],\ v_i[k]\in[0,36],\ u_i[k]\in[-10,10].\vspace{-3mm}
	\end{equation}
	These constraints must be satisfied while employing the same communication scheme as in \eqref{eq:NRF_plat}, \emph{i.e.}, the $i^\text{th}$ car can use only information available locally (from its own readings) or from the car directly in front of it.\vspace{-3mm}

	Additionally, due to the fact that the prototype NRF-based controllers discussed in \cite{plutonizare} (when properly tuned) achieve high performance, we want the second layer to intervene as little as possible in the closed-loop dynamics of the first. In other words, most of the control action should be undertaken by the $u_{fi}[k]$ signals, whereas the entries of $u_{s1i}[k]$ and $u_{s2i}[k]$ should have negligible absolute values (compared to those of $u_{fi}[k]$) whenever the constraints from \eqref{eq:ex_con} are not in danger of being violated.\vspace{-3mm}

	\subsection{Control Architecture Design}
	\label{subsec:des}\vspace{-3mm}
	
	\subsubsection{The First Layer}\vspace{-3mm}
	
	Since the application considered in this numerical example already has a dedicated NRF-based solution, we defer to \cite{plutonizare} for the procedural details of our first-layer design. In so doing, we briefly discuss the numerical implementation of these control laws before focusing our attention on the design of the second layer.\vspace{-3mm}
	
	For the homogeneous platoon whose physical parameters were given in the previous subsection, the first-layer command signals will be computed as stated in \eqref{eq:uf_implem}, via the state-space implementations of type \eqref{eq:area_NRF} given by\vspace{-3mm}
	\begin{equation}\label{eq:ex_implem1}
		u_{f1}[k]=\left[\scriptsize\begin{array}{c|c}
			a_1-z & B_{\mathbf{\Gamma}1} \\ \hline 1 & O
		\end{array}\right]\star \scriptsize\begin{array}{l}
			(x_1[k]+\zeta_1[k]+u_{s11}[k]+\beta_{s11}[k])
		\end{array},\normalsize
	\end{equation}
	and, for all $i\in\{2:N\}$, by\vspace{-3mm}
	\begin{equation}\label{eq:ex_implem2}
		\hspace{-1mm}u_{fi}[k]=\left[\scriptsize\begin{array}{c|cc}
			a_i-z & b_{\mathbf{\Phi}i} & B_{\mathbf{\Gamma}i} \\ \hline 1 & 0 & O
		\end{array}\right]\star \tiny\begin{bmatrix}
			u_{f(i-1)}[k]+\beta_{f(i-1)}[k]\\ 
			x_i[k]+\zeta_i[k]+u_{s1i}[k]+\beta_{s1i}[k]
		\end{bmatrix},\normalsize
	\end{equation}
	with $a_i$, $b_{\mathbf{\Phi}i}$ and $b_{\mathbf{\Gamma}i}$ being given in Table~\ref{tab:params} below.
	
	\begin{table}[h]
		\centering
		\begin{tabular}{r|ccc}
			$i$ & $a_i$ & $b_{\mathbf{\Phi}i}$ & $B_{\mathbf{\Gamma}i}$\\\hline\hline
			1   & 0.9690 & 0 & $\begin{bmatrix}-0.0038 & -0.0192 & 0\ \ \end{bmatrix}$ \\ 
			2   & 0.9799 & 0.0199 & $\begin{bmatrix}-0.0030 & -0.0152 & 0\ \ \end{bmatrix}$ \\
			3   & 0.9799 & 0.0200 & $\begin{bmatrix}-0.0032 & -0.0161 & 0\ \ \end{bmatrix}$ \\
			4   & 0.9798 & 0.0200 & $\begin{bmatrix}-0.0034 & -0.0171 & 0\ \ \end{bmatrix}$ \\
			5   & 0.9797 & 0.0200 & $\begin{bmatrix}-0.0036 & -0.0182 & 0\ \ \end{bmatrix}$ \\
			6   & 0.9796 & 0.0201 & $\begin{bmatrix}-0.0039 & -0.0195 & 0\ \ \end{bmatrix}$ \\
			7   & 0.9795 & 0.0201 & $\begin{bmatrix}-0.0042 & -0.0209 & 0\ \ \end{bmatrix}$ \\
			8   & 0.9794 & 0.0202 & $\begin{bmatrix}-0.0045 & -0.0224 & 0\ \ \end{bmatrix}$ \\
			9   & 0.9793 & 0.0202 & $\begin{bmatrix}-0.0049 & -0.0243 & 0\ \ \end{bmatrix}$ \\
			10  & 0.9792 & 0.0203 & $\begin{bmatrix}-0.0053 & -0.0265 & 0\ \ \end{bmatrix}$
		\end{tabular}\normalsize
		\caption{Coefficients of the first-layer controller implementations}
		\label{tab:params}
	\end{table}\stepcounter{equation}\vspace{-3mm}
	
	By using the state-space systems in \eqref{eq:plat_ss} and \eqref{eq:ex_implem1}-\eqref{eq:ex_implem2} to form the first layer's closed-loop system, we obtain a realisation of type \eqref{eq:first_layer_cl} in which the spectral radius of $A_f$ is $0.9936$, thus validating all of the design objectives associated with the first layer. In addition to this fact, the resulting closed-loop dynamics of each car in the vehicle platoon will be given by the realisations stated in \eqref{eq:ex_cl_lead}-\eqref{eq:ex_cl_fol}, located at the top of this page and where\vspace{-3mm}
	\begin{equation*}
		A_w = \left[\begin{array}{ccc}
			0  &\ -0.1 &\  0.0331\\ 
			0  &\  0   &\  0\\
			0  &\  0   &\  0\\
		\end{array}\right]\text{ and }B_w=\left[\begin{array}{c}
			-0.0381\\
			0\\
			0\\
		\end{array}\right].\vspace{-3mm}
	\end{equation*}
	It is precisely the dynamics from \eqref{eq:ex_cl_lead}-\eqref{eq:ex_cl_fol} which will be tackled by our distributed MPC-based policies.\vspace{-3mm}
	
	\subsubsection{The Second Layer}\vspace{-3mm}
	
	The first step in designing our second-layer control system, with the aim of ensuring its recursive feasibility, is to define all the sets discussed in Section~\ref{sec:MPC_des}. We begin by addressing the command signals generated by the MPC layer, for which we impose the following constraints:\vspace{-3mm}
	\begin{equation*}
		\left\{
		\begin{array}{l}
			\mathcal{U}_{s1i}=\left\{u_{s1i}\in\mathbb{R}^3\,\big\vert\,\left|e^\top_1u_{s1i}\right|\leq720,\ \left|e^\top_2u_{s1i}\right|\leq72,\right.\\ \left. \quad\qquad\qquad\qquad\qquad\qquad e^\top_3u_{s1i}=0\right\},\,\forall\,i\in\{1:N\},\\
			\mathcal{U}_{s2i}=\left\{u_{s2i}\in\mathbb{R}\,\vert\,\left|u_{s2i}\right|\leq5\right\},\,\forall\,i\in\{1:N\}.
		\end{array}
		\right.\vspace{-3mm}
	\end{equation*}
	Next, we consider the following box-type constraints for the exogenous signals:\vspace{-3mm}
	\begin{equation*}
		\left\{
		\scriptsize\begin{array}{l}
			\mathcal{V}=\left\{\nu\in\mathbb{R}^{n_x+n_w}\,\big\vert\,\left|e_j^\top \nu\right|\leq0.02,\,\forall\,j\in\{1:n_x+n_w\}\right\},\\
			\mathcal{D}_{\beta_f}=\left\{\beta_f\in\mathbb{R}^{n_u}\,\big\vert\,\left|e_j^\top \beta_f\right|\leq0.02,\,\forall\,j\in\{1:n_u\}\right\},\\
			\mathcal{D}_{\beta_{s1}}=\left\{\beta_{s1}\in\mathbb{R}^{n_x}\,\big\vert\,\left|e_j^\top \beta_{s1}\right|\leq0.01,\,\forall\,j\in\{1:n_x\}\right\},\\
			\mathcal{D}_{\beta_{s2}}=\left\{\beta_{s2}\in\mathbb{R}^{n_u}\,\big\vert\,\left|e_j^\top \beta_{s2}\right|\leq0.01,\,\forall\,j\in\{1:n_u\}\right\},\\
			\mathcal{D}_{d}=\left\{\Delta p_0\in\mathbb{R}\,\vert\,0\leq\Delta p_0\leq36T_s\right\},
		\end{array}
		\right.\vspace{-3mm}
	\end{equation*}
	and, finally, we address the constraints related to the platoon's states by first introducing the matrix denoted $W_x:=\mathrm{row}_1\left(\tiny\begin{bmatrix}
		A_{car} & B_{car}
	\end{bmatrix}\right)\mathrm{diag}(0,I_3)$ and then constructing the set $\widehat{\mathcal{X}}\subseteq\mathbb{R}^4$ given in \eqref{eq:ex_X_hat} at the top of this page.\vspace{-3mm}
	
	\begin{figure*}
		\begin{table}[H]
			\centering
			\resizebox{.975\textwidth}{!}{\scriptsize\begin{tabular}{c||cccccccccc}
					car number& 1 & 2& 3& 4 & 5&6 &7&8&9&10\\\hline
					time (seconds) &  211.740 & 757.767 & 390.860 & 396.871 & 295.800 & 235.603 & 155.952 & 204.719 &  74.838  & 71.048
			\end{tabular}}
			\caption{Computation times associated with second-layer synthesis} 
			\label{tab:design_time}
		\end{table}\vspace{-4mm}
		\hrulefill\vspace{-1mm}
	\end{figure*}
	
	By choosing $T_i=1$ for all $i\in\{1:N\}$ and by leveraging $\widehat{\mathcal{X}}$, while also accounting for the disturbance discussed in Section~\ref{subsec:ci_dist}, we proceed to implement our design procedure. {\color{black}The resulting numerical routines and all of the data sets from this numerical example are available at \texttt{https://github.com/AndreiSperila/NRMPC}. Moreover, Table~\ref{tab:design_time} (located at the top of the next page) showcases the computation times associated with the complete design of every car's second-layer subcontroller.\vspace{-3mm}
		
		{\color{black}
			\begin{remark}
				As explained in the comments that accompany the source code from the above-indicated link, our approach entails the \emph{implicit design} of the robust control invariant sets $\mathcal{C}_i\subseteq\widehat{\mathcal{X}}$, which (by their very construction; see, for example, Section~10.4 in \cite{BBM}) guarantee the recursive feasibility of our second layer. We stress the fact that (as previously indicated in Section~\ref{sec:MPC_imp}) the sets $\mathcal{C}_i$ are \emph{not computed explicitly}, with their existence being a desirable and natural consequence of our proposed approach.\vspace{-3mm}
			\end{remark}
		}
	}
	
	Finally, we choose $\overline{T}_i=0$ and we set the costs in \eqref{eq:area_prob} as\vspace{-3mm}\stepcounter{equation}
	\begin{equation}\label{eq:ex_stage_cost}
		\begin{array}{l}
			10^{-9}y_i^2[k+1]+u_{s2i}^2[k]+u_{s1i}^\top[k]u_{s1i}[k]
		\end{array}\normalsize,\vspace{-3mm}
	\end{equation}
	thereby concluding the design phase of the second layer.\vspace{-3mm}

	\subsection{Simulation Scenario and Results}\vspace{-3mm}
	\label{subsec:res}
	
	To test the obtained control laws, we simulate a scenario in which $\Delta p_0[k] = T_sv_0[k]$, where \vspace{-3mm}
	\begin{equation*}
		\scriptsize\begin{array}{l}
			v_0[k] = 10\cdot\mathds{1}[k] \,\textbf{--}\, 7\cdot\mathds{1}[k\,\textbf{--}\,400] + 30\cdot\mathds{1}[k\,\textbf{--}\,1200] \,\textbf{--}\, 30\cdot\mathds{1}[k\,\textbf{--}\,1300]
		\end{array}\normalsize\vspace{-3mm}
	\end{equation*}
	is the speed of the virtual $0^\text{th}$ car and $\mathds{1}[k]$ denotes the discrete-time Heaviside step function. As can be seen in Figure~\ref{fig:car_speed} below, in which $v_0[k]$ is depicted in orange, the speed of each car in the platoon will try to match that of the $0^{th}$ car, in order to achieve steady-state evolution. 
	
	\begin{figure}[H]
		\includegraphics[width=.9\columnwidth]{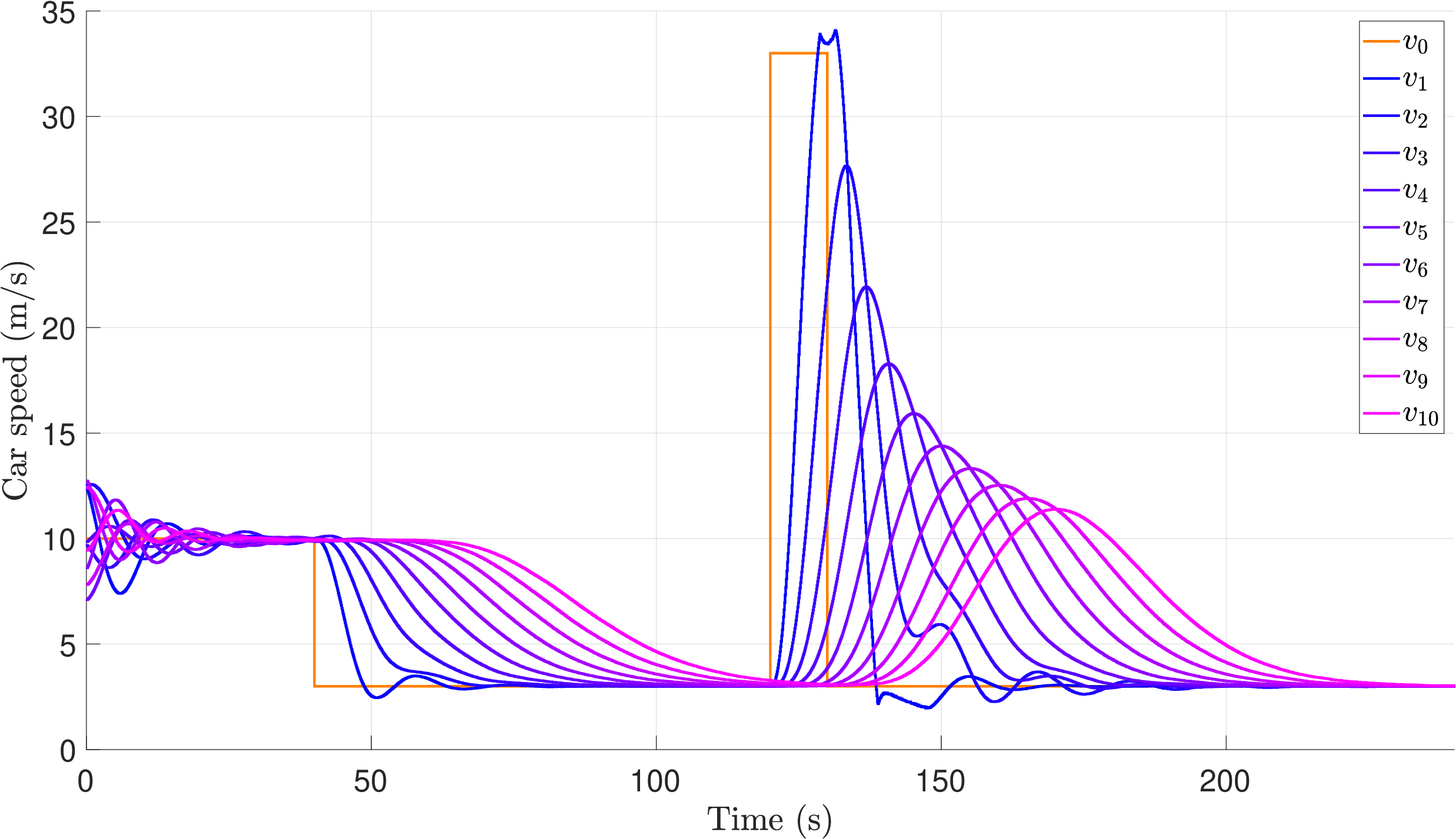}
		\caption{$0^{th}$ car speed and speeds of the platoon's cars}
		\label{fig:car_speed}
	\end{figure}\vspace{-6mm}
	
	Moreover, as shown in Figure~\ref{fig:car_inter} below, our two-layer control architecture successfully maintains all interdistance variables within their prescribed bounds, thereby avoiding collisions and preventing the cars from drifting too far apart (which can lead to traffic congestion or communication breakdown between consecutive vehicles).
	
	\begin{figure}[H]
		\includegraphics[width=.9\columnwidth]{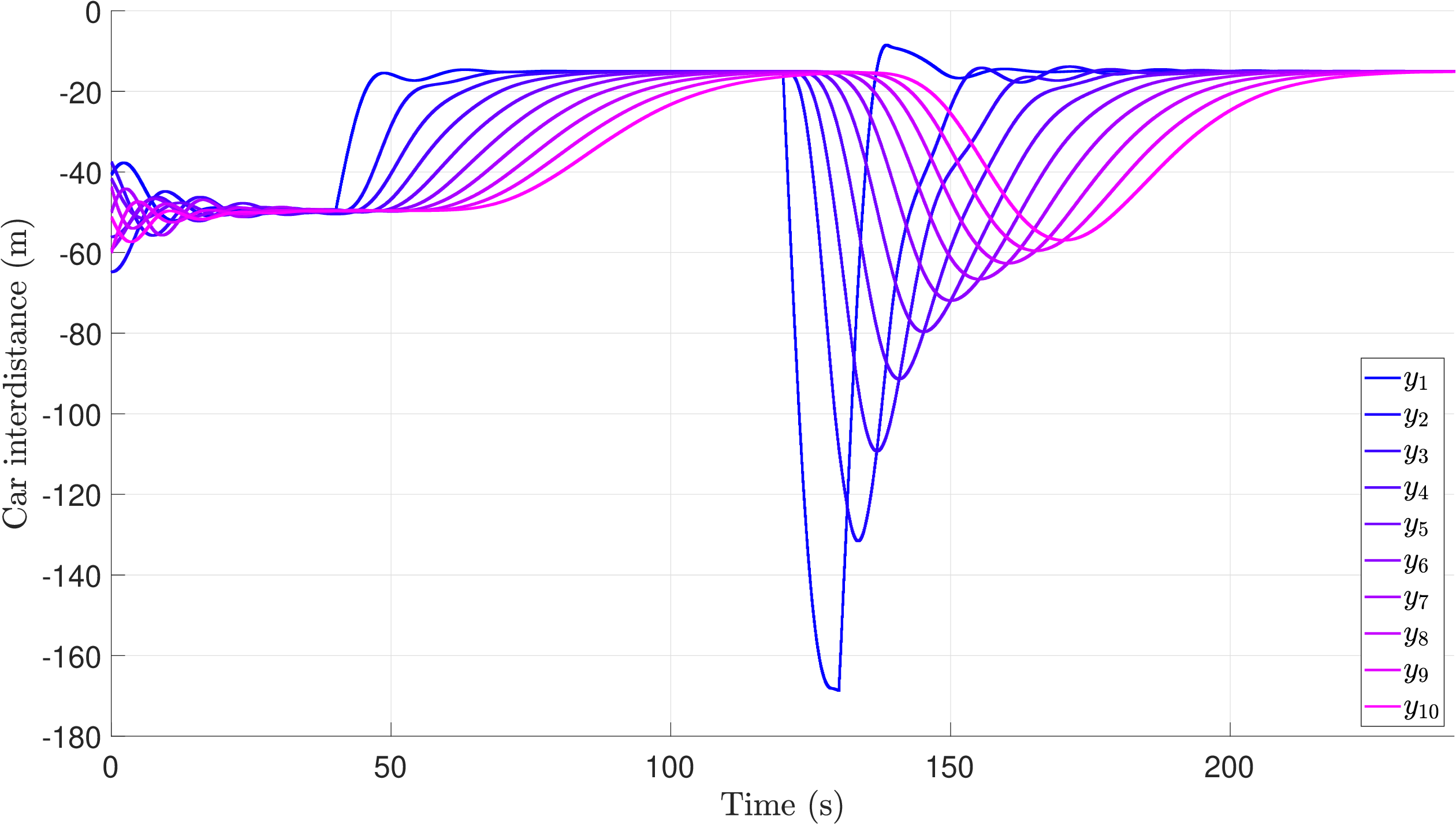}
		\caption{Length-based interdistances of the platoon's cars}
		\label{fig:car_inter}
	\end{figure}\vspace{-6mm}
	
	Given the limitations placed upon the signals sent to the car actuators, our control scheme is once again successful. As can be seen from Figures~\ref{fig:NRF_cmd}~and~\ref{fig:MPC_to_cars} below, the sum of the signals from the two aforementioned figures is well within the bounds from \eqref{eq:ex_con}, even taking into account the amplitudes of the entries in $\beta_{s2}[k]$.\vspace{-3mm}
	
	\begin{figure}[H]
		\includegraphics[width=.9\columnwidth]{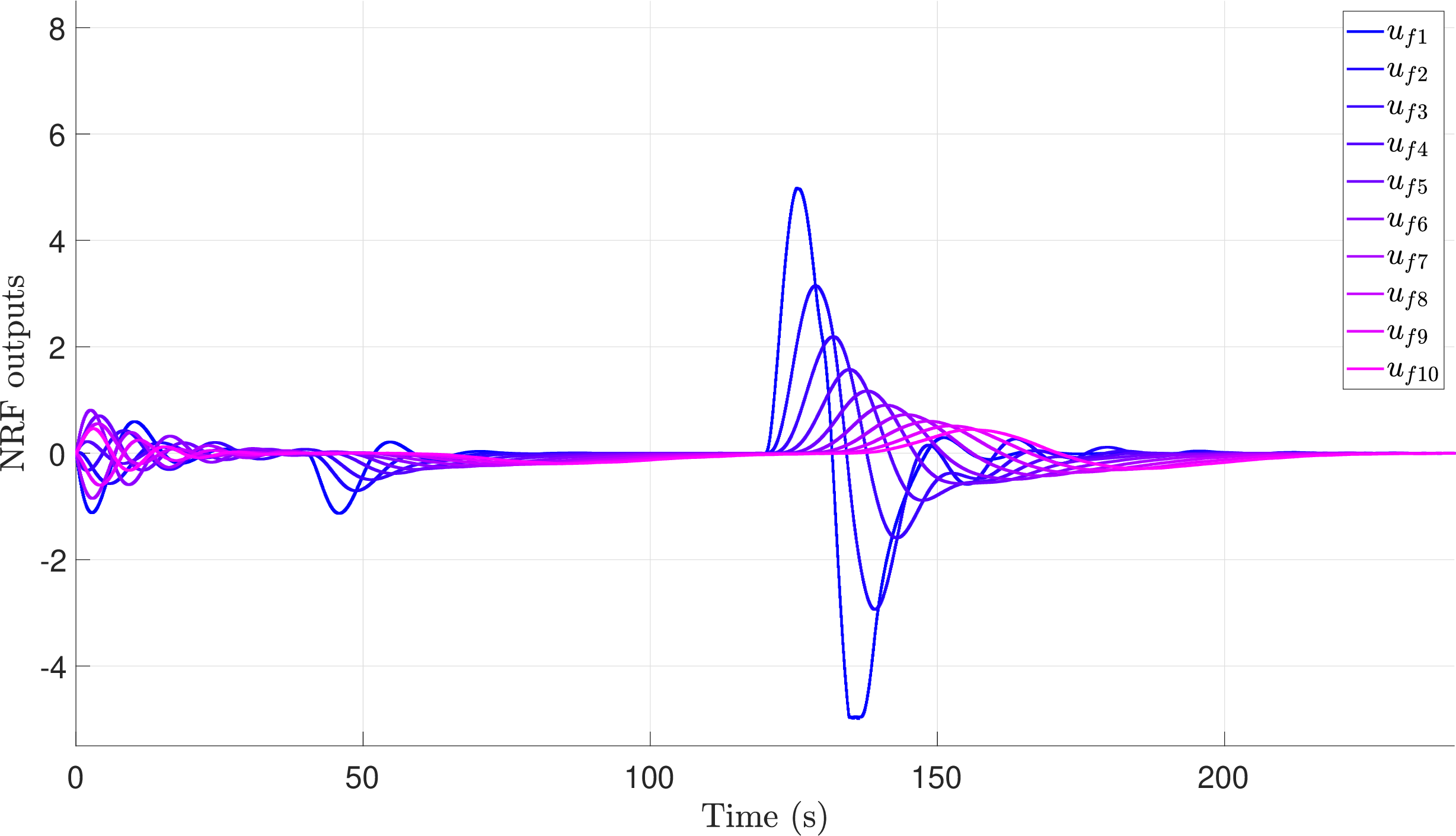}
		\caption{Outputs of the first-layer subcontrollers}
		\label{fig:NRF_cmd}
	\end{figure}\vspace{-3mm}
	
	\begin{figure}[H]
		\includegraphics[width=.9\columnwidth]{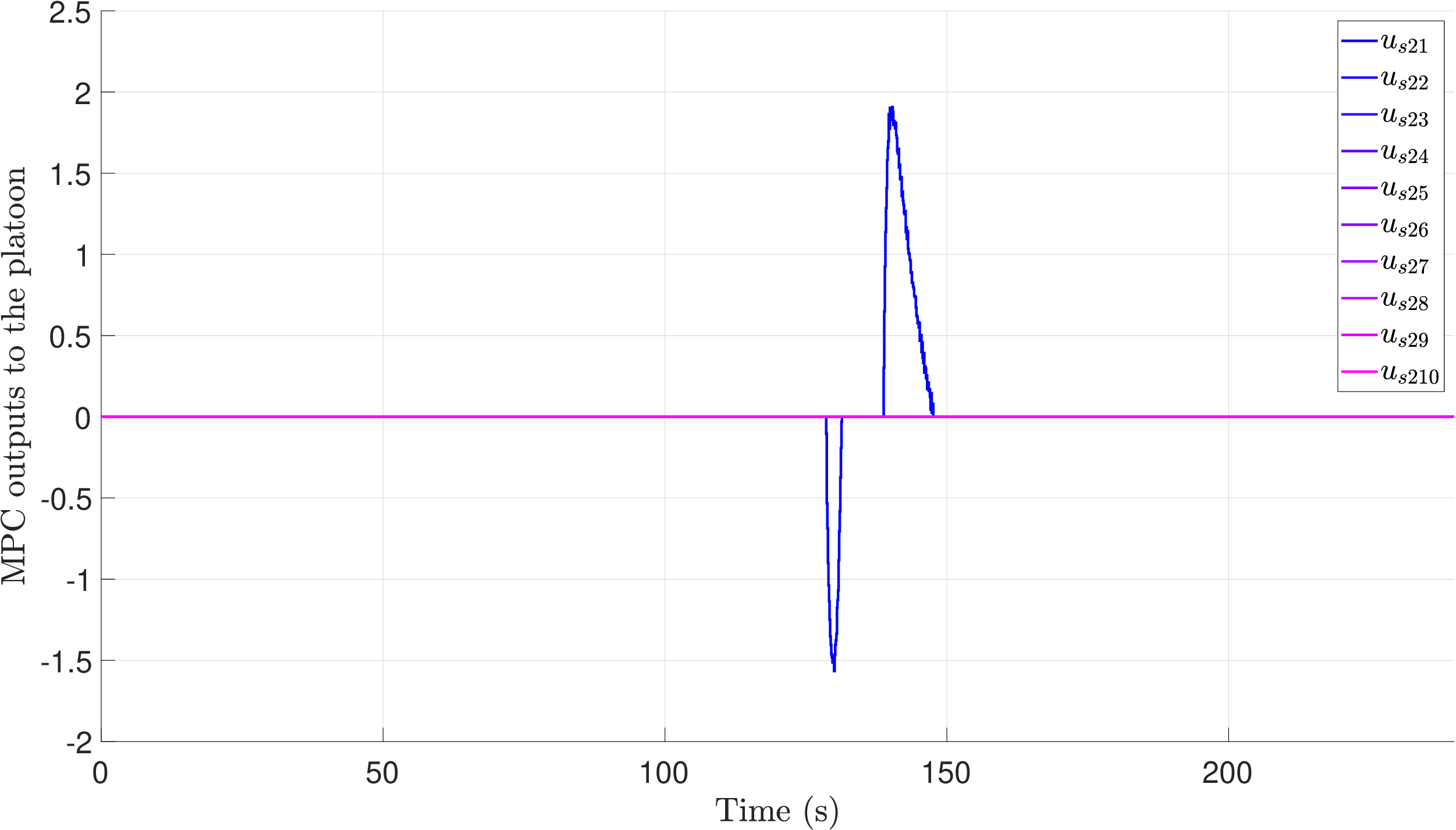}
		\caption{Second-layer outputs to the platoon}
		\label{fig:MPC_to_cars}
	\end{figure}\vspace{-6mm}
	
\begin{figure}[H]
		\includegraphics[width=.9\columnwidth]{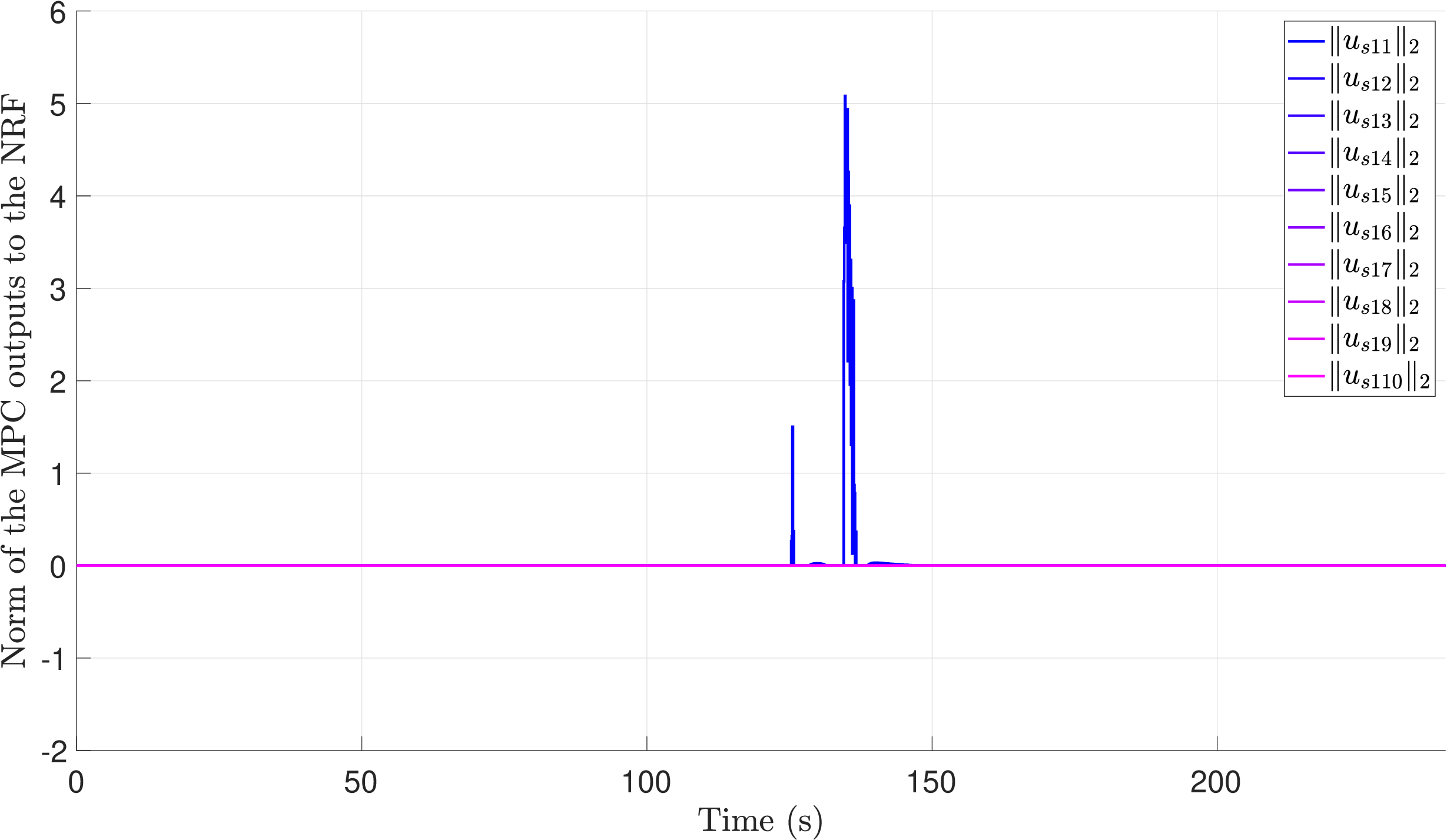}
		\caption{Norm values for second-layer outputs to the first-layer subcontrollers}
		\label{fig:MPC_to_NRF}
	\end{figure}

	Finally, due to the choice of cost functions in \eqref{eq:ex_stage_cost}, the entries of $u_{s1}[k]$ and $u_{s2}[k]$ have absolute values of at most $10^{-6}$ for most of the simulation scenario's duration, as can be seen in Figures~\ref{fig:MPC_to_cars}~and~\ref{fig:MPC_to_NRF}.
	
	\vspace{-6mm}
	
\begin{figure}[H]
		\includegraphics[width=\columnwidth]{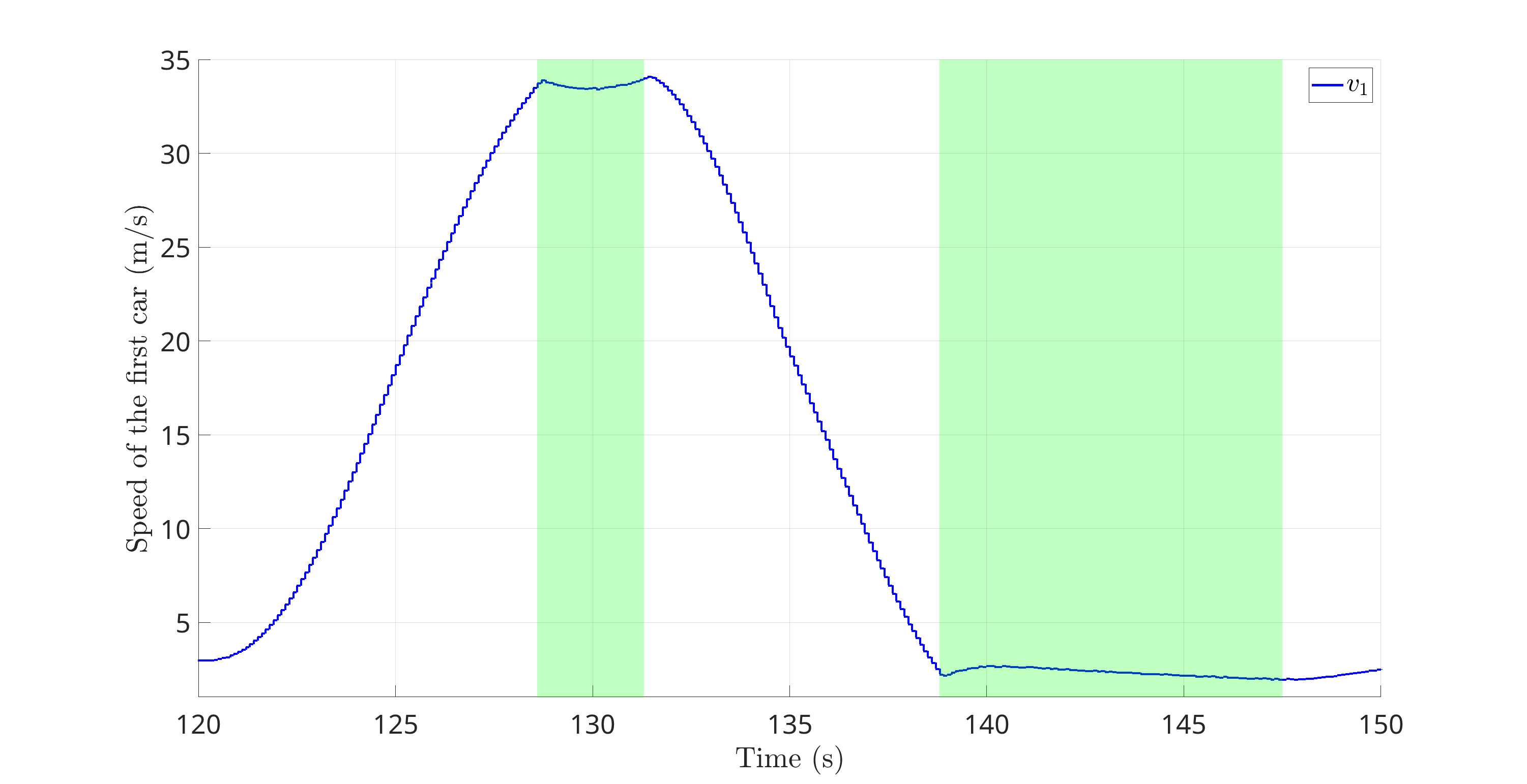}
		\caption{Constraint satisfaction for the speed of the platoon's first car (second-layer subcontroller active in the green shaded area)}
		\label{fig:speed_sat}
	\end{figure}

	The brief intervals in which the two signal vectors are of the same order of magnitude as the first-layer control signals are those in which the second layer intervenes to saturate either the output of the first car's NRF-based subcontroller (see Figure~\ref{fig:NRF_sat} below), or the speed of the lead car in the platoon (see Figure~\ref{fig:speed_sat}, also located below) with respect to these variables' prescribed bounds.\vspace{-3mm}
	
	\begin{figure}[H]
		\includegraphics[width=1\columnwidth]{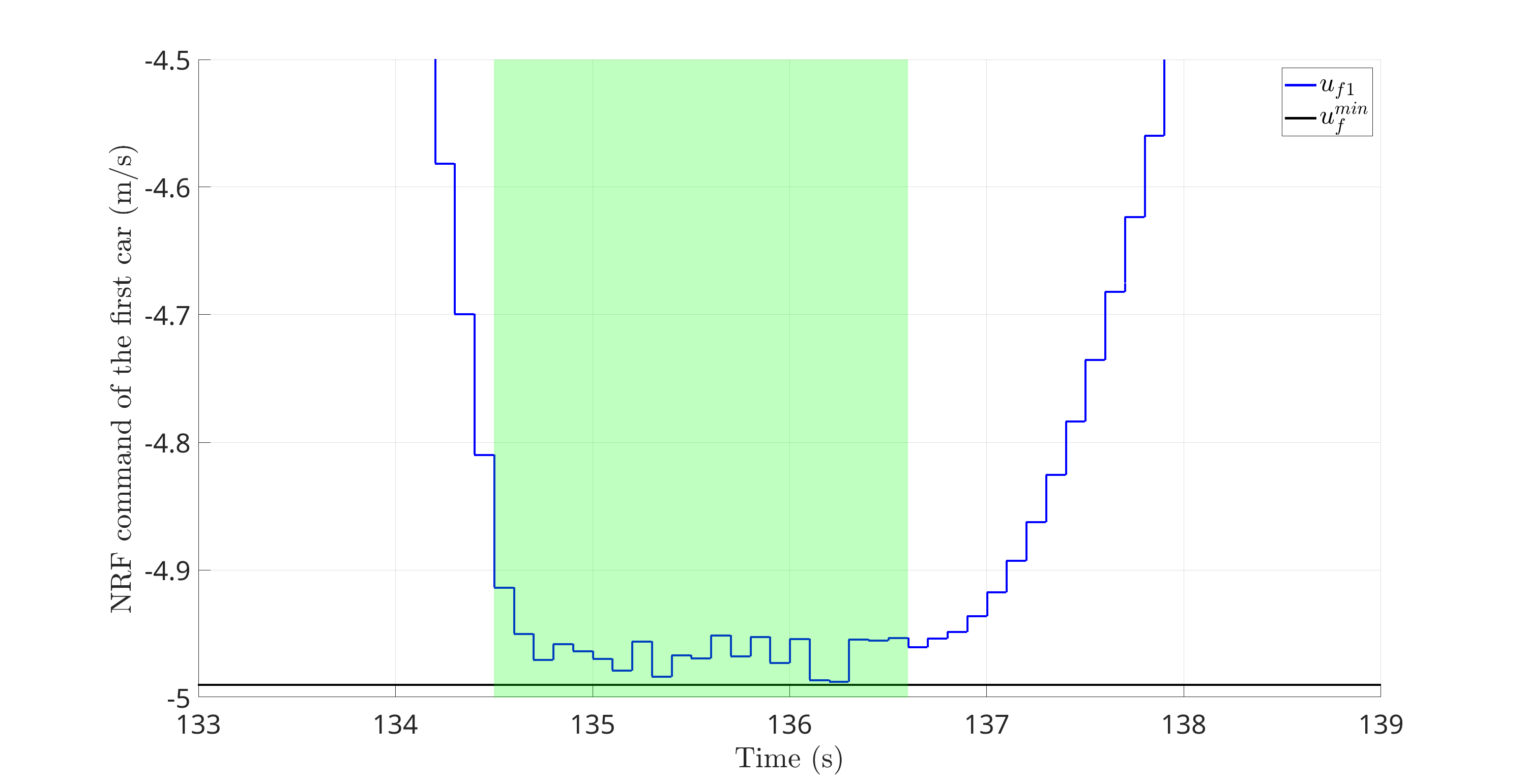}
		\caption{Constraint satisfaction for the first-layer command of the platoon's first car (second-layer subcontroller active in the green shaded area)}
		\label{fig:NRF_sat}
	\end{figure}\vspace{-3mm}
	
	\vspace{-6mm}
	
	{\color{black} Finally, as showcased by the runtime statistics displayed in Tables~\ref{tab:runtime_NRF} and \ref{tab:runtime_MPC}, which are located at the top of this page, the implementation costs associated with our solution are sufficiently modest such that command computation is well below the sampling period of our control system, namely $T_s=0.1$ seconds. Note that all simulations associated with these runtimes were performed in MATLAB R2024b on a mid-range laptop equipped with an Intel Core i9-100885H processor.\vspace{-3mm}
		
		\begin{table}[t]
			\centering
			\resizebox{.75\columnwidth}{!}
			{\scriptsize\begin{tabular}{c|ccccc}
					car number& min& max& mean& median& mode\\\hline\hline
					1  &  0.003  &  0.213  &  0.005  &  0.004  &  0.004\\
					2  &  0.001  &  0.057  &  0.003  &  0.002  &  0.002\\
					3  &  0.001  &  0.023  &  0.002  &  0.002  &  0.002\\
					4  &  0.001  &  0.022  &  0.002  &  0.002  &  0.002\\
					5  &  0.001  &  0.022  &  0.002  &  0.002  &  0.002\\
					6  &  0.001  &  0.233  &  0.002  &  0.002  &  0.002\\
					7  &  0.001  &  0.008  &  0.002  &  0.002  &  0.002\\
					8  &  0.001  &  0.010  &  0.002  &  0.002  &  0.002\\
					9  &  0.001  &  0.013  &  0.002  &  0.002  &  0.002\\
					10  &  0.001  &  0.011  &  0.002  &  0.002  &  0.002
			\end{tabular}}
			\caption{NRF-based subcontroller runtime statistics in milliseconds} 
			\label{tab:runtime_NRF}
		\end{table}
		
		\begin{table}[t]
			\centering
			\resizebox{.75\columnwidth}{!}
			{\scriptsize\begin{tabular}{c|ccccc}
					car number& min& max& mean& median& mode\\\hline\hline
					1  &  1.678  &  4.375  &  2.202  &  2.127  &  1.872\\
					2  &  1.688  &  5.028  &  2.028  &  1.955  &  1.877\\
					3  &  1.659  &  4.347  &  1.949  &  1.884  &  1.881\\
					4  &  1.661  &  3.882  &  1.930  &  1.868  &  1.742\\
					5  &  1.648  &  3.929  &  1.921  &  1.861  &  1.853\\
					6  &  1.630  &  4.936  &  1.931  &  1.864  &  1.826\\
					7  &  1.629  &  4.895  &  1.913  &  1.848  &  1.807\\
					8  &  1.624  &  4.829  &  1.924  &  1.852  &  1.852\\
					9  &  1.639  &  4.233  &  1.956  &  1.838  &  1.847\\
					10  &  1.614  &  4.680  &  1.989  &  1.838  &  1.786
			\end{tabular}}
			\caption{MPC-based subcontroller runtime statistics in milliseconds\vspace{-1mm}} 
			\label{tab:runtime_MPC}
		\end{table}
		
		{\color{black}
			\subsection{Comparison with Alternative Solutions}\vspace{-3mm}
			
			In order to conclude this numerical example, we present a comparison with several prominent alternative techniques from literature that tackle the platooning problem in an MPC-based framework. We refer here to the procedures proposed in \cite{plat_MPC_1,plat_MPC_2,plat_MPC_3,plat_MPC_4,plat_MPC_5}, and we point out that a major limitation for all of them is the fact that, unlike our proposed solution, no source of disturbance (either from exogenous sources or from model uncertainty) is considered in any of these approaches. Moreover, the control laws presented in \cite{plat_MPC_1,plat_MPC_2,plat_MPC_3} all share in common the fact that they do not explicitly support constraints for the state variables of the platoon's cars, except for a restrictive use of terminal constraints, which are meant to ensure several theoretical guarantees.\vspace{-3mm}
			
			Another common feature of the cited works is that the lead cars in \cite{plat_MPC_2,plat_MPC_3,plat_MPC_4} must broadcast information to all other vehicles in the formation, thereby placing an impractical burden on the platoon's communication infrastructure. In addition to this fact, the property of asymptotic closed-loop stability is retrieved in \cite{plat_MPC_1,plat_MPC_3,plat_MPC_4} by carefully tuning the stage costs of the MPC-based subcontrollers, rather than it being an inbuilt feature, as ensured by the NRF-based first layer (see \cite{part1} and \cite{plutonizare}). In light of these considerations, we now address the specific aspects of the five aforementioned techniques:\vspace{-3mm}
			\begin{enumerate}
				\item[a)] One of the key features of the procedure from \cite{plat_MPC_1} is the fact that the above-mentioned terminal constraint requires the computation of a \emph{regional average value}, evaluated with respect to several neighbouring cars. In addition to the demands this places on the platoon's communication infrastructure, the feasibility of the resulting control problem is notably dependent on choosing a sufficiently long prediction horizon. While the overall computational complexity of the optimisation problem to be solved is similar to that of a QP-based formulation in \eqref{eq:area_prob}, increasing the prediction horizon may have a significant computational impact which can be bypassed in our framework (recall Remark~\ref{rem:horizon}). \smallskip
				
				\item[b)] The main limitation of the technique in \cite{plat_MPC_2} is the fact that the optimisation problem associated with the MPC-based subcontrollers has to be run twice during a car's command computation routine, with said problem having a comparable cost to a QP-based formulation of our problem in \eqref{eq:area_prob}. Furthermore, we note that both the asymptotic and string stability of the closed-loop system are guaranteed only when the platoon's initial condition satisfies several restrictive assumptions.\smallskip
				
				\item[c)] In addition to the list of common restrictions detailed above, the solution proposed in \cite{plat_MPC_3} is affected by another limitation; namely, that it can only be initialised when the platoon is already in equilibrium. Despite all of these disadvantages, we point out that the procedure's computational cost is appealingly modest, roughly comparable to that of our own problem's QP-based formulation, since it does not require the double iteration from \cite{plat_MPC_2}.\smallskip
				
				\item[d)] The approach from \cite{plat_MPC_4} shares some structural similarities with the one presented in this paper, in that it includes a rudimentary PID-based lower control layer, and it also enables the use of limited, box-based constraints on the platoon's states. However, we point out that the lower-level PIDs are not only unable to ensure closed-loop string stability (in contrast to our first layer), but the NRF-based subcontrollers offer many more degrees of freedom, in terms of both sparsity patterns and closed-loop performance (see \cite{part1}). Moreover, since \cite{plat_MPC_4} is heavily inspired by the approach presented in \cite{plat_MPC_3}, the same computational benefits apply to it as well.\smallskip
				
				\item[e)] Finally, the procedure outlined in \cite{plat_MPC_5} proposes a centralised approach to the problem and, unsurprisingly, demonstrates a high degree of closed-loop performance in the presence of state constraints. In addition to being intractable in the distributed setting, its crucial shortcoming is given by the fact that closed-loop string stability is incompatible with the aforementioned constraints. In contrast to our solution, when safety-critical constraints become active, the technique in \cite{plat_MPC_5} sacrifices the platoon's string stability to guarantee constraint satisfaction.
			\end{enumerate}\vspace{-4mm}
		}
		
		\section{Conclusions}\vspace{-4mm}
		\label{sec:outro}
		
		As a contribution to the synthesis of distributed constraint management techniques, we have shown that it is possible to elegantly decouple structural design constraints (which represent the focal point of the distributed control paradigm) from the time-based constraints which are ubiquitous in MPC-oriented literature. By leveraging a promising class of distributed control strategies \cite{NRF,part1,aug_sparse} and by adapting a series of set-theoretical notions \cite{STMC} to the distributed setting, we obtain a computationally efficient control architecture which provides \emph{powerful theoretical guarantees}, and whose components can be \emph{designed fully offline}.\vspace{-4mm}
		
		\begin{ack}\vspace{-4mm}
			The authors would like to thank both \c Serban Sab\u au and Cristian Oar\u a for the many fruitful discussions which ultimately led to the elaboration of this manuscript.\vspace{-4mm}
		\end{ack}
		
		\bibliographystyle{plain}
		\bibliography{manuscript}
		\vspace{-4mm}
		
		\appendix
		
		\begin{figure*}
			\begin{subequations}
				\scriptsize\begin{align}\label{eq:NRF_state_obs_a}\tag{B.1a}
					\mathrm{row}_j(w_{r\ell}[k+1])=& -a_{j\ell}\,\mathrm{row}_1(w_{r\ell}[k])+\mathrm{row}_{j+1}(w_{r\ell}[k])+K_{j\ell}\tiny\begin{bmatrix}
						u_f[k]+\beta_f[k]\\ x[k]+\zeta[k]+u_{s1}[k]+\beta_{s1}[k]
					\end{bmatrix},\,\forall\,j\in\{1:n_{r\ell}-1\},\\\label{eq:NRF_state_obs_b}\tag{B.1b}
					\hspace{-1mm}\mathrm{row}_{n_{r\ell}}(w_{r\ell}[k+1])=& -a_{n_{r\ell}\ell}\,\mathrm{row}_1(w_{r\ell}[k])+K_{n_{r\ell}\ell}\tiny\begin{bmatrix}
						u_f[k]+\beta_f[k]\\ x[k]+\zeta[k]+u_{s1}[k]+\beta_{s1}[k]
					\end{bmatrix},\\
					\mathrm{row}_{\ell}(u_f[k])=&\ \mathrm{row}_1(w_{r\ell}[k]),\,\forall\,k\geq k_0,\,\ell\in\mathcal{A}_{ui}.\label{eq:NRF_state_obs_c}\tag{B.1c}
				\end{align}\normalsize
			\end{subequations}\phantom{ }
			
			\vspace{-9mm}
			\hrulefill\vspace{-2mm}
			\begin{subequations}
				\scriptsize\begin{align}\nonumber
					x_i[k_0+1] =&\ C_{xi}\xi_i[k_0+1]+ (\psi_{x i}[k_0+1] + \theta_{x i}[k_0+1] + \delta_{x i}[k_0+1]),\,\psi_{x i}[k_0+1]\in\Psi_{xi1},\,\theta_{x i}[k_0+1]\in\Theta_{xi1},\,\delta_{x i}[k_0+1]\in\Delta_{xi1},\label{eq:next_step_a}\tag{B.2a}\\\nonumber
					\mathrm{row}_j(u_{fi}[k_0+1]) =&\ \mathrm{row}_j(C_{ui})\xi_i[k_0+1]+ (\mathrm{row}_j(\psi_{u i}[k_0+1]) + \mathrm{row}_j(\theta_{u i}[k_0+1]) + \mathrm{row}_j(\delta_{u i}[k_0+1])),\,\mathrm{row}_j(\psi_{u i}[k_0+1])\in \left(e_j^\top\Psi_{ui1}\right),\\
					&\ \mathrm{row}_j(\theta_{u i}[k_0+1])\in\left(e_j^\top\Theta_{ui1}\right),\,\mathrm{row}_j(\delta_{u i}[k_0+1])\in\left(e_j^\top\Delta_{ui1}\right),\,\forall\,j\in\{1:n_{ui}\}.\label{eq:next_step_b}\tag{B.2b}
				\end{align}\normalsize
			\end{subequations}\phantom{ }
			
			\vspace{-9mm}
			\hrulefill\vspace{-2mm}
		\end{figure*}
		
		{\color{black}\section{Summary of NRF-based Control}\label{app:NRF}\vspace{-3mm}
			
			We include here several key definitions and properties associated with the class of control laws from \cite{NRF}.\vspace{-3mm}
			
			Fundamentally, the concept of an NRF-based representation arises when attempting to represent the TFM of a centralised controller, denoted henceforth as $\mathbf{K}(z)$ and assumed to be of dimension $n_u\times n_x$, via the expression\vspace{-2mm}
			\begin{equation*}
				\mathbf{K}(z) = (I_{n_u}-\mathbf{\Phi}(z))^{-1}\mathbf{\Gamma}(z),\vspace{-2mm}
			\end{equation*}
			for which $\mathrm{det}(I_{n_u}-\mathbf{\Phi}(z))\not\equiv 0$ and $\mathrm{elm}_{ii}({\Phi}(z))\equiv 0$, for all $i\in\{1:n_u\}$. Therefore, the pair $(\mathbf{\Phi}(z),\mathbf{\Gamma}(z))$ enables\newpage\noindent the computation of the signal $u[k]$ in \eqref{eq:ss_a}-\eqref{eq:ss_c} as follows\vspace{-2mm}
			\begin{equation}\label{eq:NRF_general}
				u[k] = \mathbf{\Phi}(z)\star u[k]+\mathbf{\Gamma}(z)\star x[k].\vspace{-2mm}
			\end{equation}
			Since the diagonal elements of $\mathbf{\Phi}(z)$ are all \emph{identically zero} this enables a distinct feedback-feedforward expression of the resulting control laws, in which an arbitrary command signal $u_i[k]$ may be computed via the use of a feedback component $x[k]$ alongside the feedforward of the command signals $u_j[k]$ belonging to other nodes in the network ($j\neq i$). It is precisely this feature which, when expressed via \eqref{eq:NRF_general}, yields the identities from \eqref{eq:uf_implem}.\vspace{-3mm}
			
			In the NRF framework, the problem of imposing sparsity on $\mathbf{K}(z)$ for a distributed implementation (a highly challenging objective, see \cite{NRF} and \cite{aug_sparse}) is relaxed into ensuring sparsity for its NRF pair $(\mathbf{\Phi}(z),\mathbf{\Gamma}(z))$. The latter goal is far more tractable, due to the manner in which the pair may be computed. To this end, we now introduce the notion of \emph{Doubly Coprime Factorisation} (DCF) over a set $\mathbb{C}_g$, which can be employed as a design parameter.\vspace{-3mm}
			
			\begin{remark}\label{rem:NRF_stab}
				The choice of $\mathbb{C}_g$ underlines several key properties of the resulting NRF-based control laws. For example, choosing\, $\mathbb{C}_g\subseteq\{z\in\mathbb{C}\,\vert\,|z|\hspace{-0.5mm}<\hspace{-0.5mm}1\}$ ensures that the effect of \emph{bounded} exogenous disturbance on the signals arising in closed-loop configuration \emph{remains bounded}.\vspace{-3mm}
			\end{remark}
			
			With respect to the realisation from \eqref{eq:ss_a}-\eqref{eq:ss_c}, we define $\mathbf{G}_u(z):=(zI_{n_x}-A)^{-1}B_u$ and we say that any collection of eight TFMs $(\mathbf{N}(z),$ $\mathbf{M}(z),\mathbf{X}(z),\mathbf{Y}(z),\widetilde{\mathbf{N}}(z),$ $\widetilde{\mathbf{M}}(z),\widetilde{\mathbf{X}}(z),\widetilde{\mathbf{Y}}(z))$ forms a DCF over $\mathbb{C}_g$ of $\mathbf{G}_u(z)$ if:\vspace{-3mm}
			\begin{enumerate}
				\item[a)] $\mathbf{M}(z)$ along with $\widetilde{\mathbf{M}}(z)$ are invertible and they satisfy $\mathbf{G}_u(z)=\mathbf{N}(z)\mathbf{M}^{-1}(z)=\widetilde{\mathbf{M}}^{-1}(z)\widetilde{\mathbf{N}}(z)$;\medskip
				
				\item[b)] all eight TFMs are proper (taking $|z|\rightarrow\infty$ results in a matrix having only finite entries) and have no poles outside of $\mathbb{C}_g$;\medskip
				
				\item[c)] the eight TFMs satisfy the (B\' ezout-like) identity\vspace{-2mm}
				\begin{equation*}
					\scriptsize\begin{bmatrix}
						\phantom{-}\widetilde{\mathbf{Y}}(z) & -\widetilde{\mathbf{X}}(z)\\
						-\widetilde{\mathbf{N}}(z) & \phantom{-}\widetilde{\mathbf{M}}(z)
					\end{bmatrix}
					\begin{bmatrix}
						\mathbf{M}(z) & \mathbf{X}(z) \\\mathbf{N}(z) & \mathbf{Y}(z)
					\end{bmatrix}=\begin{bmatrix}
						I_{n_u} & O \\ O & I_{n_x}
					\end{bmatrix}.\normalsize\vspace{-2mm}
				\end{equation*}
			\end{enumerate}
			Using just such a DCF over $\mathbb{C}_g$, it is possible to obtain a proper (and, thus, implementable) NRF pair for a controller of the system in \eqref{eq:ss_a}-\eqref{eq:ss_c} as follows\vspace{-2mm}
			\begin{equation*}
				\hspace{-2mm}\left\{\begin{aligned}
					\widetilde{\mathbf{Y}}_{\mathbf{Q}}(z):=&\ \widetilde{\mathbf{Y}}(z)+{\mathbf{Q}}(z)\widetilde{\mathbf{N}}(z),\\
					\widetilde{\mathbf{X}}_{\mathbf{Q}}(z):=&\ \widetilde{\mathbf{X}}(z)+{\mathbf{Q}}(z)\widetilde{\mathbf{M}}(z),\\
					\widetilde{\mathbf{Y}}_{\mathbf{Q}}^{\mathrm{diag}}(z):=&\ \scriptsize\begin{array}{l}
						\mathrm{diag}\left(\mathrm{elm}_{11}\left(\widetilde{\mathbf{Y}}_{\mathbf{Q}}(z)\right),\dots,\mathrm{elm}_{n_un_u}\left(\widetilde{\mathbf{Y}}_{\mathbf{Q}}(z)\right)\right)
					\end{array},\\
					\mathbf{\Phi}(z):=&\ I_{n_x}-\left(\widetilde{\mathbf{Y}}_{\mathbf{Q}}^{\mathrm{diag}}(z)\right)^{-1}\widetilde{\mathbf{Y}}_{\mathbf{Q}}(z),\\
					\mathbf{\Gamma}(z):=&\ \left(\widetilde{\mathbf{Y}}_{\mathbf{Q}}^{\mathrm{diag}}(z)\right)^{-1}\widetilde{\mathbf{X}}_{\mathbf{Q}}(z).
				\end{aligned}\right.\vspace{-2mm}
			\end{equation*}
			for a TFM $\mathbf{Q}(z)$ that is proper, that has no poles outside of $\mathbb{C}_g$, and which ensures that $\widetilde{\mathbf{Y}}_{\mathbf{Q}}(z)$ and $\widetilde{\mathbf{Y}}_{\mathbf{Q}}^{\mathrm{diag}}(z)$ have proper inverses. For systems of type \eqref{eq:ss_a}-\eqref{eq:ss_c}, the DCF over $\mathbb{C}_g$ can always be chosen so that such a $\mathbf{Q}(z)$ exists.\vspace{-3mm}\newpage
			
			One of the main features of this representation is the fact that, by computing the NRF pair as above, its two terms \emph{inherit the sparsity patterns} of $\widetilde{\mathbf{Y}}_{\mathbf{Q}}(z)$ and $\widetilde{\mathbf{X}}_{\mathbf{Q}}(z)$. Indeed, since the two TFMs are affine terms of $\mathbf{Q}(z)$, ensuring desired sparsity patterns on these expressions is a particularly tractable problem (see \cite{aug_sparse}, for example). A further benefit of employing this control formalism is the fact that all closed-loop maps from exogenous disturbance to the interconnection's internal signals (recall Remark~\ref{rem:NRF_stab}) are similarly simple expressions of the $\mathbf{Q}(z)$ parameter. For example, the TFM $\mathbf{F}_{\mathbf{Q}}(z)\scriptsize\begin{bmatrix}
				I_{(n_x+n_u)} & O
			\end{bmatrix}^\top$ from \eqref{eq:cl_dyn} can be expressed as follows\vspace{-3mm}
			\begin{equation*}
				\mathbf{F}_{\mathbf{Q}}(z)\scriptsize\begin{bmatrix}
					I_{(n_x+n_u)} \\ O
				\end{bmatrix} = \left[\begin{array}{cc}
					\mathbf{N}(z)\widetilde{\mathbf{X}}_{\mathbf{Q}}(z) &
					\mathbf{N}(z)\widetilde{\mathbf{Y}}_{\mathbf{Q}}(z)\\
					\mathbf{M}(z)\widetilde{\mathbf{X}}_{\mathbf{Q}}(z) &
					\mathbf{M}(z)\widetilde{\mathbf{Y}}_{\mathbf{Q}}(z)-I_{n_u}
				\end{array}\right],\vspace{-3mm}
			\end{equation*}
			with the other TFMs from \eqref{eq:cl_dyn} being given explicitly in the main result of our companion paper \cite{part1}. Consequently, systemic norm-based procedures (such as the ones proposed in \cite{aug_sparse}) can be used to manipulate these amenable expressions and to dynamically decouple the areas of the network from \eqref{eq:ss_a}-\eqref{eq:ss_c}, as per Remark~\ref{rem:decup}.\vspace{-3mm}
			
		}

		\section{Proofs of the Main Results}\label{app:proofs}\vspace{-3mm}

		We begin by proving the result, which enables us to constrain the state variables of the first-layer subcontrollers.\vspace{-3mm}
		
		\textbf{Proof of Proposition~\ref{prop:NRF_state_manip}}\vspace{-3mm}
		
		The result follows by straightforward algebraic manipulations of the state-space dynamics given in \eqref{eq:area_NRF}. For the sake of clarity, we write these equations out in an explicit manner, and we obtain the identities from \eqref{eq:NRF_state_obs_a}-\eqref{eq:NRF_state_obs_c}, which are located at the top of this page.\vspace{-3mm}
		
		First, we tackle the inclusions showcased in point i). Assumption~A2) implies the fact that $\mathrm{row}_1(w_{r\ell}[k_0])\in\mathcal{W}_{\ell 1}$ and, due to Assumption~A1), using \eqref{eq:NRF_state_sets_a} and \eqref{eq:NRF_state_obs_b} yields the inclusion $\mathrm{row}_{n_{r\ell}}(w_{r\ell}[k_0+1])\in\mathcal{W}_{\ell n_{r\ell}}$. Moreover, by employing the same arguments for \eqref{eq:NRF_state_sets_b} and by effecting successive substitutions in \eqref{eq:NRF_state_obs_a}, we retrieve the fact that $\mathrm{row}_{j}(w_{r\ell}[k_0+1])\in\mathcal{W}_{\ell j}$, for all $j\in\{2:n_{r\ell}-1\}$. To get that $\mathrm{row}_1(w_{r\ell}[k_0+1])\in\mathcal{W}_{\ell 1}$, we simply use Assumption~A3) in conjunction with \eqref{eq:NRF_state_obs_c}. To retrieve point ii), recall Assumption~A2) and then use \eqref{eq:NRF_state_obs_c} along with \eqref{eq:Wi_sets} to get $u_{fi}[k_0]\in{\mathcal{U}}_{fi}$. The desired inclusion follows from Assumption~A1), \eqref{eq:app_cmd} and \eqref{eq:NRF_cmd_sets}.\vspace{-3mm}
		
		Moving on, we now prove the result which serves as the cornerstone of our constraint management technique.\vspace{-3mm}
		
		\begin{figure*}
			\begin{subequations}
				\scriptsize\begin{align}\nonumber
					x_i[k_0+1] =&\left(\left(C_{xi}\xi_i[k_0+1]+ \widetilde\theta_{x i}[k_0+1]\right)+ \left(\psi_{x i}[k_0+1]  + \widetilde\delta_{x i}[k_0+1]\right)\right)+(-\eta_{xi}[k_0+1]),\\&\qquad\qquad\qquad\qquad\qquad\qquad\psi_{x i}[k_0+1]\in\Psi_{xi1},\,\widetilde\theta_{x i}[k_0+1]\in\widetilde\Theta_{xi1},\,\widetilde\delta_{x i}[k_0+1]\in\widetilde\Delta_{xi1},\,\eta_{x i}[k_0+1]\in\mathcal{H}_{xi1},\label{eq:next_step_tilde_a}\tag{B.4a}\\\nonumber
					\mathrm{row}_j(u_{fi}[k_0+1]) =&\left(\left(\mathrm{row}_j(C_{ui})\xi_i[k_0+1]+ \mathrm{row}_j\left(\widetilde\theta_{u i}[k_0+1]\right)\right)\hspace{-0.5mm}+\hspace{-0.5mm} \left(\mathrm{row}_j(\psi_{u i}[k_0+1])  \hspace{-0.5mm}+\hspace{-0.5mm} \mathrm{row}_j\left(\widetilde\delta_{u i}[k_0+1]\right)\right)\right)+(-\mathrm{row}_j(\eta_{u i}[k_0+1])),\\\nonumber
					&\quad\ \mathrm{row}_j(\psi_{u i}[k_0+1])\in \left(e_j^\top\Psi_{ui1}\right),\,\mathrm{row}_j\left(\widetilde\theta_{u i}[k_0+1]\right)\in\left(e_j^\top\widetilde\Theta_{ui1}\right),\,\mathrm{row}_j\left(\widetilde\delta_{u i}[k_0+1]\right)\in\left(e_j^\top\widetilde\Delta_{ui1}\right),\\
					&\quad\ \mathrm{row}_j(\eta_{u i}[k_0+1])\in\left(e_j^\top\mathcal{H}_{ui1}\right),\forall\,j\in\{1:n_{ui}\}.\label{eq:next_step_tilde_b}\tag{B.4b}
				\end{align}\normalsize
			\end{subequations}\phantom{ }
			
			\vspace{-9mm}
			\hrulefill\vspace{-2mm}
		\end{figure*}
		
		\textbf{Proof of Proposition~\ref{prop:one_step_feas}}\vspace{-3mm}

		The proof is constructive in nature and, to begin with, notice that Assumption~A2) enables the local computation of the second-layer command pairs $u_{s1i}[k_0]\in\mathcal{U}_{s1i}$ and $u_{s2i}[k_0]\in\mathcal{U}_{s2i}$, for each $i\in\{1:N\}$, as in \textbf{Step~3} of Algorithm~\ref{alg:MPC_implem}. Since the algebraic conditions from \eqref{eq:dist_implem} are satisfied by each of the first layer's subcontrollers, then there is no need to pass on the values of $u_{s1j}[k_0]$, for any $j\in\{1:N\}\setminus\,\mathcal{N}_i$, to the $i^\text{th}$ subcontroller from the first layer. Therefore, it follows that by communicating and by applying these partial command signals which originate from the second layer, as indicated in \textbf{Steps~4a}~and~\textbf{4b} of Algorithm~\ref{alg:MPC_implem}, the implemented command scheme is equivalent to the one shown in Figure~\ref{fig:NRF_implem}. Consequently, the implementation described in Algorithm~\ref{alg:MPC_implem} ensures the fact that the closed-loop dynamics of our control scheme's first layer are those in \eqref{eq:disc_cl_dyn_a}-\eqref{eq:disc_cl_dyn_d}.\vspace{-3mm}
		
		By applying the above-mentioned commands, the $i^\text{th}$ area's state and its first-layer subcontroller's outputs at time $k=k_0+1$ will be given by the identities showcased in \eqref{eq:next_step_a}-\eqref{eq:next_step_b}, located at the top of this page, where\vspace{-3mm}\stepcounter{equation}\stepcounter{equation}
		\begin{equation}\label{eq:xi_next_step}
			\xi_i[k_0+1] = B_{s1i} u_{s1i}[k_0]+B_{s2i} u_{s2i}[k_0],\vspace{-3mm}
		\end{equation}
		for all $i\in\{1:N\}$. Allowing for the state measurements received by each second-layer subcontroller to be subject, at $k=k_0$, to a disturbance of the type described in \eqref{eq:noise_init}, it follows due to Assumption~A1) and to \eqref{eq:Wi_suf_sets} that these measurements can be expressed as in \eqref{eq:pert_ci_sets_a}-\eqref{eq:pert_ci_sets_b}.\vspace{-3mm}

		By employing now the identities from \eqref{eq:noise_pert_1}-\eqref{eq:noise_pert_2} and \eqref{eq:noise_pert_3}, it follows that the variables described in \eqref{eq:next_step_a}-\eqref{eq:next_step_b} can be rewritten as in  \eqref{eq:next_step_tilde_a}-\eqref{eq:next_step_tilde_b}, located at the top of the next page. However, since $u_{s1i}[k_0]$ and $u_{s2i}[k_0]$ are obtained by solving \eqref{eq:area_prob}, it follows that the vectors located on the left-hand side of \eqref{eq:xi_next_step} must satisfy\vspace{-3mm}
		\begin{equation*}
			\xi_i[k_0+1]\in\Xi_{i1},\,\forall\,i\in\{1:N\}.\vspace{-3mm}
		\end{equation*}
		By considering the expressions of the sets defined in \eqref{eq:inner_appr_1}-\eqref{eq:state_sets}, it follows by simple algebraic manipulations that\vspace{-3mm}
		\begin{equation*}
			\left\{\scriptsize\begin{array}{r}
				\left(C_{xi}\xi_i[k_0+1]+ \widetilde\theta_{x i}[k_0+1]\right)
				\in\mathcal{P}_{1i1},\\
				\left(\mathrm{row}_j(C_{ui})\xi_i[k_0+1]+ \mathrm{row}_j\left(\widetilde\theta_{u i}[k_0+1]\right)\right)
				\in\mathcal{P}_{2(j+\alpha_{ui}) 1},\end{array}\right.\vspace{-3mm}
		\end{equation*}
		for all $i\in\{1:N\}$, along with all $j\in\{1:n_{ui}\}$. Moreover, by using the identities which form \eqref{eq:next_step_tilde_a}-\eqref{eq:next_step_tilde_b} in tandem with the set inclusions from \eqref{eq:inner_appr_1}-\eqref{eq:inner_appr_2} and with the set definitions introduced in \eqref{eq:tilde_sets} and \eqref{eq:Wi_tilde_sets}, we get that $x_i[k_0+1]\in\mathcal{X}_i$ and that $\mathrm{row}_j(u_{fi}[k_0+1])\in\mathcal{W}_{(j+\alpha_{ui})1}$ for all $i\in\{1:N\}$ and all $j\in\{1:n_{ui}\}$.\vspace{-3mm}
		
		The first conclusion we can draw, at this point, is the fact that $x[k_0+1]\in\mathcal{X}$, by virtue of \eqref{eq:set_equiv}. Now, to retrieve the remaining inclusions given in our result's statement, notice that all of the assumptions associated with Proposition~\ref{prop:NRF_state_manip} hold. Thus, by applying said result, we get that $\mathrm{row}_j(w_{r\ell}[k_0+1])\in\mathcal{W}_{\ell j},$ for all $\,i\in\{1:N\},\,\ell\in\mathcal{A}_{ui}$ and $\,j\in\{1:n_{r\ell}\}$, along with the fact that $u_i[k_0]\in\mathcal{U}_i,\,\forall\,i\in\{1:N\}$. Finally, the latter inclusions yield the fact that $u[k_0]\in\mathcal{U}$, by way of \eqref{eq:set_equiv}.\vspace{-3mm}
		
		We now proceed to employ the arguments made in the proof of Proposition~\ref{prop:one_step_feas} to obtain guarantees of recursive feasibility for our architecture's second layer.\vspace{-3mm}

		\textbf{Proof of Theorem~\ref{thm:ongoing_feas}}\vspace{-3mm}
		
		The central idea, which underpins the entire proof of this result, is to show that all of the problems described in \eqref{eq:area_prob} are guaranteed to be feasible at an arbitrary time instant $k_0$, and considering some prediction horizons $T_i>0$ and $\overline T_i\geq0$. Once these guarantees are obtained, the proof boils down to a series of induction-based arguments in which Proposition~\ref{prop:one_step_feas} is repeatedly applied to ensure the sought-after inclusions from the result's statement.\vspace{-3mm}
		
		{\color{black}
			We begin by choosing an arbitrary index $i\in\{1:N\}$, and by investigating its associated area of the network. Due to Assumption~A2), we have that \eqref{eq:feas_sets_include} holds for some $\rho_i\geq 1$. Thus, for any $\widetilde\theta_{it}\in\widetilde{\Theta}_{it}$ and any $t\in\{1:\rho_i\}$, there must exist $p_{it}\in\widetilde{\mathcal{P}}_{it}$, $u_{1it}\in\mathcal{U}_{s1i}$ along with $u_{2it}\in\mathcal{U}_{s2i}$ such that the following identities hold:\vspace{-3mm}\stepcounter{equation}
			\begin{equation}\label{eq:feas_exp_1}
				\begin{array}{l}
					\widetilde\theta_{it} = \left(\begin{bmatrix}
						-C_{xi}\\-C_{ui}
					\end{bmatrix}\displaystyle\sum_{j=1}^t A_{si}^{j-1}(B_{s1i}u_{1ij}+B_{s2i}u_{1ij})\right)+{p}_{it}.
				\end{array}
		\end{equation}}\noindent
		Now, by exploiting \eqref{eq:feas_sets_def_1} and \eqref{eq:feas_sets_def_2}, it becomes possible to rewrite \eqref{eq:feas_exp_1} in a more explicit manner, which yields\vspace{-3mm}
		\begin{equation}\label{eq:feas_exp_2}
			\left\{
			\scriptsize\begin{array}{l}
				\widetilde\theta_{xit}+C_{xi}\displaystyle\sum_{j=1}^t A_{si}^{(j-1)}\scriptsize\begin{bmatrix}
					B_{s1i} & B_{s2i}
				\end{bmatrix}\hspace{-1mm}\scriptsize\begin{bmatrix}
					u_{1ij} \\ u_{2ij}
				\end{bmatrix}\hspace{-1mm} = {p}_{xit},\\
				\widetilde\theta_{u\ell t}+ \mathrm{row}_{(\ell-\alpha_{ui})}(C_{ui})\displaystyle\sum_{j=1}^t A_{si}^{(j-1)}\scriptsize\begin{bmatrix}
					B_{s1i} & B_{s2i}
				\end{bmatrix}\hspace{-1mm}\scriptsize\begin{bmatrix}
					u_{1ij} \\ u_{2ij}
				\end{bmatrix}\hspace{-1mm} = {p}_{u\ell t},
			\end{array}
			\right.\vspace{1mm}
		\end{equation}
		for all $t\in\{1:\rho_i\}$ and all $\ell\in\mathcal{A}_{ui}$, where $\widetilde\theta_{xit}\in\widetilde{\Theta}_{xit}$, $\widetilde\theta_{u\ell t}\in \left(e_{(\ell-\alpha_{ui})}^\top\widetilde{\Theta}_{uit}\right)$, ${p}_{xit}\in\mathcal{P}_{1it}$ and ${p}_{u\ell t}\in\mathcal{P}_{2\ell t}$.
		We now employ the expressions deduced in \eqref{eq:feas_exp_2} to prove the recursive feasibility of the problems stated in \eqref{eq:area_prob}.\vspace{-3mm} 
		
		To do so, choose any desired initial time instant $k_0$ for the optimisation problem associated with the $i^\text{th}$ area index. For all $t\in\{1:\rho_i\}$, consider any $\widetilde{\theta}_{xi}[k_0+t]\in\widetilde{\Theta}_{xit}$ and any $\widetilde{\theta}_{ui}[k_0+t]\in\widetilde{\Theta}_{uit}$ which may arise in the context of the aforementioned optimisation problem, to define $\widetilde\theta_{xit}:=\widetilde{\theta}_{xi}[k_0+t]$ and $\widetilde\theta_{u\ell t}:=\mathrm{row}_{(\ell-\alpha_{ui})}\left(\widetilde{\theta}_{ui}[k_0+t]\right)$, where $\ell\in\mathcal{A}_{ui}$. Due to the arguments made in the previous paragraph of this proof, it follows that there must exist some appropriately constrained vectors denoted $u_{1it}$ and $u_{2it}$ along with $p_{xit}$ and $p_{u\ell t}$, with $t\in\{1:\rho_i\}$ and $\ell\in\mathcal{A}_{ui}$, such that the identities from \eqref{eq:feas_exp_2} hold. Thus, by setting $T_i=\rho_i$ and by selecting an arbitrary $\overline T_i\in\mathbb{N}$, it is always possible to choose the command signal vectors appearing in \eqref{eq:area_prob} as $u_{s1i}[k_0+t-1]:=u_{1ij}\in\mathcal{U}_{s1i}$ and as $u_{s2i}[k_0+t-1]:=u_{2ij}\in\mathcal{U}_{s2i}$, for all $t\in\{1:T_i\}$.\vspace{-3mm}
		
		A direct consequence of all these choices is the fact that\vspace{-4mm}
		\begin{equation}\label{eq:f_resp}
			\xi_i[k_0+t]=\sum_{j=1}^t A_{si}^{(j-1)}(B_{s1i}u_{1ij}+B_{s2i}u_{2ij}),\vspace{-4mm}
		\end{equation}
		for all $t\in\{1:T_i\}$. When employed in tandem with \eqref{eq:feas_exp_2}, the identities obtained in \eqref{eq:f_resp} yield the fact that\vspace{-2mm}
		\begin{equation}\label{eq:equiv_incl}
			\left\{
			\scriptsize\begin{array}{r}
				C_{xi}\xi_i[k_0+t]+\widetilde{\theta}_{xi}[k_0+t]\in\mathcal{P}_{1it},\\
				\mathrm{row}_{(\ell-\alpha_{ui})}(C_{ui})\xi_i[k_0+t]+\mathrm{row}_{(\ell-\alpha_{ui})}\left(\widetilde{\theta}_{xi}[k_0+t]\right)\in\mathcal{P}_{2\ell t},
			\end{array}
			\right.
		\end{equation}
		for all $t\in\{1:T_i\}$ and all $\ell\in\mathcal{A}_{ui}$. It follows by direct inspection of the inequalities used to define the sets introduced in \eqref{eq:inner_appr_1}-\eqref{eq:state_sets} that the inclusions obtained in \eqref{eq:equiv_incl} are equivalent to $\xi_i[k_0+t]\in\Xi_{it}$, for all $t\in\{1:T_i\}$.\vspace{-3mm}
		
		At this point in the proof, we have shown that by setting $T_i=\rho_i$ and by choosing an arbitrary $\overline T_i\in\mathbb{N}$, there exists at least one sequence of command vectors which satisfies the constraints of the problems formulated in \eqref{eq:area_prob}, thus rendering them feasible. Moreover, recall that the time instant $k_0$ and the area index $i\in\{1:N\}$ were chosen arbitrarily at the beginning of the proof, thus making the feasibility of all the aforementioned problems a generic property. By recalling now Assumption~A1), we employ Proposition~\ref{prop:one_step_feas} to compute the second-layer command values at time instant $k=k_0$ and, by iterating the procedure detailed in Algorithm~\ref{alg:MPC_implem}, we arrive at time instant $k=k_0+1$, in which Assumption~A1) is still valid and in which all $N$ local optimisation problems remain feasible. It follows that this procedure may be successfully repeated at any subsequent instant $k\geq k_0$, thus guaranteeing recursive feasibility, \emph{i.e.}, point i) of our result, while also ensuring the satisfaction of the constraints which make up point ii) of our result.
		
	\end{document}